\documentclass[10pt,journal,twoside]{IEEEtran}
\usepackage[utf8]{inputenc}
\usepackage{amsmath,amssymb,amsthm}
\usepackage{amsfonts}

\usepackage{subfiles}
\usepackage{blindtext}

\usepackage{mathtools}
\usepackage{color}
\usepackage[nohead,margin=1.0in]{geometry}
\usepackage{booktabs}
\usepackage{graphicx,caption,subcaption}
\usepackage{algorithm,algorithmic}
\usepackage{verbatim}
\usepackage{url}
\usepackage{grffile}
\usepackage{float}
\usepackage{diagbox}
\usepackage[table]{xcolor}
\definecolor{lightgray}{gray}{0.9}
\usepackage{color, colortbl}
\usepackage{placeins}
\usepackage{rotating}

\theoremstyle{plain}

\numberwithin{equation}{section}

\newcommand{\specialcell}[2][c]{%
  \begin{tabular}[#1]{@{}c@{}}#2\end{tabular}}

\newcommand{\N}{\mathbb{N}}
\newcommand{\R}{\mathbb{R}}

\renewcommand{\d}{\mathrm{d}}

\title{Deep image prior for 3D magnetic particle imaging: A quantitative comparison of regularization techniques on \texttt{Open MPI dataset}}
\author{S\"oren Dittmer, Tobias Kluth, Mads Thorstein Roar Henriksen and Peter Maass 
\thanks{Manuscript received May 8th, 2020.}%
\thanks{S. Dittmer, T. Kluth, and P. Maass are with Center for Industrial Mathematics, University of Bremen, Bibliothekstr. 5, 28357 Bremen, Germany ($\{$\texttt{sdittmer}, \texttt{tkluth}, \texttt{pmaass} $\}$\texttt{@math.uni-bremen.de}).
S. Dittmer and T. Kluth acknowledge funding by the Deutsche Forschungsgemeinschaft
(DFG, German Research Foundation) - project number 281474342/GRK2224/1 ``Pi$^3$ : Parameter Identification
- Analysis, Algorithms, Applications'' and T. Kluth further acknowledges support by the project ``MPI$^2$'' funded by the Federal Ministry of Education and Research (BMBF, project no. 05M16LBA). }%
}

\begin{document}

\maketitle

\begin{abstract}
Magnetic particle imaging (MPI) is an imaging modality exploiting the nonlinear magnetization behavior of (super-)paramagnetic nanoparticles to obtain a space- and often also time-dependent concentration of a tracer consisting of these nanoparticles. MPI has a continuously increasing number of potential medical applications. One prerequisite for successful performance in these applications is a proper solution to the image reconstruction problem. More classical methods from inverse problems theory, as well as novel approaches from the field of machine learning, have the potential to deliver high-quality reconstructions in MPI. We investigate a novel reconstruction approach based on a deep image prior, which builds on representing the solution by a deep neural network. Novel approaches, as well as variational and iterative regularization techniques, are compared quantitatively in terms of peak signal-to-noise ratios and structural similarity indices on the publicly available \texttt{Open MPI dataset}.
\end{abstract}
\begin{IEEEkeywords}
magnetic particle imaging, deep image prior, iterative/variational regularization, image quality 
\end{IEEEkeywords}

\section{Introduction}
The imaging modality magnetic particle imaging (MPI) was invented by Gleich and Weizenecker in 2005 ~\cite{gleich2005tomographic}. The goal of the technique is to reconstruct a concentration of (super-)paramagnetic iron oxide nanoparticles by applying a dynamic magnetic field. 
MPI benefits from high temporal resolution and potentially high spatial resolution which make it suitable for several {\it in-vivo} applications like
imaging blood flow \cite{weizenecker2009three}, long-term monitoring by utilizing a circulating tracer \cite{khandhar2017evaluation}, flow estimation \cite{franke2017}, tracking/guiding medical instruments \cite{haegele2012magnetic}, cancer detection \cite{Yu2017}, and cancer treatment by hyperthermia \cite{murase2015usefulness}. 
Moreover, the list of potential medical applications exploiting MPI is still growing, e.g.,  
the recent prototype development of a human-sized scanner suitable for scanning a human's head \cite{graeser2019human} illustrates the potential of MPI for cerebral applications like stroke detection and monitoring \cite{Ludewig2017}. 
For further technical background information on MPI, we refer to \cite{knopp2012magnetic,knopp2017magnetic,kluth2018mathematical}.

A critical prerequisite for potential medical applications is a proper solution to the image reconstruction problem in MPI, which aims for the determination of the space- and in some instances time-dependent tracer concentration.  
Neglecting concentration-dependent effects \cite{loewa2016}, likely to be caused by particle-particle interactions, the reconstruction is a linear ill-posed inverse problem \cite{Kluth2018b}, which for a given system matrix is typically solved by applying Tikhonov regularization in combination with the algebraic reconstruction technique and a nonnegativity constraint \cite{weizenecker2009three, Knopp2010e}.
An alternative reconstruction approach for MPI is a direct inversion method exploiting Cartesian excitation patterns, the so-called x-space reconstruction \cite{Goodwill2010,Goodwill2011a}.
More generally, reconstruction methods taking into account further prior information, e.g., fused lasso regularization and directional total variation, have been applied to experimental \cite{storath2016edge} and simulated data \cite{Bathke2017}.
More sophisticated approaches explicitly taking into account operator uncertainty have been proposed by using a total-least-squares approach combined with standard Tikhonov regularization as well as a sparsity-promoting penalty term \cite{Kluth2017}.
Further efforts to improve the reconstruction in terms of accuracy and efficiency have been made, e.g., low-rank approximations and whitening \cite{KluthJin2019a} motivated by the non-polynomial decay of singular values \cite{Kluth2018b}, problem reduction via postprocessing \cite{Knopp_Online_2016}, introducing an $\ell^1$-data fidelity term \cite{KluthJin2019b_preprint}, and many more (see \cite{knopp2017magnetic} for an extensive overview).

Solving inverse problems by using techniques from the field of machine learning has received increased interest during the last years, and the number of their successful applications is continuously increasing \cite{arridge2019}.
One specific class of methods, based on the training of deep neural network architectures, is often summarized by the term deep learning-based methods.  Image reconstruction specific examples for such methods include fully learned approaches~\cite{zhu2018image,he2019radon}, unrolled iterative algorithms~\cite{adler2017solving,adler2018learned}, learned penalty terms~\cite{lunz2018adversarial}, or postprocessing of classical reconstructions~\cite{chen2017low,jin2017deep,yang2018low}. All of these methods have in common that they rely on training data. Often this training data is required in the form of tuples consisting of a given ground truth reconstruction and a corresponding measurement. In general, and also for MPI, this reliance on data leads to a chicken-and-egg type problem, since one relies on the assumption that sufficient ground truth data is available.
In contrast, a deep image prior (DIP) approach \cite{DittmerKluthMaassOtero2019} exploiting generative neural networks has been proposed recently for solving inverse problems in general. The DIP~\cite{ulyanov2017dip} is a novel regularization technique based on untrained neural network architectures. The basic idea is to use a neural network architecture/setting, that implicitly encodes a prior which encourages plausible reconstructions and relies on one single measurement only.

In the context of reconstruction methods for MPI, the investigation in the present paper is twofold, i.e.,
\begin{enumerate}
    \item we introduce a novel three-dimensional deep image prior (DIP) for MPI, and
    \item we compare various numerical reconstruction methods quantitatively as well as qualitatively in the context of the \texttt{Open MPI dataset}~\cite{knopp2020openmpidata}.
\end{enumerate}
The subsequent part of the paper is structured as follows:
In Section \ref{sec:methods}, we provide a description of the used system matrix approach in MPI, variational reconstruction methods, a description of DIP, and the experimental data set. 
In Section \ref{sec:results}, we continue with numerical results, including a quantitative comparison between DIP and classical methods, including variational and iterative approaches to image reconstruction.
We conclude with a discussion in Section \ref{sec:discussion}.

\section{Methods}
\label{sec:methods}
\subsection{Magnetic particle imaging and preprocessing chain for derivation of the linear system of equations}
\label{sec:mpi}
In the following, we sketch the system matrix approach commonly used in MPI, see also \cite{Knopp2017} for further reading. We begin with some basic notation. Let $\Omega \subset \R^3$ be a bounded domain. Further, let $T>0$ denote the maximal data acquisition time and $I:=(0,T)$ the time interval during which the measurement process takes place. 

The measured voltage signals $v_\ell : I \rightarrow \R$, $\ell=1,\hdots,L$, obtained at $L\in \N$ receive coil units, is given by a superposition of a signal $v_{\mathrm{P},\ell}$ caused by the particles and the direct feedthrough $v_{\mathrm{E},\ell}$ (background signal) caused mainly by the applied magnetic field. 
The inverse problem is thus to find the concentration $c:\Omega\to \mathbb{R}^+\cup\{0\}$ from $\{v_\ell\}_{\ell=1}^L$:
\begin{equation} \label{eq:general-problem}
 v_\ell(t) =  \int_\Omega c(x)  s_\ell(x,t) \d x + v_{\mathrm{E},\ell}  = S_\ell c(t) + v_{\mathrm{E},\ell},
\end{equation}
where $S_\ell: L^2(\Omega) \rightarrow L^2(I)$ is the forward operator and where $s_\ell \in L^2(\Omega \times I)$ is the background-corrected system function.

The calibration procedure obtains single measurements from a small "delta" sample at predefined positions $\{x^{(i)}\}_{i=1,\hdots, N} \in \Omega^N$ which builds the basis for the commonly used system matrix approach.
For this purpose let $\Gamma \subset \R^3$ be a reference volume placed at the origin. 
The concentration phantoms are given by $c^{(i)} = c_0 \chi_{x^{(i)}+\Gamma}$ for some reference concentration $c_0>0$. Typical choices for $\Gamma$ are small cubes. 
If $\{x^{(i)}+\Gamma\}_{i=1,\hdots,N}$ form a partition of the domain $\Omega$, the background-corrected measurements $v_{\ell}^{(i)} - v_{0,\ell}^{(i)} =  S_\ell c^{(i)}$, $i=1,\hdots,N$, can then be used directly to characterize the system matrix $S$ for $L$ receive coil units ($v_{0,\ell}^{(i)}$, $\ell =1,\hdots,L$, are background measurements used for system matrix correction). 
For given phantom measurements $v_\ell$, $\ell=1,\hdots,L$, we build the measurement vector $v$ analogously. Both are then given by
\begin{equation}\label{eq:SM_freq}
 S= \frac{1}{c_0} \left[ \begin{array}{c}
     \mathrm{Re}((\langle v_{1}^{(i)} - v_{0,1}^{(i)}, \psi_j \rangle)_{j\in J_1, i=1,\hdots,N }) \\ \hline
          \mathrm{Im}((\langle v_{1}^{(i)} - v_{0,1}^{(i)}, \psi_j \rangle)_{j\in J_1, i=1,\hdots,N }) \\ \hline
\vdots \\ \hline
\mathrm{Re}((\langle v_{L}^{(i)} - v_{0,L}^{(i)} , \psi_j \rangle)_{j\in J_L, i=1,\hdots,N }) \\ \hline
\mathrm{Im}((\langle v_{L}^{(i)} - v_{0,L}^{(i)} , \psi_j \rangle)_{j\in J_L, i=1,\hdots,N })
    \end{array}\right]\in \R^{M\times N},
    \end{equation}
    \begin{equation}
    \quad 
     v= \left[ \begin{array}{c}
     \mathrm{Re}((\langle v_1 , \psi_j \rangle)_{j\in J_1 }) \\ \hline
          \mathrm{Im}((\langle v_1 , \psi_j \rangle)_{j\in J_1 }) \\ \hline
\vdots \\ \hline
\mathrm{Re}((\langle v_L , \psi_j \rangle)_{j\in J_L }) \\ \hline
\mathrm{Im}((\langle v_L , \psi_j \rangle)_{j\in J_L }) 
    \end{array}\right] \in \R^M,
\end{equation}
where $\{ \psi_j \}_{j\in \mathbb{Z}}$ is the Fourier basis of time-periodic signals of $L^2(I)$, i.e., $\psi_j(t)=1/\sqrt{T} (-1)^j e^{i2\pi j t/T}$, $j \in \mathbb{Z}$. For the purpose of preprocessing prior to reconstruction the sets $J_\ell\subset \mathbb{Z}$, $\ell=1,\hdots,L$ are restrictions to certain frequency indices, which also yield $M=2\sum_\ell^L | J_\ell |$.
Two frequency-selection approaches which are commonly combined result in the index sets $J_\ell$, $\ell=1,\hdots,L$: a bandpass approach and SNR-type thresholding with threshold $\tau\geq 0$ (see, for example, \cite{KluthJin2019a} for a more detailed description).

Let $v_0$ be the analogous measurement vector of the direct feedthrough, i.e., of an empty scanner.
Then one obtains a measured signal from the $L$ receive coils by $v^\delta=v+v_0+\eta$ with noise vector $\eta$, $\|\eta\|\leq \delta$.

We thus obtain a linear system of equations
$Sc= v^\delta-v_0.$ 
Also, we include two processing steps, whitening, and low-rank approximation, which have been successfully used to improve reconstruction quality and computation times in MPI \cite{KluthJin2019a}.
The linear equation system is multiplied with a whitening matrix $W$ obtained from the diagonal covariance matrix of multiple background measurements. 
For the low-rank approximation step, let $(\tilde U_K,\tilde \Sigma_K,\tilde V_K)$ be the randomized singular value decomposition (rSVD) for the $K\leq \min(M,N)$ largest singular values of the matrix $WS$.

This leaves us with the linear system
\begin{equation}\label{eq:main_lin_system}
Ac=y^\delta
\end{equation}
with the processed matrix $A=\tilde{U}_K^t W S\in\mathbb{R}^{K\times N}$ and measurements $y^\delta=\tilde{U}_K^t W(v-v_0)\in\mathbb{R}^{K}$ ($\cdot^t$ denotes the transpose matrix). 

In summary, we apply the following preprocessing steps to derive the final system of equations in \eqref{eq:main_lin_system} for a given SNR threshold $\tau$ (note that this is the only parameter in the preprocessing chain which is varied in the subsequent results section):
\begin{enumerate}
    \item frequency selection by bandpass filtering and SNR-type thresholding,
    \item concatenation of multiple receive coil units and splitting real and imaginary part,
    \item weighting by diagonal whitening matrix, and
    \item system reduction by projection on subspace corresponding to $K$ largest singular values of the whitened system matrix. 
\end{enumerate}

\subsection{Classical reconstruction methods}
\label{sec:classical}
Before describing the DIP method in the next subsection, we give a brief description of "classical reconstruction methods" to address the MPI problem which is given by an ill-posed operator equation ("classical" is to be understood in contrast to emerging reconstruction methods from the field of machine/deep learning). 

Given some suitable spaces $X$ and $Y$ commonly assumed to be general Banach or Hilbert spaces, the reconstruction task is to compute a concentration $c\in X$ that agrees with a noisy measurement $y^\delta\in Y$ obtained by the linear measurement operator $A:X\to Y$, i.e., we want to find a ``plausible'' $c$ such that
\begin{equation}
    Ac\approx y^\delta.
\end{equation}
In line with the MPI setup specified in Section \ref{sec:mpi}, we consider the Hilbert spaces $X=\R^N$ and $Y=\R^K$ in the present paper. 

For many inverse problems and also in MPI one follows a variational approach by minimizing a Tikhonov-type functional, i.e., a superposition of data fidelity and additive penalty term, which for the finite-dimensional system in the present work is given by 
\begin{equation} \label{eq:variational_problem}
    J_\lambda (c)=\frac1p \| Ac - y^\delta \|_p^p + \lambda R(c),
\end{equation}
where the regularization parameter $\lambda \geq 0$ and penalty term $R:\R^N \rightarrow \R_+$ include prior knowledge on the solution to obtain a stable reconstruction from noisy measurements.
Common values for $p$ are $1$ and $2$. Well-studied choices for $R$ are, for example, $\frac{1}{2}\|\cdot\|_2^2$, $\|\cdot\|_1$, and total-variation terms TV~\cite{rudin1992nonlinear}.
The functionals are then minimized using suitable optimization techniques like, projected gradient descent type methods or incremental gradient descent method -- which also include the Kaczmarz-type methods \cite{kaczmarz1937angenaherte,dax1993row} preferably used in MPI.
In the latter case, one needs to be aware of the iterative nature of the algorithms, which can introduce an additional kind of regularization if a small iteration number is chosen not being sufficient for reaching convergence (early stopping).

\subsection{Deep image prior}
\label{sec:dip}
We will now discuss the idea behind the Deep image prior (DIP). The concept of a DIP was first introduced in~\cite{ulyanov2017dip}, and it has been quickly adopted by fields like compressed sensing~\cite{van2018compressed} and inverse problems ~\cite{DittmerKluthMaassOtero2019}.
The core idea is to use the architecture of an untrained neural network $\varphi_\theta: \R^Z \rightarrow \R^N$ as a regularization for the reconstruction and to determine the network parameters $\theta \in \R^Q$ for a random but fixed input $z\in \R^Z$. 
More formally, in DIP one tries to minimize the data fidelity
\begin{equation}
    \label{eq:dip_functional}
    J_\mathrm{DIP}(\theta)= \|A\varphi_\theta(z)-y^\delta\|_p^p,
\end{equation}
with respect to $\theta$, where usually $p=2$, but throughout this paper we will use $p=1$. The minimization is done iteratively by Adam~\cite{kingma2014adam}.
This is in contrast to common approaches relying on the minimization of a Tikhonov-type functional as given in \eqref{eq:variational_problem}.
One crucial prerequisite for the DIP is the proper choice of network architecture. While the original DIP uses an architecture for 2D images, we use an architecture suitable for the 3D MPI problem.

As our regularizing architecture, we use an autoencoder based on the architecture of the original DIP paper's ``skip-architecture''\cite{ulyanov2017dip}. In detail, our architecture differs in the following aspects:
\begin{itemize}
    \item We do not use skip connections. This means the U-net~\cite{ronneberger2015u} reduces to the aforementioned autoencoder.
    \item Naturally, we replaced the two-dimensional convolutional layers by three-dimensional ones.
    \item Our final activation function is $\mbox{ReLU}$, not $\mbox{Sigmoid}$. We use $\mbox{ReLU}$ because we know that our particle concentration is non-negative, but do not know an upper bound.
\end{itemize}

We conclude this subsection with a discussion of the relationship between DIP and the variational approach, i.e.,
 \textit{rewriting variational regularizations as deep image priors}:
As demonstrated in~\cite{DittmerKluthMaassOtero2019}, the DIP can be seen as a variational approach to inverse problems by assuming continuous differentiability of the network with respect to its parameters $\theta$ and applying the idea of Lagrange multipliers~\cite{bertsekas2014constrained} to the constrained optimization problem
\begin{equation}
    \min_c \|Ac-y^\delta\|_2^2 \text{ s.t.\ } \|c - \varphi_\theta(z)\|_2^2 = 0.
\end{equation}
We would like to point out that it is also possible to write most of the common regularization functionals $R$ in the context of the variational problem 
\begin{equation}
    \label{eq:variational_problem_l2}
    \min_c \|Ac-y^\delta\|_2^2 + \lambda R(c)
\end{equation}
as a DIP -- although this can require exotic architectures. One way to show this is to assume continuous differentiability of $R$ and utilize Lagrange multipliers to rewrite the Expression~\eqref{eq:variational_problem_l2} as the constrained optimization problem
\begin{equation}
    \label{eq:variational_problem_constrained}
    \min_c \|Ac-y^\delta\|_2^2 \text{ s.t.\ } R(c) = \tau(\lambda).
\end{equation}
If we further assume the existence of an $R$ specific functional $g_{\tau}:\mbox{domain}(R)=\mathbb{R}^Q=\mathbb{R}^N\to\mathbb{R}$ s.t. $R(g_{\tau}(\theta)\theta) = \tau$, we can rewrite \eqref{eq:variational_problem_constrained} in terms of the problem
\begin{equation}
    \label{eq:variational_problem_as_dip}
    \min_{\theta} \|Ag_\tau(\theta)\theta-y^\delta\|_2^2.
\end{equation}
The solution $\theta$ of this problem yields the solution to the original problem in~\eqref{eq:variational_problem_l2} via $g_{\tau}(\theta)\theta$.
The assumption that such an $g_{\tau}$ exists is fulfilled by most common penalty terms in use, like total-variation~\cite{rudin1992nonlinear}, and $\ell^p$-norms which fulfil this property in the form of positive homogeneity. E.g., for a penalty term of the form
\[R(c) = \|c\|_p^p,\]
and $c\ne0$, we have the functional $g_\tau(c) = \frac{\sqrt[p]{\tau}}{\|c\|_p}$ and therefore the DIP
\begin{equation}
    \varphi_\theta(z) = \frac{\sqrt[p]{\tau}}{\|\theta\|_p}\theta.
\end{equation}

\subsection{Experimental data and image quality assessment}
\label{sec:dataset}
For the experimental evaluation we use the 3D \texttt{open MPI dataset} \cite{knopp2020openmpidata} (downloaded
from \url{https://www.tuhh.de/ibi/research/open-mpi-data.html}, last accessed on April 02, 2020) provided
in the MPI Data Format (MDF) \cite{2016arXiv160206072K}. The system matrix is measured using a cuboid sample of size 2 mm $\times$ 2 mm $\times$ 1 mm and a 3D Lissajous-type FFP excitation while obtaining averaged empty scanner measurements every 19 calibration scans. The calibration is carried out with a Perimag tracer having a concentration of 100 mmol/l.
The field-of-view has a size of 38 mm $\times$ 38 mm $\times$ 19 mm, and the sample positions have a distance
of 2 mm in $x$- and $y$-direction and 1 mm in $z$-direction, resulting in $19\times19\times19=6859=N$ voxels. Measurements
are averaged over 1000 repetitions, and with each phantom, an empty measurement with 1000 repetitions is provided.
By taking the mean of the empty measurements provided with the phantom, we obtain $v_{0,\ell}$, $\ell=1,\hdots,L$, respectively $v_0$, in Section \ref{sec:mpi}. For the background removal in the system matrix we used $v_{0,\ell}^{(i)}=v_{0,\ell}$ for any $i=1,\hdots,N$ and, if not mentioned differently, use $K=2000$ for the low-rank approximation step discussed in Section~\ref{sec:mpi}.

We use the ``shape'' and ``resolution'' phantoms provided in the dataset. We state the detailed description for the sake of completeness. It can also be found on at \url{https://www.tuhh.de/ibi/research/open-mpi-data.html} or in \cite{KluthJin2019b_preprint}. The ``shape''
phantom is a cone defined by a 1 mm radius tip, an apex angle of 10 degrees, and a height of 22 mm. The total
volume is 683.9 $\mu$l. Perimag tracer with a concentration of 50 mmol/l is used. See \cite[Figure 3]{KluthJin2019b_preprint}
for a schematic illustration, where the plots are adapted from the \texttt{Open MPI dataset}.
The ``resolution'' phantom consists of 5 tubes filled with Perimag
tracer with a concentration of 50 mmol/l. The 5 tubes have a common origin on one side of the phantom,
and extend in different angles from the origin within the $x$-$y$- and $y$-$z$-planes.
In the $z$-direction, the angles in the $y$-$z$-plane are chosen smaller (10 deg and 15 deg) than in
$x$-$y$-plane (20 deg and 30 deg); see \cite[Figure 4]{KluthJin2019b_preprint} for the illustration. 

For image quality assessment, we exploit the approach in \cite{KluthJin2019b_preprint}, where the authors extracted voxel images from the CAD drawings and included position uncertainty in peak-signal-to-noise-ratio (PSNR) and structural similarity measure (SSIM) \cite{HoreZiou:2010}.
More precisely, a reference image $c_0: \mathbb{R}^3
\to \mathbb{R}_+$ is used in combination with shifts $\Delta r \in
\mathcal{R}$, where $\mathcal{R}$ is the set of all possible position shifts in the neighborhood $[-3\text{mm},3\text{mm}]^3$
with a step size $0.5$mm in each direction ($|\mathcal{R}|=2197$). 
Then the $L^2$-scalar products of
$c_0(r+\Delta r)$ and the piecewise constant basis functions $\{\psi_j\}_j$ with respect to the
voxel grid yields the reference image $x_{\mathrm{ref},\Delta r}\in\R^N$.  Following the approach in \cite{KluthJin2019b_preprint} we define the following conservative image quality measures
\begin{align}
 \epsilon_\mathrm{PSNR}(x)&= \max_{\Delta r \in \mathcal{R}} \mathrm{PSNR}(x,x_{\mathrm{ref},\Delta r})\\
 \epsilon_\mathrm{SSIM}(x)&= \max_{\Delta r \in \mathcal{R}} \mathrm{SSIM}(x,x_{\mathrm{ref},\Delta r})
\end{align}
which are used for quantitative comparison in the following. Here, we use a data range of $100$ for the SSIM computations.

\section{Results}
\label{sec:results}
In the following quantitative and qualitative comparison, we distinguish three classes of reconstruction methods which we evaluate on the two phantoms of the 3D \texttt{Open MPI dataset} described in Section \ref{sec:dataset}:
\begin{enumerate}
    \item \textbf{DIP}: Reconstructions based on the DIP approach minimizing a data fidelity term as described in Section \ref{sec:dip} using Adam with varying numbers of iterations.
    \item \textbf{VAR}: Reconstructions based on the minimization of a Tikhonov-type functional consisting of a data-fidelity and a penalty term as outlined in Section \ref{sec:classical}. Minimization is performed by AMSGrad~\cite{j.2018on}  until convergence is reached such that results rely on the \textit{variational} regularization only.
    \item \textbf{KACZ}: Analogous to VAR, but the minimization realized by a \textit{Kaczmarz}-type algorithm with varying numbers of iterations. Effectively yields reconstructions, which are results of a hybrid regularization, mixing iterative and variational elements.
\end{enumerate}
We will encode our methods based on the class, the respective data fidelity, and the penalty terms, i.e., \texttt{Method class D data fidelity + P penalty}.
All methods rely on a certain number of parameters controlling the degree of regularization. We performed a discrete optimization with respect to these parameters to obtain optimal results in terms of PSNR and SSIM. The methods of interest in this comparison are specified in more detail in the following list:
\begin{itemize}
\item \textbf{DIP D$\ell^1$ + P-}: We minimize the functional \eqref{eq:dip_functional} with $p=1$ only as it led to superior reconstruction performance when compared to the case $p=2$. 
\begin{itemize}
    \item Our fixed random input to the network, $z$, has entries coming from the uniform distribution between $0$ and $0.7$ and has, like the output, the shape $(1, 19, 19, 19)$ where $1$ is the number of channels and $19$ is the edge length of the concentration cube we aim to reconstruct.
    \item The encoder steps down-sample by a factor of $2$ and have $64$, $128$ and $256$ channels respectively. The decoder is symmetric to this.
    \item We minimize the functional for $20000$ iterations with Adam~\cite{kingma2014adam} and for the different learning rates $\alpha_i=10^{-i}$, for $i=3,4,5$, and the standard momenta settings $\beta = (0.9, 0.999)$.
\end{itemize}
For the purpose of parameter optimization we extracted reconstructions after iterations $s\in \{$1, 2, $\hdots$, 10, 12, $\hdots$, 30, 35,  $\hdots$, 50, 60, $\hdots$, 150, 175, $\hdots$, 500, 600, $\hdots$, 2000, 2500, $\hdots$, 5000, 6000, $\hdots$, 20000$\}$.

\item \textbf{KACZ D$\ell^2$ + P$\ell^2$}: This is one of the commonly used techniques in MPI which minimizes the functional $\frac12 \|Ac-y^\delta\|_2^2 + \rho\frac{1}{2} \|c\|_2^2$ with non-negativity constraints by a row action method for inconsistent systems of equations~\cite{herman1980storage} (see \cite[Alg. 1]{KluthJin2019a} or \cite[Sec. 4]{dax1993row} for the detailed algorithm). For the parameter optimization $\rho \in \{ 0.5^{i-1}|i=1,\cdots 40\}=:\mathcal{P}$ were used. Like all the following KACZ reconstructions, we used 500 sweeps (i.e., going 500 times over all rows in order) and consider all reconstructions after the same sweeps/iterations $s$ as for the DIP up to $500$.

\item \textbf{KACZ D$\ell^2$ + P($\ell^1$+$\ell^2$)}: This is an extension of the previous algorithm for inconsistent systems of equations taking into account an additional sparsity-promoting $\ell^1$-penalty term when minimizing the functional $\|Ac-y^\delta\|_2^2 + \rho\frac{1}{2} \|c\|_2^2+ \lambda \|c\|_1 $ with non-negativity constraints. We exploit a splitting method \cite[Sec. 9.4.1]{scherzer2010handbook} first applying one sweep over the matrix as in the previous method followed by applying a soft shrinkage operator before starting a new sweep over the matrix. A similar approach applying the soft shrinkage operator in each row action can be found in \cite{Lorenz2014} for consistent systems. Both regularization parameters are taken from $\mathcal{P}$ as above (i.e., resulting in $40^2$ parameter combinations).

\item \textbf{KACZ D$\ell^2$ + P$\ell^1$}: We consider the previous functional for $\rho=0$, i.e., $\frac12 \|Ac-y^\delta\|_2^2 + \lambda_i \|c\|_1$ with non-negativity constraints via the previous algorithm (see also \cite[Alg. 2]{lieb2020wavelet} for an example implementation). We ran reconstructions for all parameter choices $\lambda \in \mathcal{P}$. 

\item \textbf{KACZ+TSVD D$\ell^2$ + P$\ell^1$}: When using $\rho=0$ in the previous method, it is not guaranteed that a consistent system of equations is fed into the Kaczmarz algorithm which can result in unstable reconstructions. Due to the low-rank approximation and the ill-posedness of the problem, the matrix likely contains rows close to zero. We thus also used only the 32, 64, 128, 256, 512, and 1024 rows with the largest norms in the previous method (and corresponding entries of the measurement), i.e., we set $K$ in Section~\ref{sec:mpi} to these values which is equivalent to a truncated SVD (TSVD). Reconstructions for $\lambda\in\mathcal{P}$ are computed.

\item \textbf{VAR}: Here, we minimize a functional of the form~\ref{eq:variational_problem}, where we chose the penalty parameter as above and the penalty function as $\ell^1$, $\ell^2$ and $\mbox{TV}$. We also compare $p=1$ and $p=2$. Minimization of the functional is performed up to convergence by applying AMSGrad~\cite{j.2018on} with a learning rate of $10^{-2}$ and again $\beta = (0.9, 0.999)$ (\texttt{pyTorch}~\cite{paszke2017automatic} standard settings) for $500$ iterations, after each iteration we project $c$ to be non-negative.
\end{itemize}

All methods are implemented in \texttt{python}~\cite{10.5555/1593511} mainly using the packages \texttt{numpy}~\cite{oliphant2006guide} and \texttt{pyTorch}~\cite{paszke2017automatic}.
\begin{table*}
\centering
\rowcolors{1}{}{lightgray}
\begin{tabular}{ |l|||c|c|c|c||c|c|c|c|  }
 \hline
 Methods & \multicolumn{4}{c||}{PSNRs} & \multicolumn{4}{c|}{SSIMs} \\
 \hline
 \rowcolor{white}
 \diagbox[dir=NE]{}{} & $\tau=0$ & $\tau=1$ & $\tau=3$ & $\tau=5$ & $\tau=0$ & $\tau=1$ & $\tau=3$ & $\tau=5$\\
 \hline \hline
DIP D$\ell^1$ + P-                    &     28.70 &     29.36 &     \textbf{29.81} &     \underline{\textbf{29.94}} &     \textbf{0.968} &     \textbf{0.968} &     \underline{\textbf{0.973}} &     \underline{\textbf{0.973}} \\
KACZ D$\ell^2$ + P$\ell^2$            &     29.12 &     29.31 &     28.89 &     28.79 &     0.957 &     0.958 &     0.955 &     0.953 \\
KACZ D$\ell^2$ + P$\ell^1$            &     26.21 &     26.63 &     18.01 &     17.09 &     0.886 &     0.888 &     0.534 &     0.534 \\
KACZ+TSVD D$\ell^2$ + P$\ell^1$       &     28.84 &     29.62 &     28.76 &     28.53 &     0.953 &     0.953 &     0.953 &     0.947 \\
KACZ D$\ell^2$ + P$(\ell^1$+$\ell^2)$ &     \textbf{29.52} &     \textbf{29.70} &     29.43 &     29.15 &     0.963 &     0.964 &     0.961 &     0.958 \\
VAR D$\ell^1$ + P$\ell^1$             &     21.45 &     21.58 &     21.45 &     22.64 &     0.856 &     0.865 &     0.840 &     0.876 \\
VAR D$\ell^1$ + P$\ell^2$             &     27.57 &     27.89 &     27.50 &     27.81 &     0.932 &     0.910 &     0.876 &     0.885 \\
VAR D$\ell^1$ + P TV                  &     25.98 &     25.84 &     26.29 &     27.31 &     0.932 &     0.917 &     0.930 &     0.938 \\
VAR D$\ell^2$ + P$\ell^1$             &     17.70 &     21.28 &     23.03 &     24.06 &     0.646 &     0.807 &     0.882 &     0.897 \\
VAR D$\ell^2$ + P$\ell^2$             &     20.14 &     24.53 &     26.11 &     26.33 &     0.684 &     0.854 &     0.913 &     0.906 \\
VAR D$\ell^2$ + P TV                  &     18.96 &     22.78 &     25.75 &     25.58 &     0.648 &     0.818 &     0.915 &     0.914 \\

 \hline
\end{tabular}
\caption{PSNR and SSIM values as achieved by the different SNR thresholds $\tau$ settings for the shape phantom. Largest values for each column are in bold font. The overall largest value is underlined.}
\label{tab:whitened_shape_metrics}
\end{table*}

\begin{table*}
\centering
\rowcolors{1}{}{lightgray}
\begin{tabular}{ |l|||c|c|c|c||c|c|c|c|  }
 \hline
 Methods & \multicolumn{4}{c||}{PSNRs} & \multicolumn{4}{c|}{SSIMs} \\
 \hline
 \rowcolor{white}
 \diagbox[dir=NE]{}{} & $\tau=0$ & $\tau=1$ & $\tau=3$ & $\tau=5$ & $\tau=0$ & $\tau=1$ & $\tau=3$ & $\tau=5$\\
 \hline \hline
DIP D$\ell^1$ + P-                    &     31.73 &     32.46 &     \underline{\textbf{32.92}} &     \textbf{32.33} &     0.945 &     \underline{\textbf{0.952}} &     \textbf{0.951} &     0.940 \\
KACZ D$\ell^2$ + P$\ell^2$            &     31.58 &     32.08 &     31.89 &     31.71 &     0.946 &     0.947 &     0.949 &     0.945 \\
KACZ D$\ell^2$ + P$\ell^1$            &     31.82 &     32.08 &     30.24 &     29.39 &     0.943 &     0.944 &     0.887 &     0.829 \\
KACZ+TSVD D$\ell^2$ + P$\ell^1$       &     \textbf{31.94} &     \textbf{32.57} &     31.80 &     31.35 &     0.947 &     0.948 &     0.947 &     0.942 \\
KACZ D$\ell^2$ + P$(\ell^1$+$\ell^2)$ &     31.82 &     32.38 &     32.10 &     31.76 &     \textbf{0.949} &     0.949 &     0.949 &     \textbf{0.946} \\
VAR D$\ell^1$ + P$\ell^1$             &     30.43 &     30.82 &     29.91 &     30.25 &     0.915 &     0.919 &     0.901 &     0.904 \\
VAR D$\ell^1$ + P$\ell^2$             &     31.42 &     31.59 &     31.42 &     31.14 &     0.945 &     0.947 &     0.942 &     0.936 \\
VAR D$\ell^1$ + P TV                  &     31.22 &     31.61 &     31.47 &     31.24 &     0.940 &     0.948 &     0.933 &     0.928 \\
VAR D$\ell^2$ + P$\ell^1$             &     29.00 &     28.93 &     30.08 &     30.93 &     0.803 &     0.860 &     0.910 &     0.905 \\
VAR D$\ell^2$ + P$\ell^2$             &     29.51 &     30.23 &     30.88 &     31.24 &     0.839 &     0.889 &     0.932 &     0.931 \\
VAR D$\ell^2$ + P TV                  &     29.26 &     29.54 &     30.44 &     30.79 &     0.823 &     0.878 &     0.925 &     0.925 \\
 \hline
\end{tabular}
\caption{PSNR and SSIM values as achieved by the different SNR thresholds $\tau$ settings for the resolution phantom. Largest values for each column are in bold font. The overall largest value is underlined.}
\label{tab:whitened_resolution_metrics}
\end{table*}

\subsection*{Method comparison by image quality}
First, we compare the results quantitatively based on the optimal PSNR and SSIM values, which are presented in Tables~\ref{tab:whitened_shape_metrics} and~\ref{tab:whitened_resolution_metrics}, for the shape and resolution phantom, respectively. The corresponding optimal parameters that led to these values can be found in the supplementary material in Tables~\ref{tab:whitened_shape_parameters} and~\ref{tab:whitened_resolution_parameters}. Note that all the data discussed in this section relates to the whitened case discussed in the preprocessing part of Section~\ref{sec:mpi}. The supplementary material contains the analogous parts of the non-whitened case, which qualitatively leads to similar results, though, with PSNRs being on average $0.5$ lower and SSIMs being on average $0.12$ lower for peak values.

\begin{figure}
\centering
\begin{tabular}{c c c}
 \hline
  & phantom & reconstruction  \\
 \hline\\[-2.0ex]
Shape &
\raisebox{-\totalheight/2}{\includegraphics[trim={100mm 43mm 100mm 43mm},clip,width=0.35\linewidth]{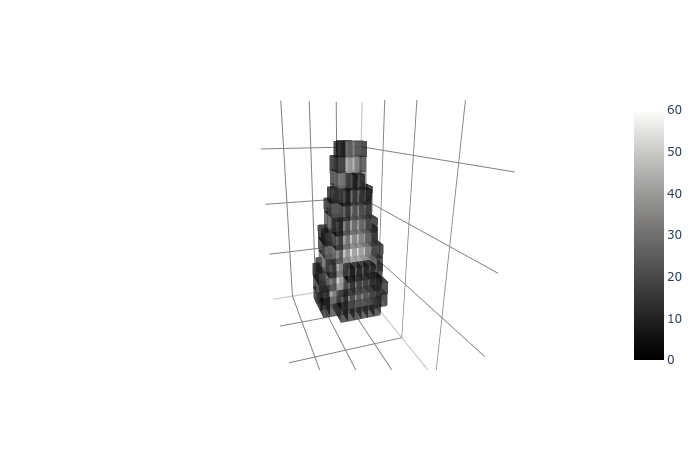}} &
\raisebox{-\totalheight/2}{\includegraphics[trim={100mm 43mm 100mm 43mm},clip,width=0.35\linewidth]{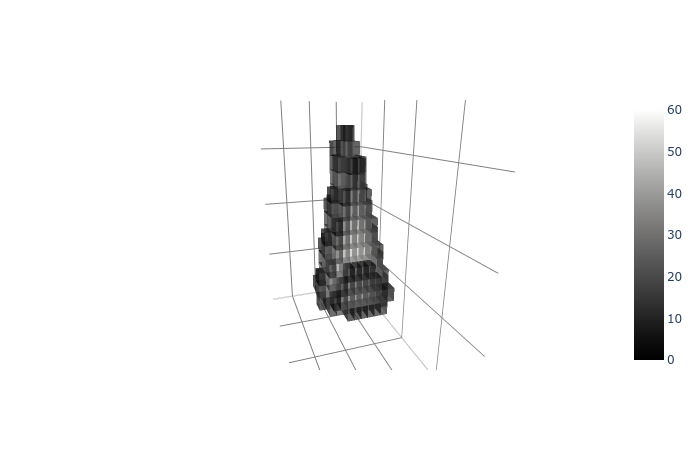}} \\
\\
Resolution &
\raisebox{-\totalheight/2}{\includegraphics[trim={85mm 30mm 65mm 50mm},clip,width=0.35\linewidth]{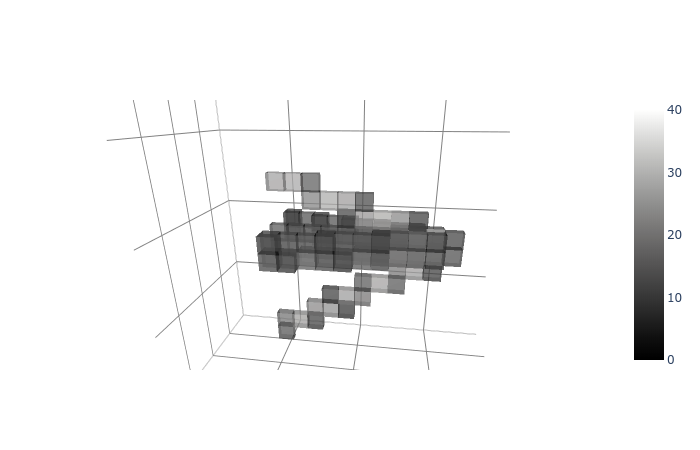}} &
\raisebox{-\totalheight/2}{\includegraphics[trim={85mm 30mm 65mm 50mm},clip,width=0.35\linewidth]{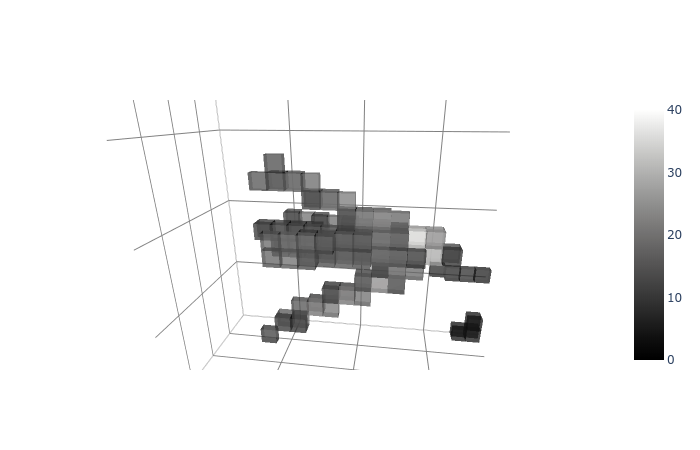}} 
\end{tabular}
\caption{Qualitative comparison of the phantoms and the best reconstructions in terms of PSNR (all achieved by our DIP). The gray scale goes from $0$ (black) to $60$ or $40$ (white), for the shape and resolution phantom respectively.}
\label{tab:3dTable}
\end{figure}

\begin{figure*}
\centering
\rowcolors{1}{}{lightgray}
\begin{tabular}{c   c c c c  c c c c}
 \hline
 Methods & \multicolumn{4}{c  }{PSNRs} & \multicolumn{4}{c }{SSIMs} \\
 \hline
 \rowcolor{white}
  & $\tau=0$ & $\tau=1$ & $\tau=3$ & $\tau=5$ & $\tau=0$ & $\tau=1$ & $\tau=3$ & $\tau=5$\\
 \hline\\[-2.0ex]

\vspace{.5mm} DIP D$\ell^1$ + P- &
\raisebox{-\totalheight/2}{\includegraphics[width=0.06\linewidth]{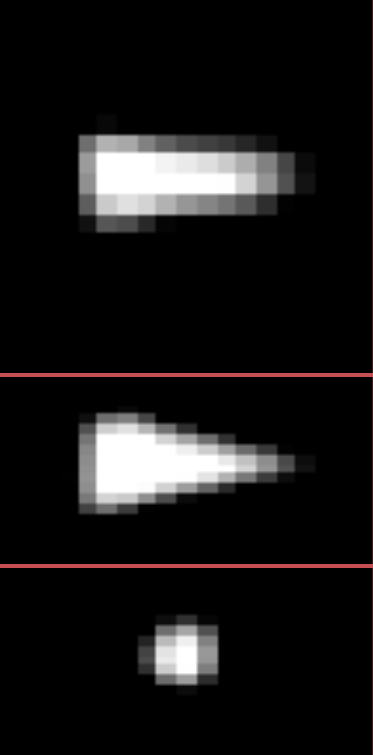}} &
\raisebox{-\totalheight/2}{\includegraphics[width=0.06\linewidth]{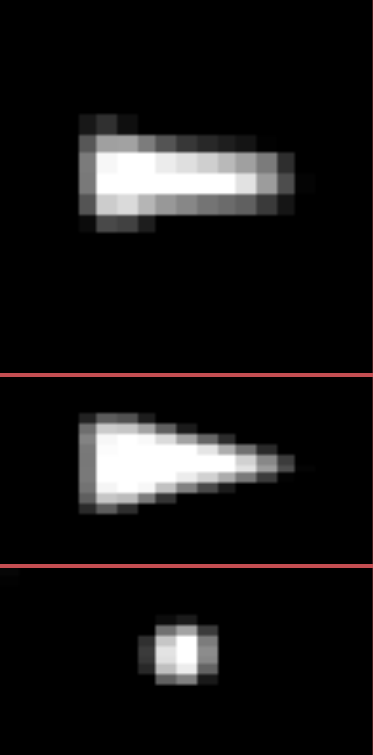}} &
\raisebox{-\totalheight/2}{\includegraphics[width=0.06\linewidth]{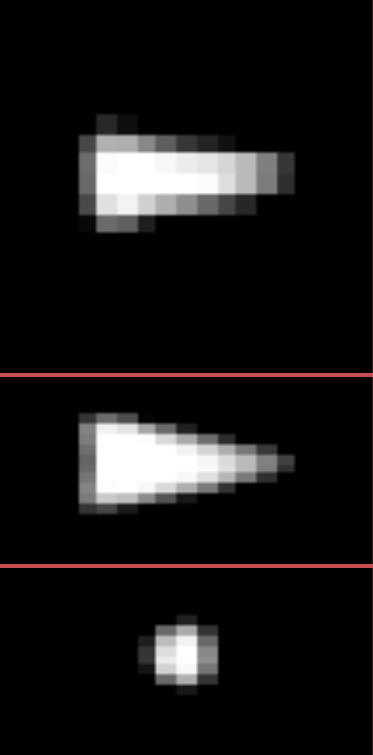}} &
\raisebox{-\totalheight/2}{\includegraphics[width=0.06\linewidth]{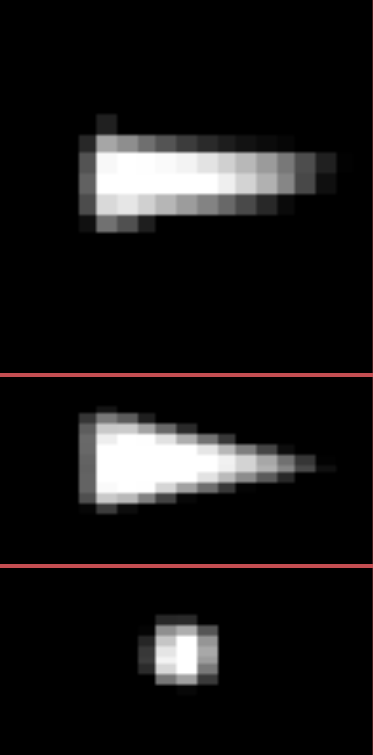}} &
\raisebox{-\totalheight/2}{\includegraphics[width=0.06\linewidth]{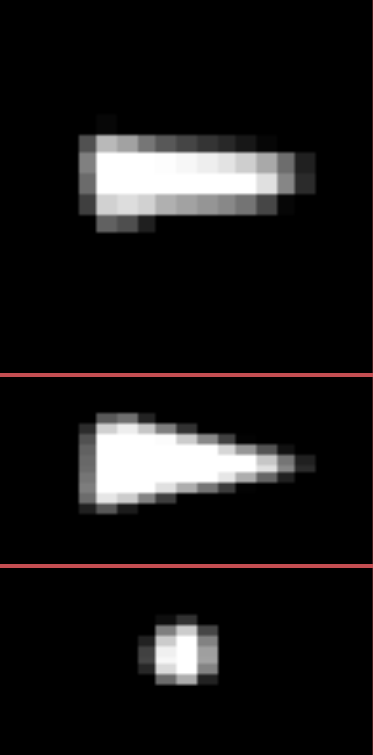}} &
\raisebox{-\totalheight/2}{\includegraphics[width=0.06\linewidth]{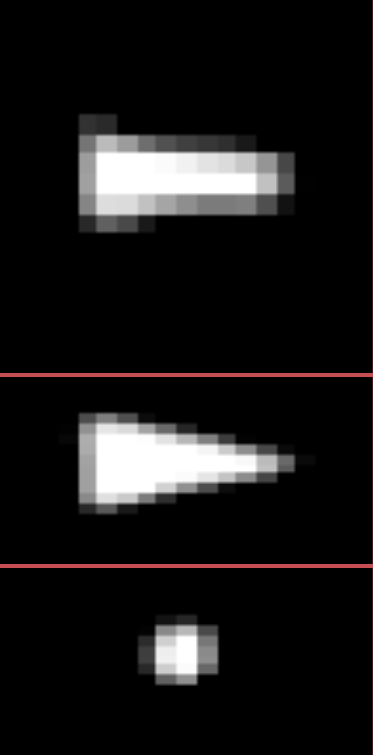}} &
\raisebox{-\totalheight/2}{\includegraphics[width=0.06\linewidth]{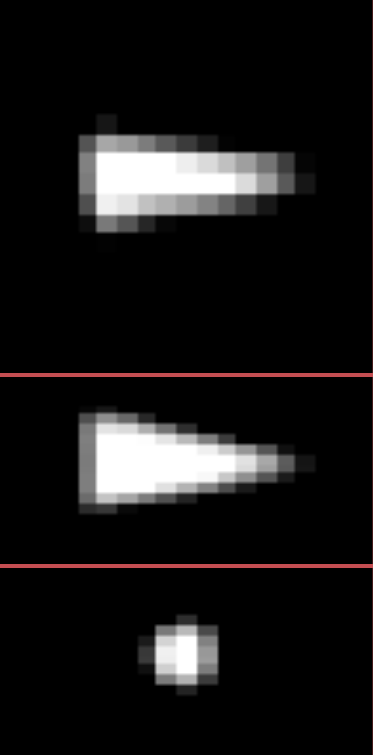}} &
\raisebox{-\totalheight/2}{\includegraphics[width=0.06\linewidth]{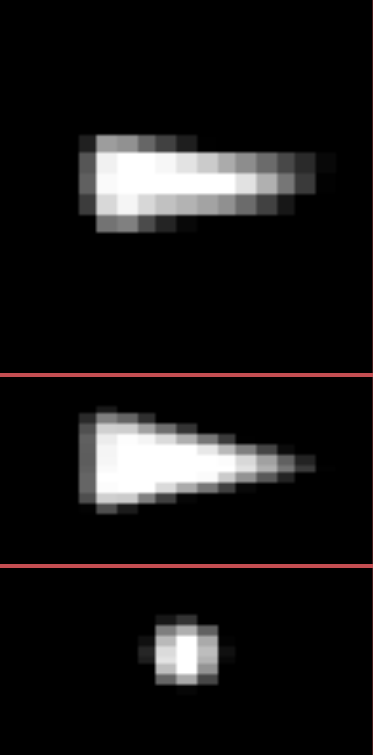}}\\

\vspace{.5mm} KACZ D$\ell^2$ + P$\ell^2$ &
\raisebox{-\totalheight/2}{\includegraphics[width=0.06\linewidth]{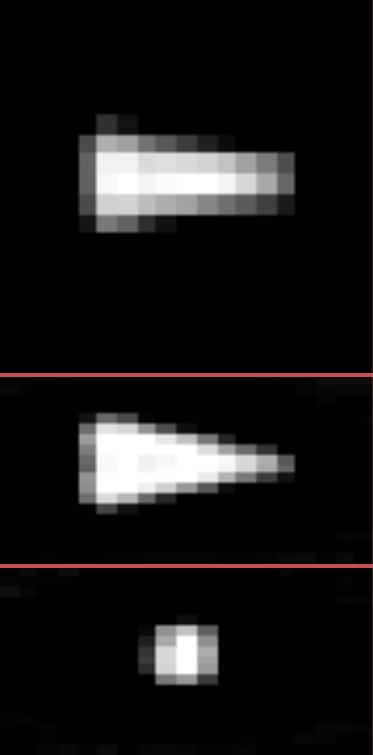}} &
\raisebox{-\totalheight/2}{\includegraphics[width=0.06\linewidth]{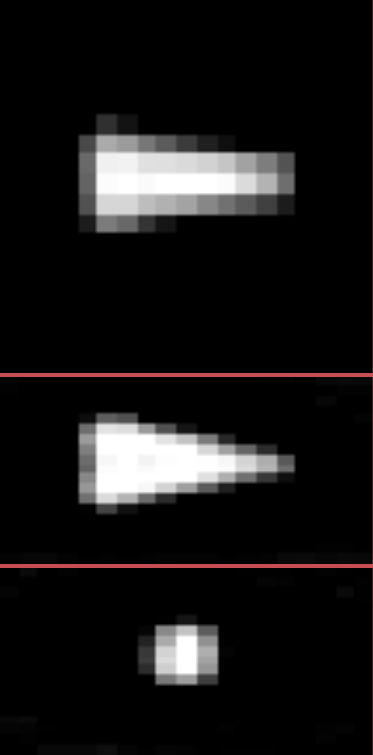}} &
\raisebox{-\totalheight/2}{\includegraphics[width=0.06\linewidth]{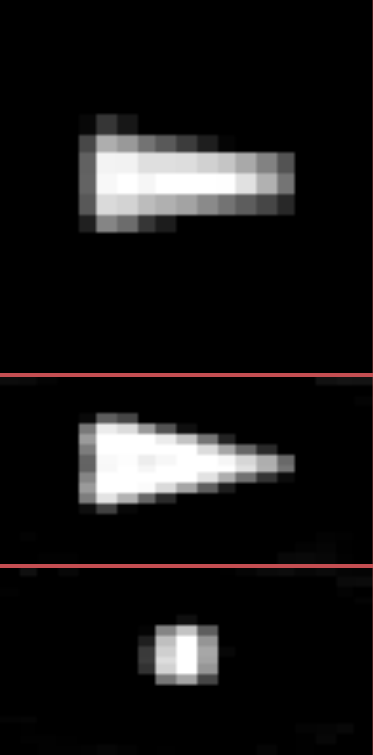}} &
\raisebox{-\totalheight/2}{\includegraphics[width=0.06\linewidth]{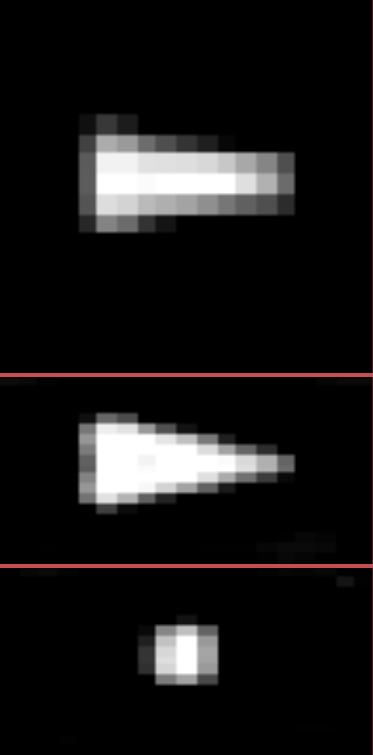}} &
\raisebox{-\totalheight/2}{\includegraphics[width=0.06\linewidth]{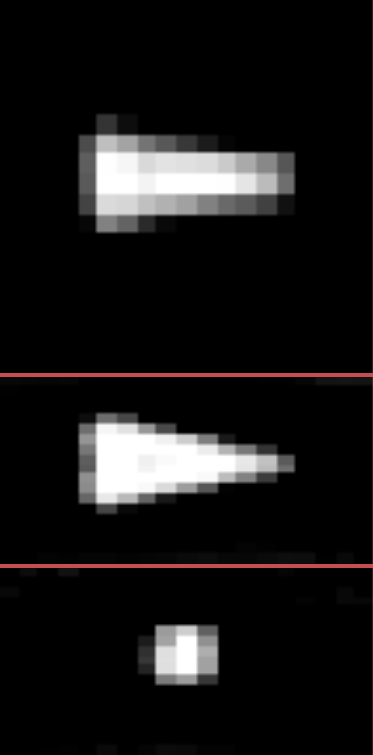}} &
\raisebox{-\totalheight/2}{\includegraphics[width=0.06\linewidth]{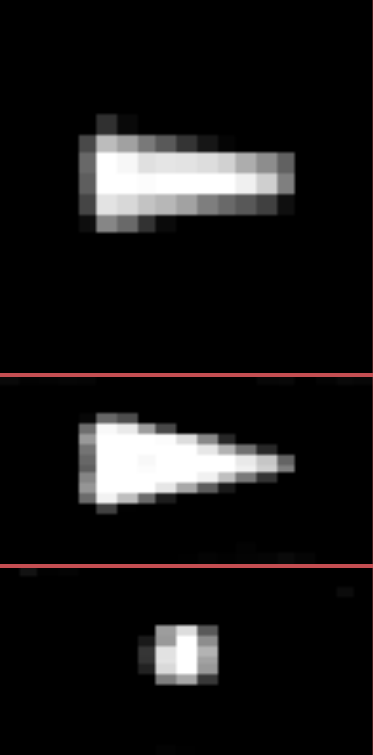}} &
\raisebox{-\totalheight/2}{\includegraphics[width=0.06\linewidth]{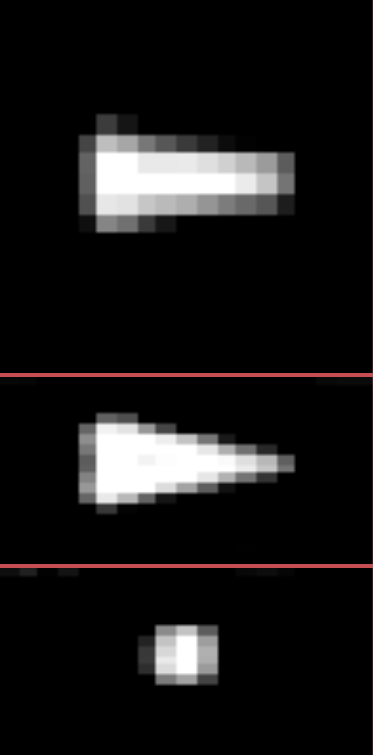}} &
\raisebox{-\totalheight/2}{\includegraphics[width=0.06\linewidth]{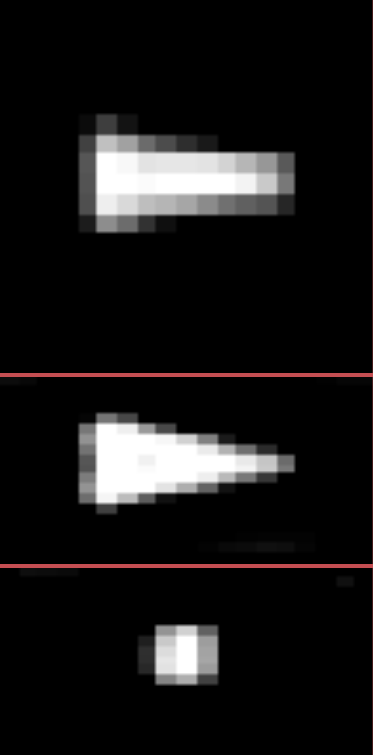}}\\

\vspace{.5mm} KACZ D$\ell^2$ + P$\ell^1$ &
\raisebox{-\totalheight/2}{\includegraphics[width=0.06\linewidth]{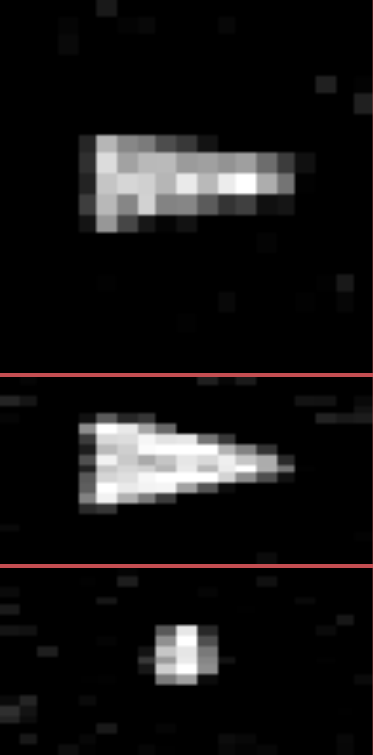}} &
\raisebox{-\totalheight/2}{\includegraphics[width=0.06\linewidth]{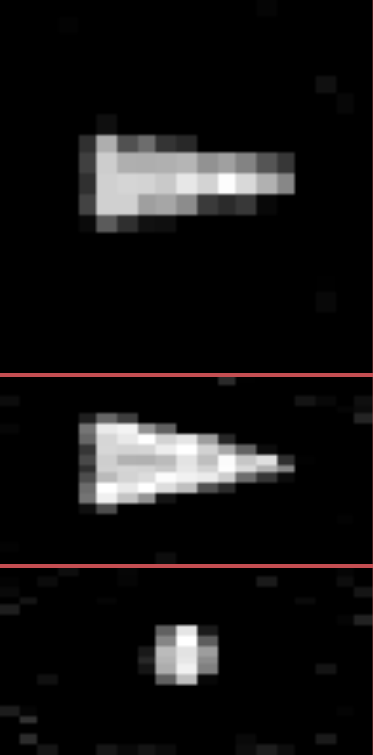}} &
\raisebox{-\totalheight/2}{\includegraphics[width=0.06\linewidth]{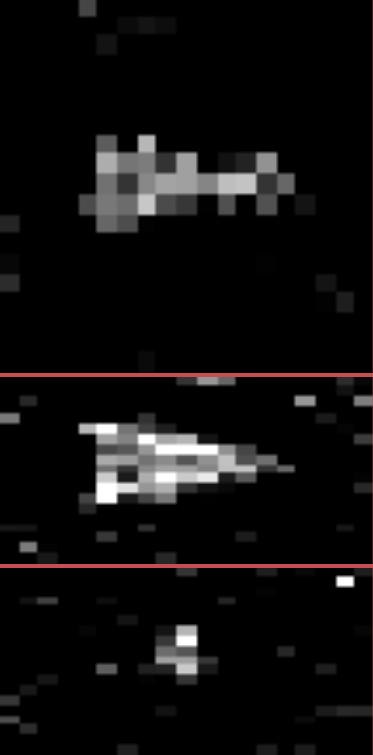}} &
\raisebox{-\totalheight/2}{\includegraphics[width=0.06\linewidth]{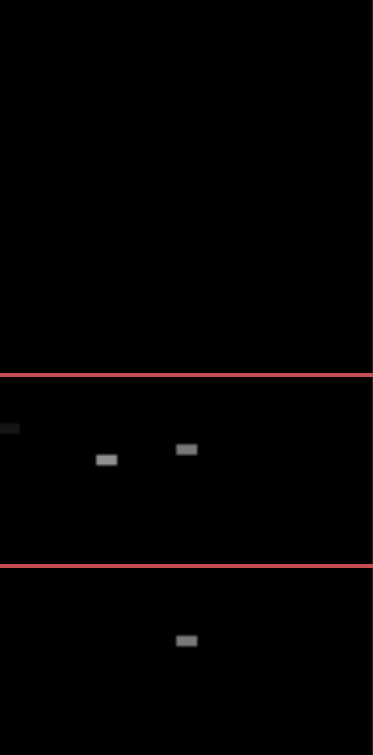}} &
\raisebox{-\totalheight/2}{\includegraphics[width=0.06\linewidth]{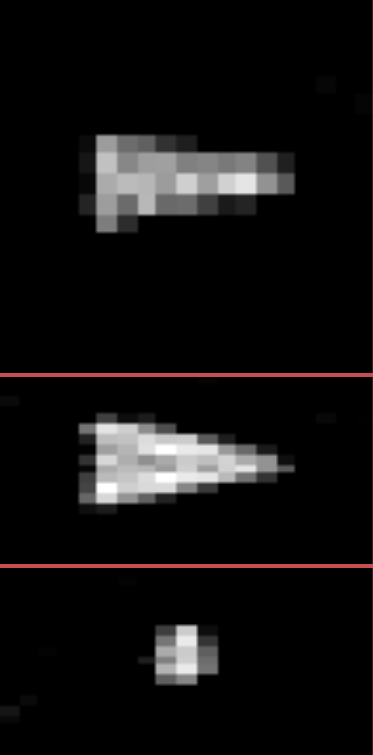}} &
\raisebox{-\totalheight/2}{\includegraphics[width=0.06\linewidth]{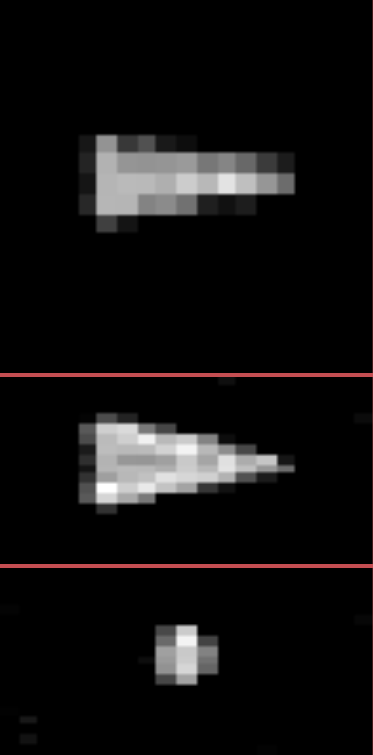}} &
\raisebox{-\totalheight/2}{\includegraphics[width=0.06\linewidth]{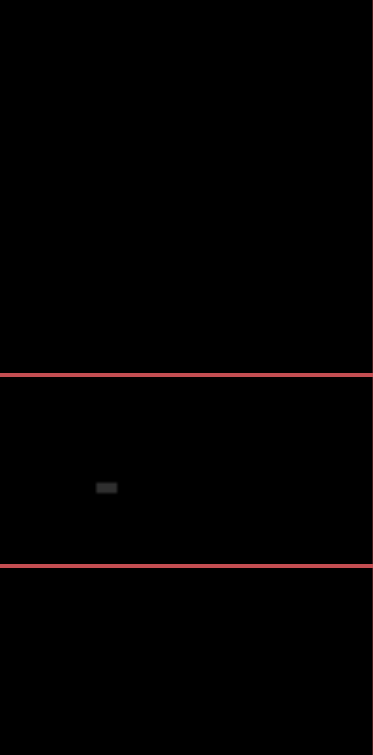}} &
\raisebox{-\totalheight/2}{\includegraphics[width=0.06\linewidth]{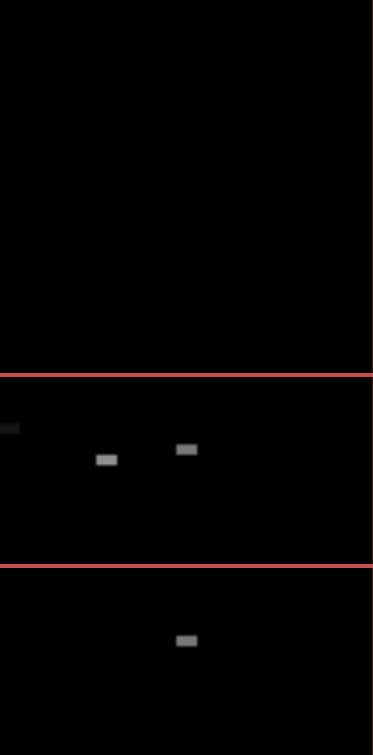}}\\

\vspace{.5mm} KACZ+TSVD D$\ell^2$ + P$\ell^1$ &
\raisebox{-\totalheight/2}{\includegraphics[width=0.06\linewidth]{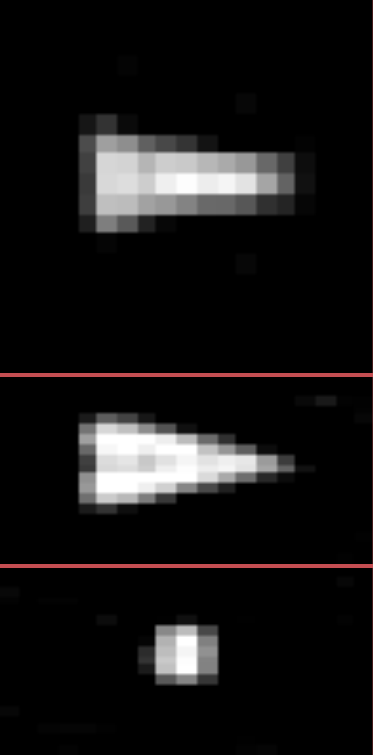}} &
\raisebox{-\totalheight/2}{\includegraphics[width=0.06\linewidth]{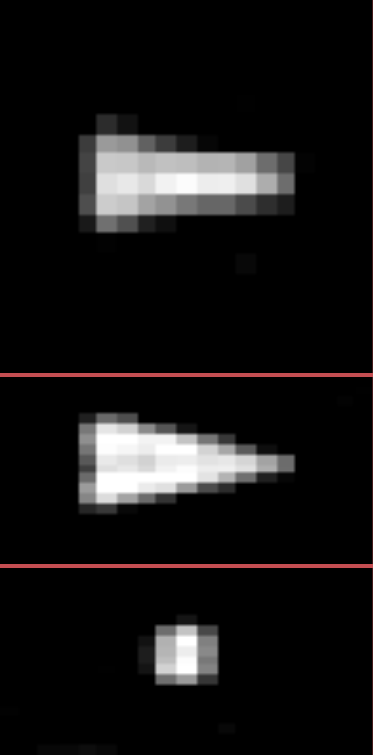}} &
\raisebox{-\totalheight/2}{\includegraphics[width=0.06\linewidth]{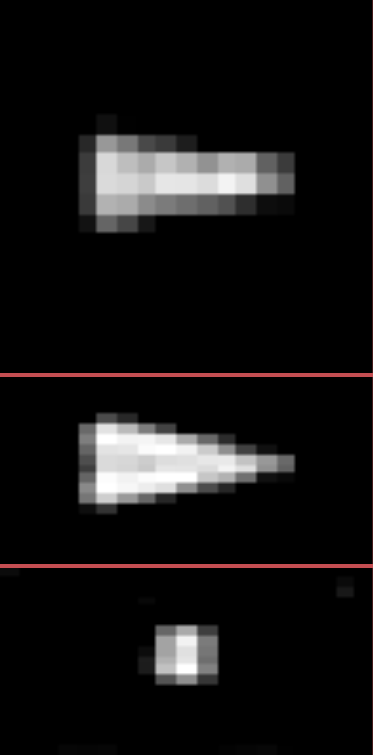}} &
\raisebox{-\totalheight/2}{\includegraphics[width=0.06\linewidth]{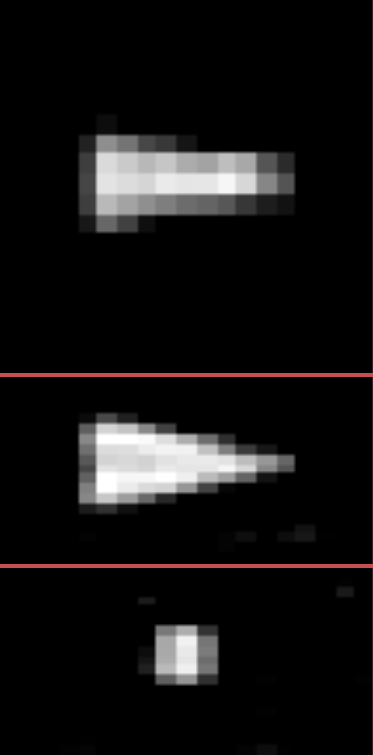}} &
\raisebox{-\totalheight/2}{\includegraphics[width=0.06\linewidth]{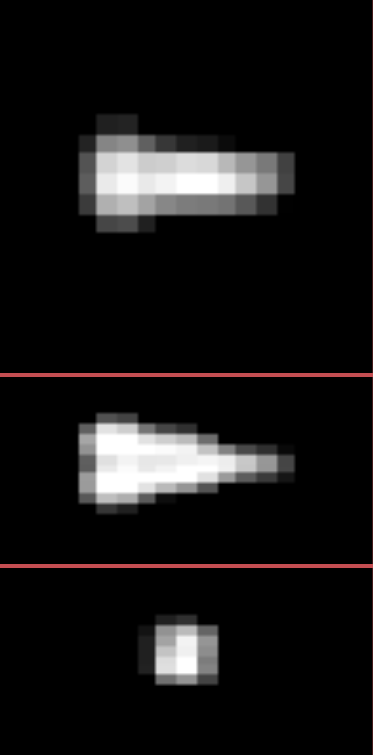}} &
\raisebox{-\totalheight/2}{\includegraphics[width=0.06\linewidth]{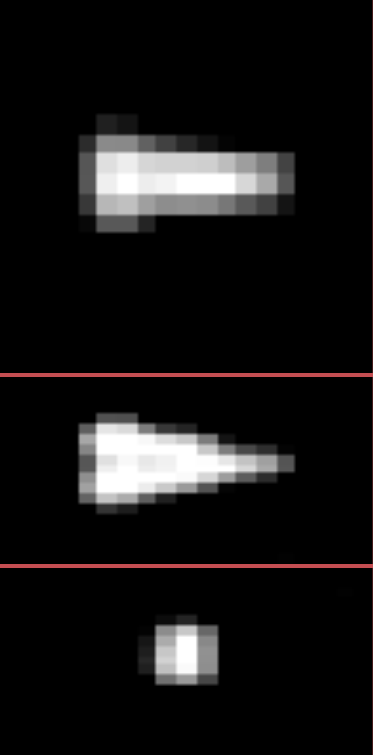}} &
\raisebox{-\totalheight/2}{\includegraphics[width=0.06\linewidth]{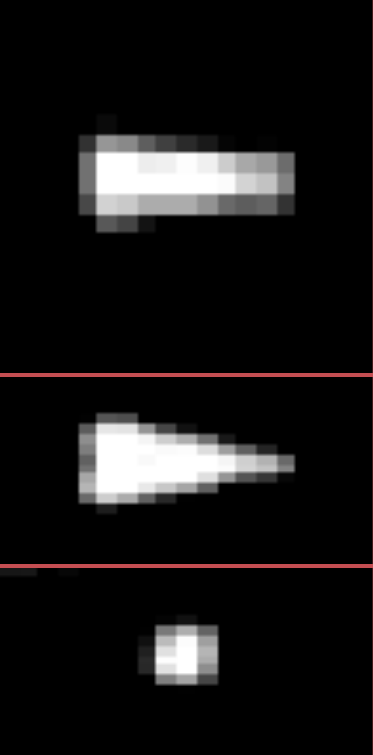}} &
\raisebox{-\totalheight/2}{\includegraphics[width=0.06\linewidth]{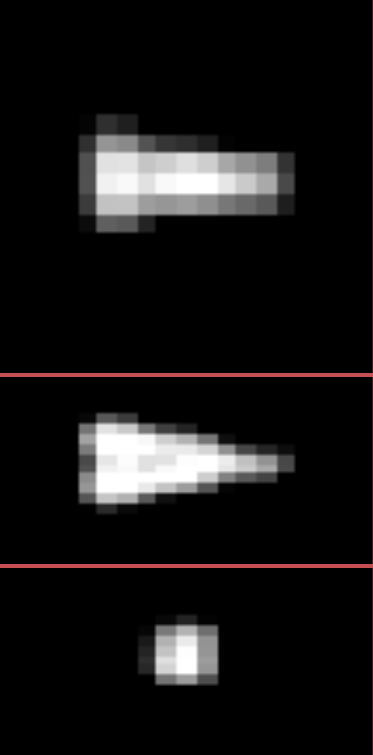}}\\

\vspace{.5mm} KACZ D$\ell^2$ + P$(\ell^1$+$\ell^2)$ &
\raisebox{-\totalheight/2}{\includegraphics[width=0.06\linewidth]{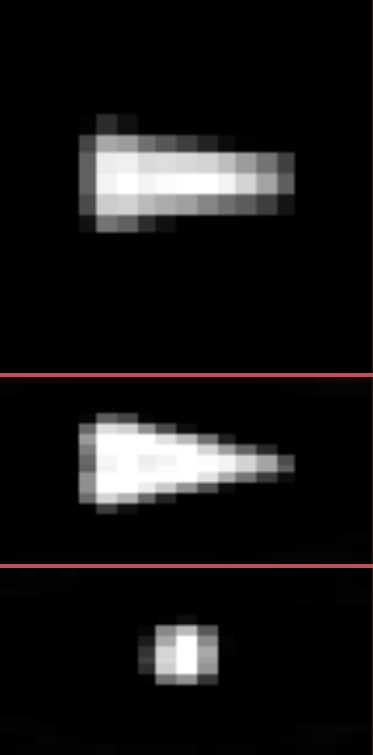}} &
\raisebox{-\totalheight/2}{\includegraphics[width=0.06\linewidth]{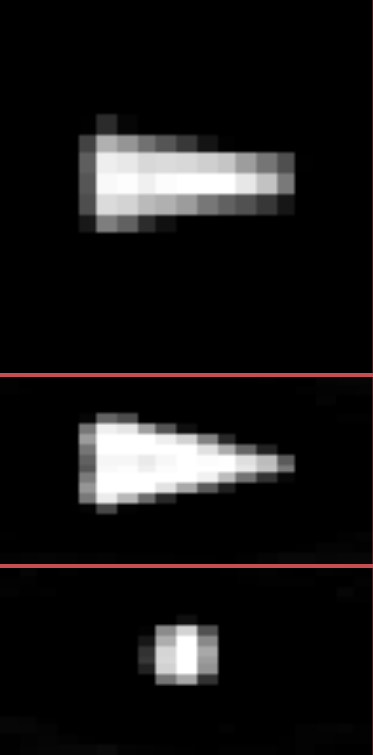}} &
\raisebox{-\totalheight/2}{\includegraphics[width=0.06\linewidth]{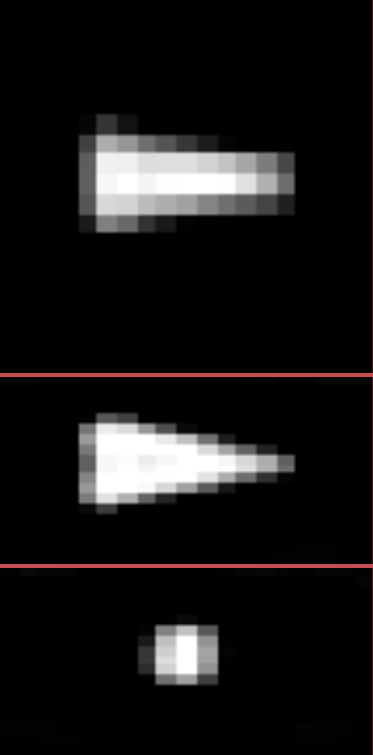}} &
\raisebox{-\totalheight/2}{\includegraphics[width=0.06\linewidth]{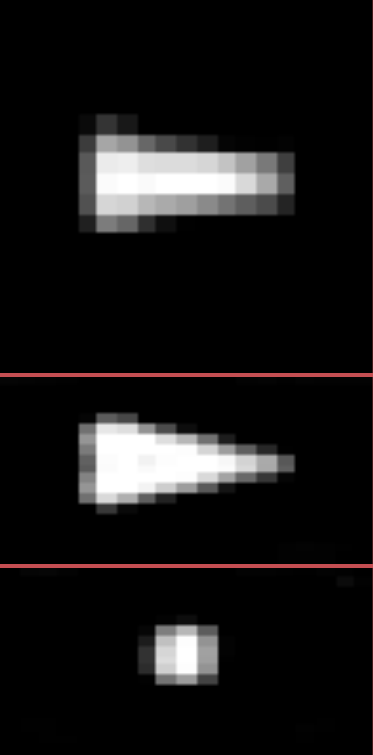}} &
\raisebox{-\totalheight/2}{\includegraphics[width=0.06\linewidth]{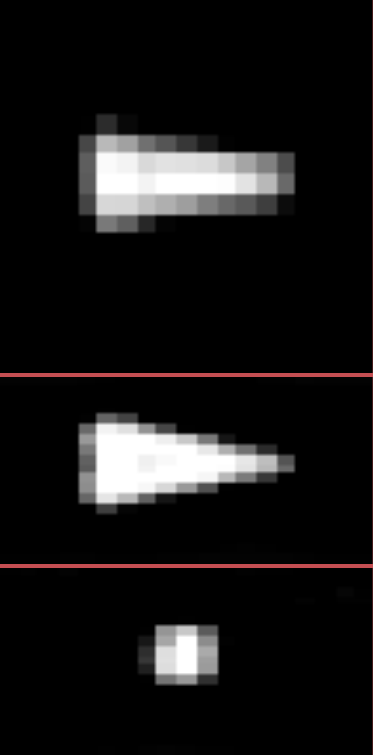}} &
\raisebox{-\totalheight/2}{\includegraphics[width=0.06\linewidth]{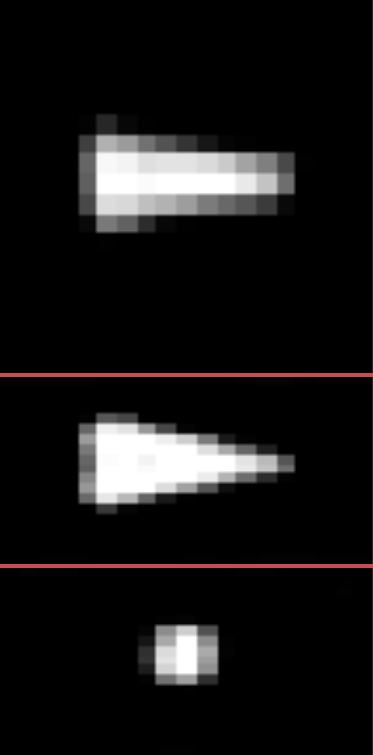}} &
\raisebox{-\totalheight/2}{\includegraphics[width=0.06\linewidth]{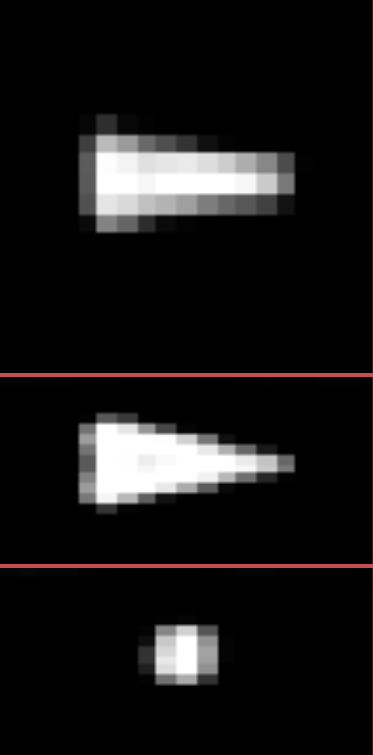}} &
\raisebox{-\totalheight/2}{\includegraphics[width=0.06\linewidth]{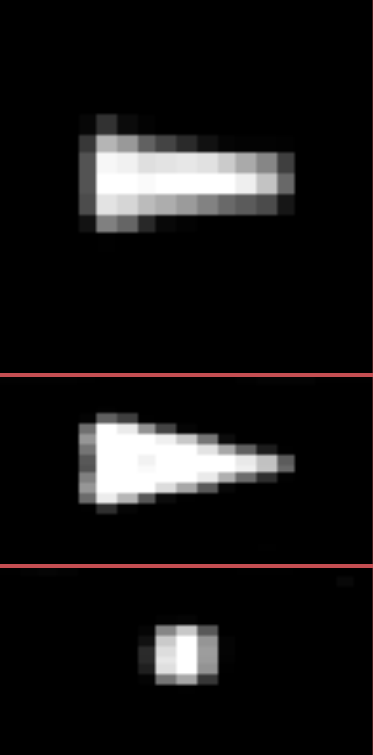}}\\

\vspace{.5mm} VAR D$\ell^1$ + P$\ell^1$  &
\raisebox{-\totalheight/2}{\includegraphics[width=0.06\linewidth]{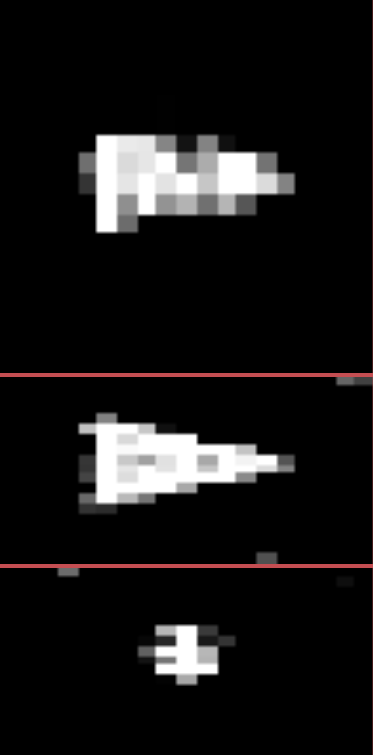}} &
\raisebox{-\totalheight/2}{\includegraphics[width=0.06\linewidth]{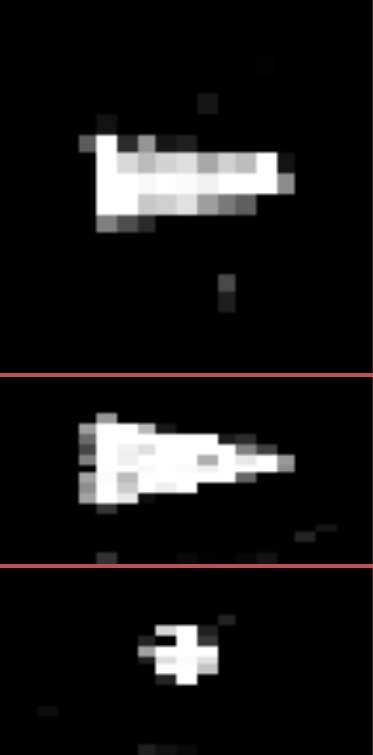}} &
\raisebox{-\totalheight/2}{\includegraphics[width=0.06\linewidth]{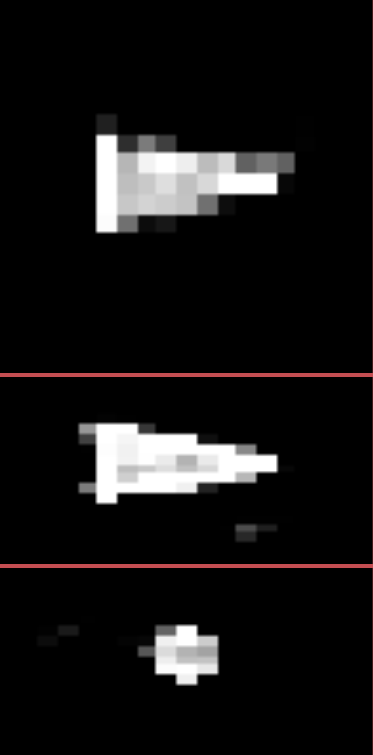}} &
\raisebox{-\totalheight/2}{\includegraphics[width=0.06\linewidth]{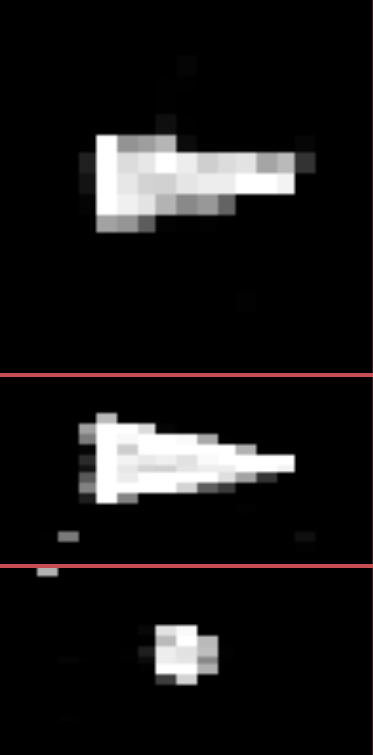}} &
\raisebox{-\totalheight/2}{\includegraphics[width=0.06\linewidth]{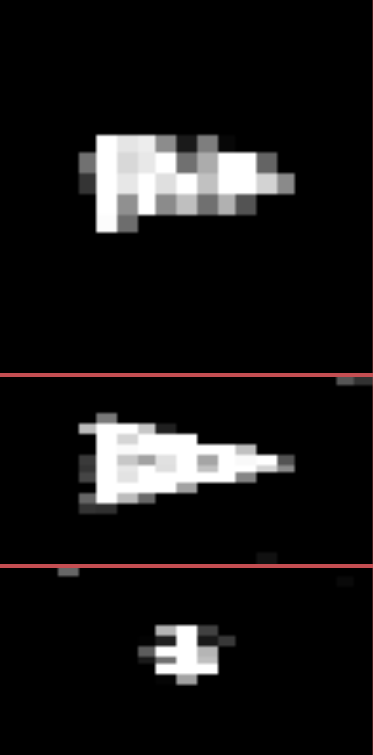}} &
\raisebox{-\totalheight/2}{\includegraphics[width=0.06\linewidth]{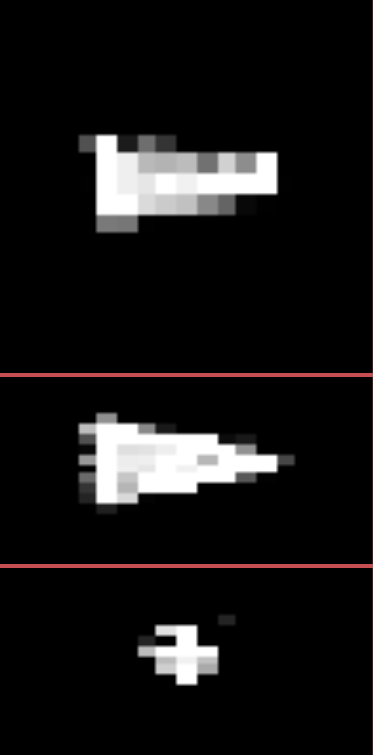}} &
\raisebox{-\totalheight/2}{\includegraphics[width=0.06\linewidth]{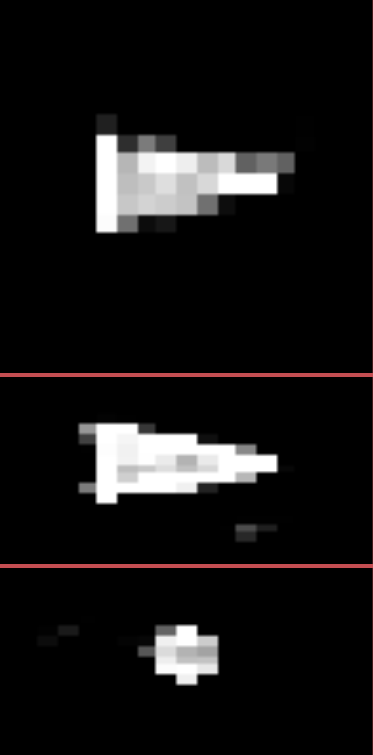}} &
\raisebox{-\totalheight/2}{\includegraphics[width=0.06\linewidth]{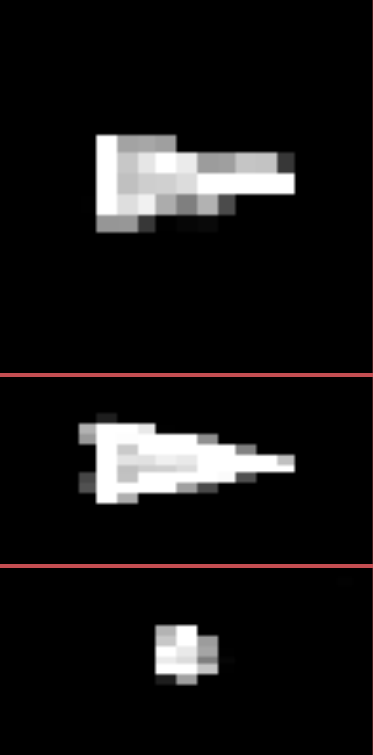}}\\

\vspace{.5mm} VAR D$\ell^1$ + P$\ell^2$ &
\raisebox{-\totalheight/2}{\includegraphics[width=0.06\linewidth]{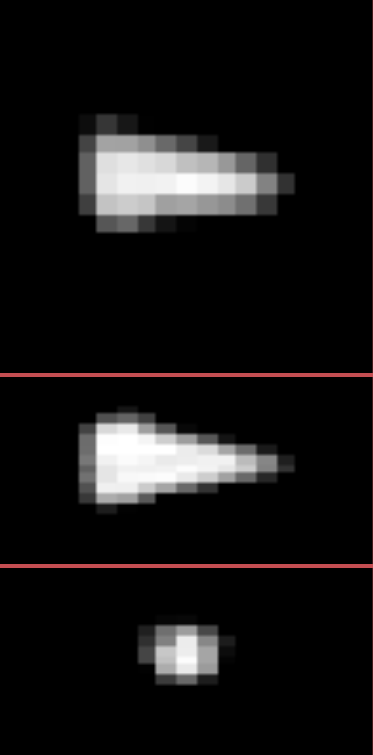}} &
\raisebox{-\totalheight/2}{\includegraphics[width=0.06\linewidth]{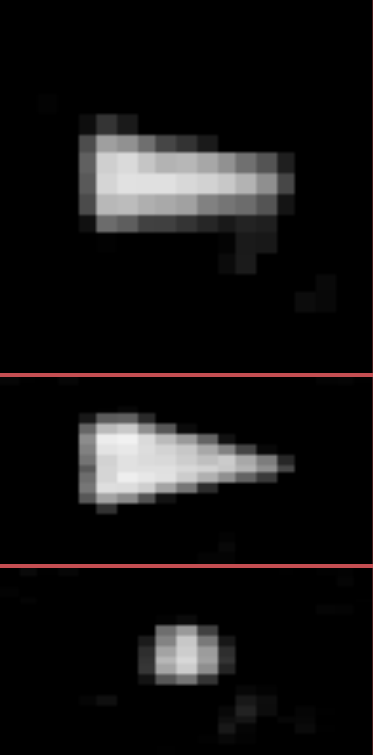}} &
\raisebox{-\totalheight/2}{\includegraphics[width=0.06\linewidth]{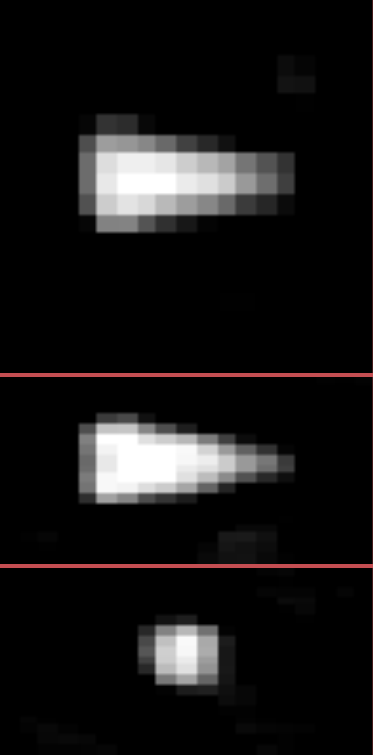}} &
\raisebox{-\totalheight/2}{\includegraphics[width=0.06\linewidth]{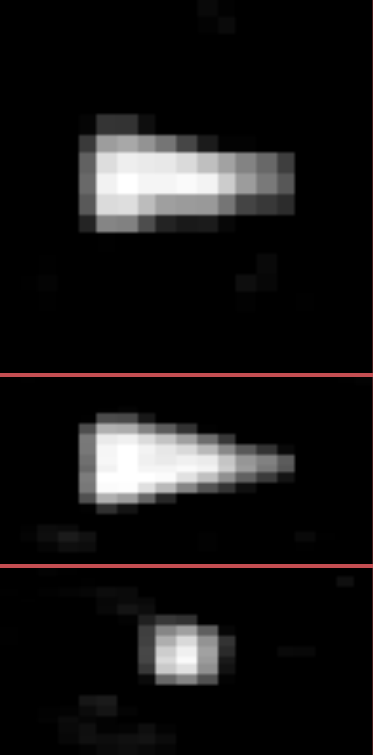}} &
\raisebox{-\totalheight/2}{\includegraphics[width=0.06\linewidth]{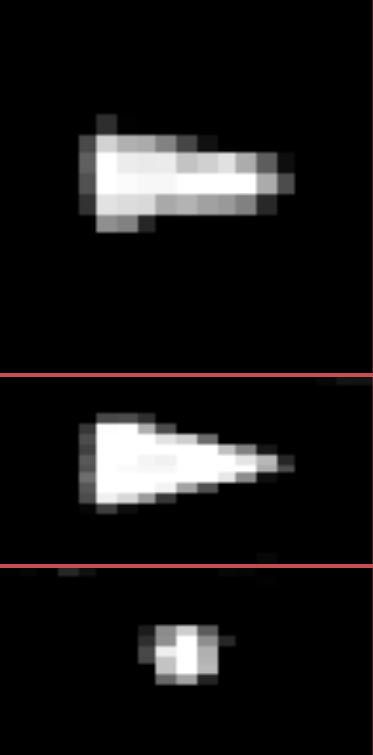}} &
\raisebox{-\totalheight/2}{\includegraphics[width=0.06\linewidth]{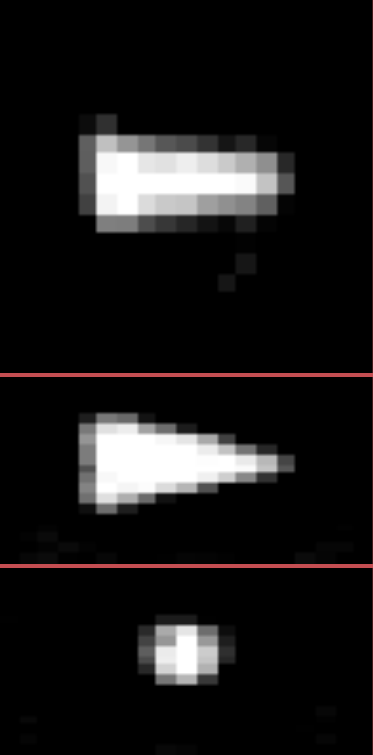}} &
\raisebox{-\totalheight/2}{\includegraphics[width=0.06\linewidth]{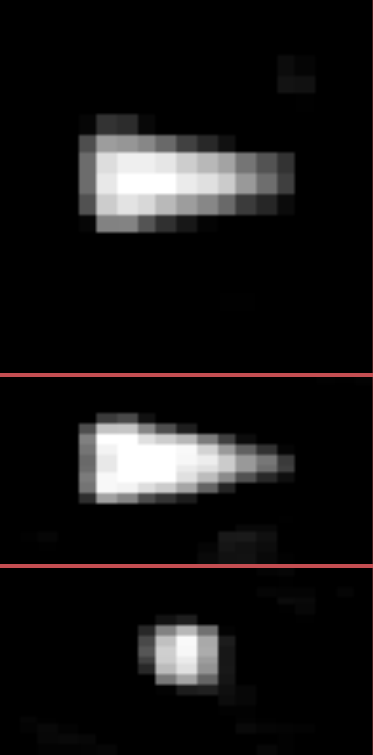}} &
\raisebox{-\totalheight/2}{\includegraphics[width=0.06\linewidth]{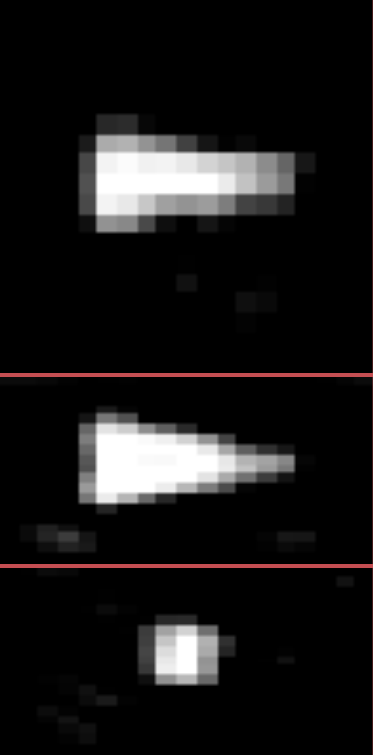}}\\

\vspace{.5mm} VAR D$\ell^1$ + P TV & 
\raisebox{-\totalheight/2}{\includegraphics[width=0.06\linewidth]{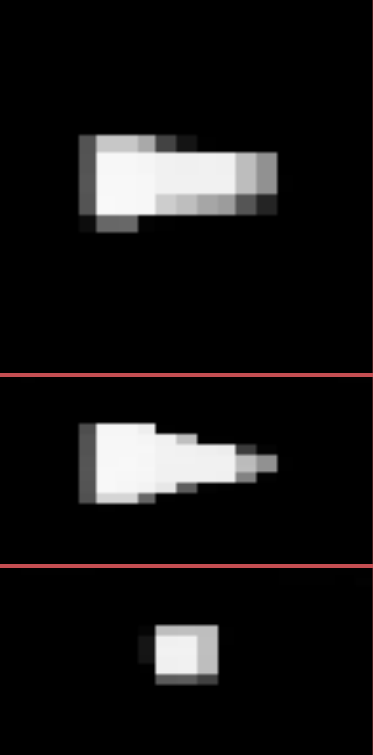}} &
\raisebox{-\totalheight/2}{\includegraphics[width=0.06\linewidth]{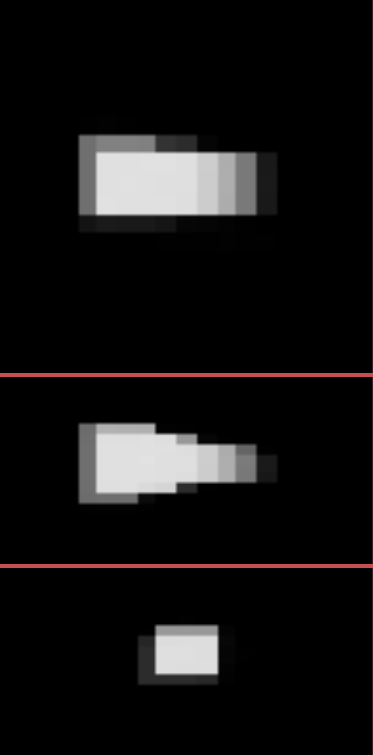}} &
\raisebox{-\totalheight/2}{\includegraphics[width=0.06\linewidth]{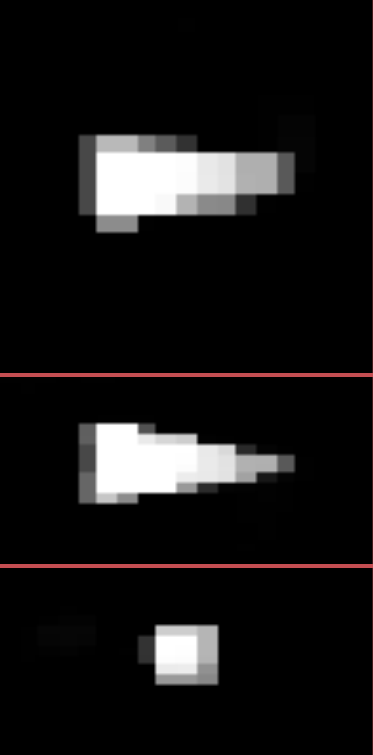}} &
\raisebox{-\totalheight/2}{\includegraphics[width=0.06\linewidth]{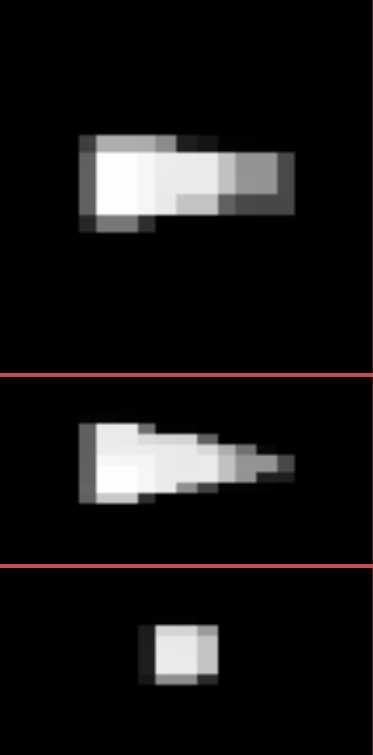}} &
\raisebox{-\totalheight/2}{\includegraphics[width=0.06\linewidth]{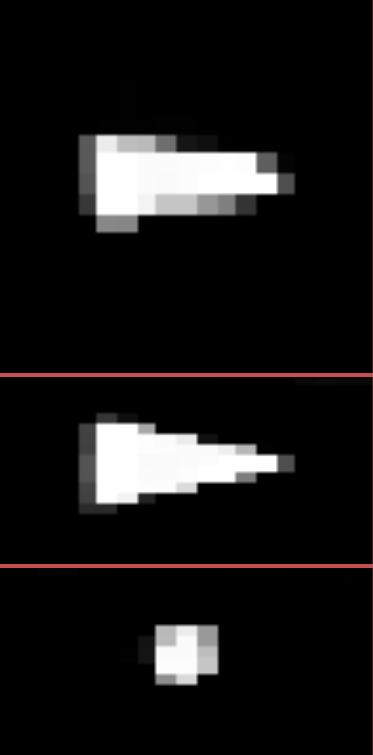}} &
\raisebox{-\totalheight/2}{\includegraphics[width=0.06\linewidth]{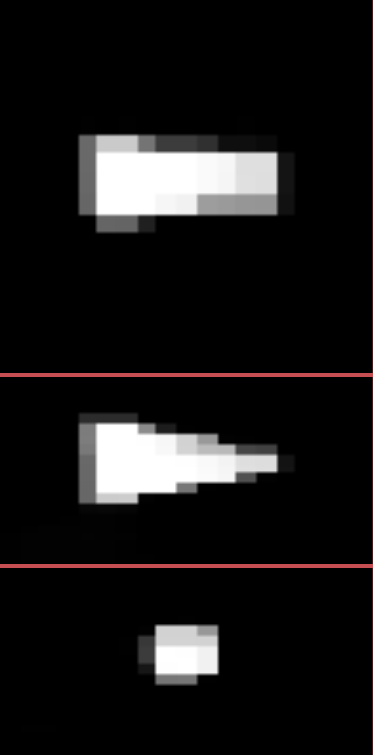}} &
\raisebox{-\totalheight/2}{\includegraphics[width=0.06\linewidth]{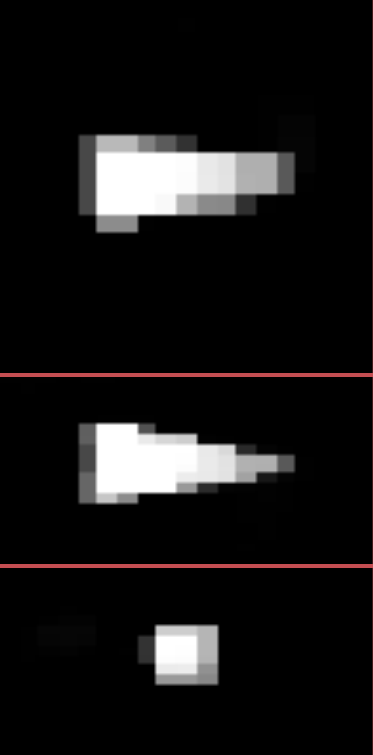}} &
\raisebox{-\totalheight/2}{\includegraphics[width=0.06\linewidth]{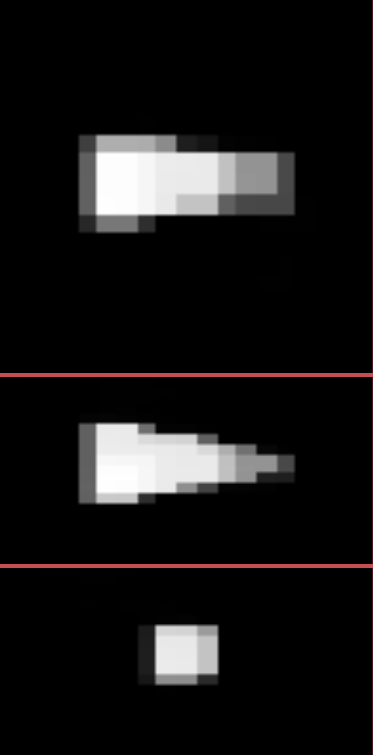}}\\

 \hline
\end{tabular}
\caption{The best reconstructions for the shape phantom, corresponding to the values in Tables~\ref{tab:whitened_shape_metrics} and~\ref{tab:whitened_shape_parameters}. The color scale goes from $0$ (black) to $60$ (white). Each image is separated into three vertically stacked parts, which are separated by red lines. Each of these parts represents one of the three central slices/planes of the three-dimensional reconstruction (x-y-, x-z-, and y-z-plane from top to bottom).}
\label{tab:whitened_shape_images}
\end{figure*}

\begin{figure*}
\centering
\rowcolors{1}{}{lightgray}
\begin{tabular}{c   c c c c  c c c c}
 \hline
 Methods & \multicolumn{4}{c  }{PSNRs} & \multicolumn{4}{c }{SSIMs} \\
 \hline
 \rowcolor{white}
  & $\tau=0$ & $\tau=1$ & $\tau=3$ & $\tau=5$ & $\tau=0$ & $\tau=1$ & $\tau=3$ & $\tau=5$\\
 \hline\\[-2.0ex]

\vspace{.5mm} DIP D$\ell^1$ + P- &
\raisebox{-\totalheight/2}{\includegraphics[width=0.06\linewidth]{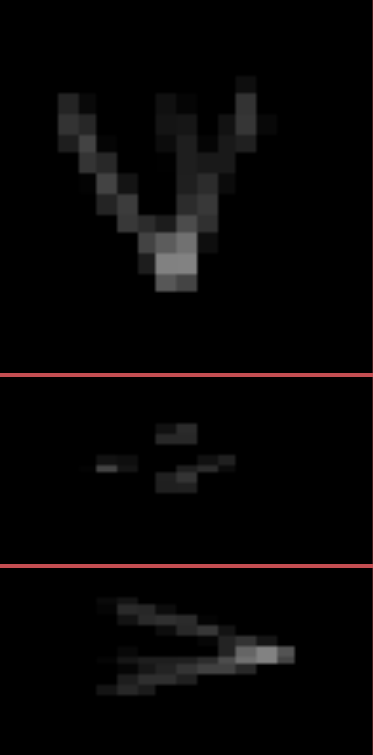}} &
\raisebox{-\totalheight/2}{\includegraphics[width=0.06\linewidth]{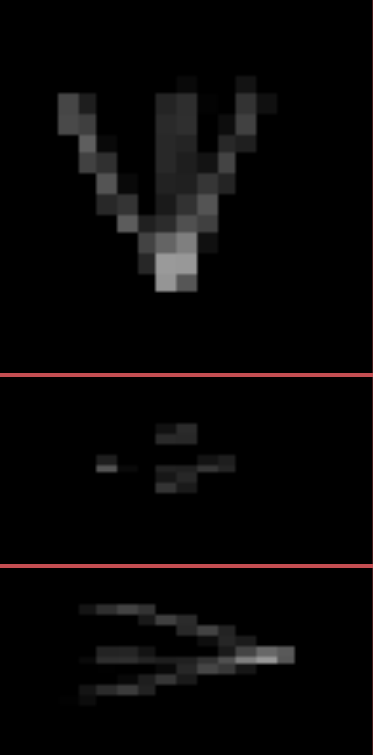}} &
\raisebox{-\totalheight/2}{\includegraphics[width=0.06\linewidth]{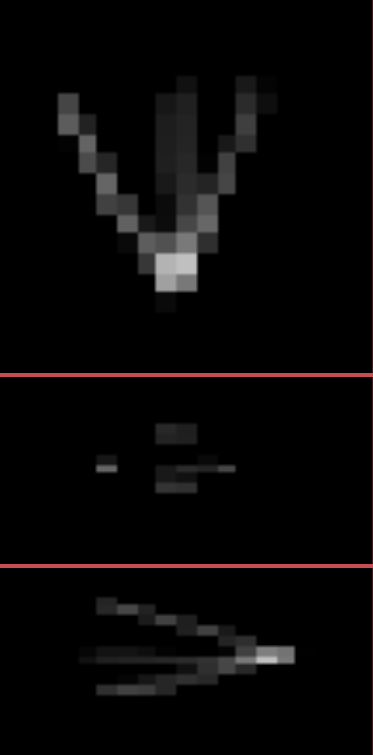}} &
\raisebox{-\totalheight/2}{\includegraphics[width=0.06\linewidth]{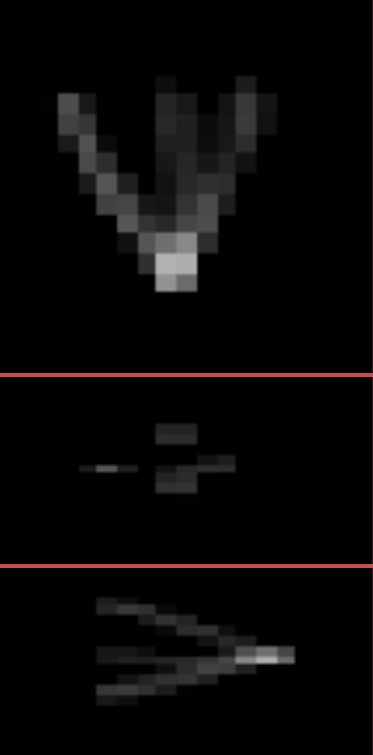}} &
\raisebox{-\totalheight/2}{\includegraphics[width=0.06\linewidth]{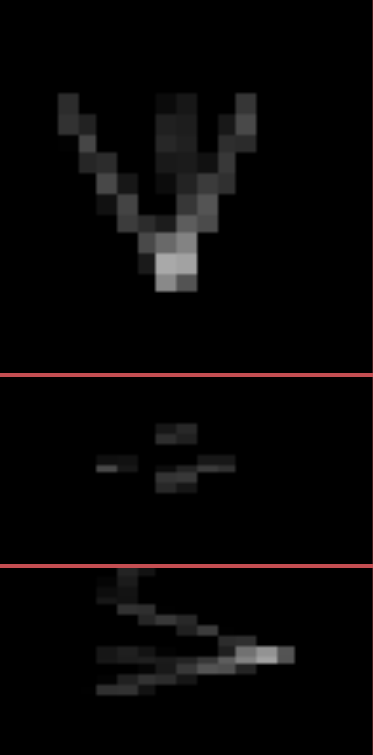}} &
\raisebox{-\totalheight/2}{\includegraphics[width=0.06\linewidth]{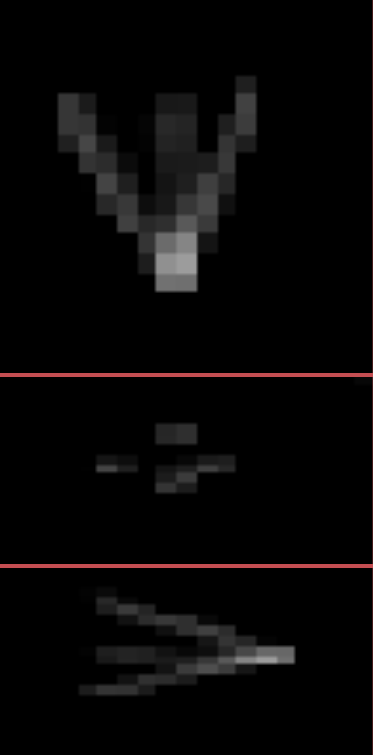}} &
\raisebox{-\totalheight/2}{\includegraphics[width=0.06\linewidth]{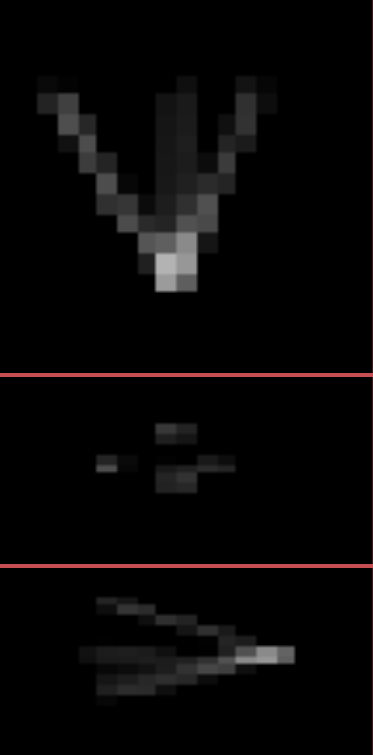}} &
\raisebox{-\totalheight/2}{\includegraphics[width=0.06\linewidth]{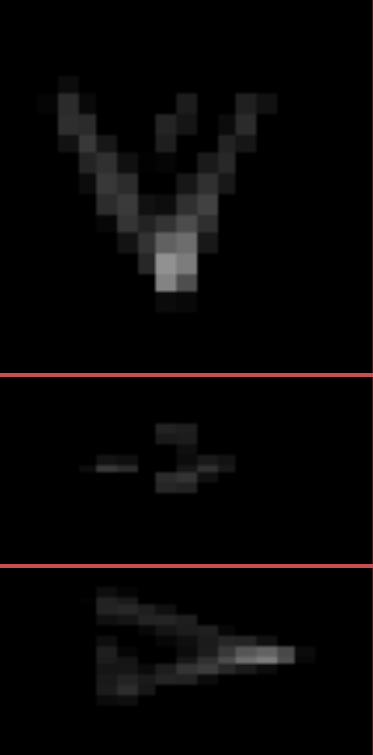}}\\

\vspace{.5mm} KACZ D$\ell^2$ + P$\ell^2$ &
\raisebox{-\totalheight/2}{\includegraphics[width=0.06\linewidth]{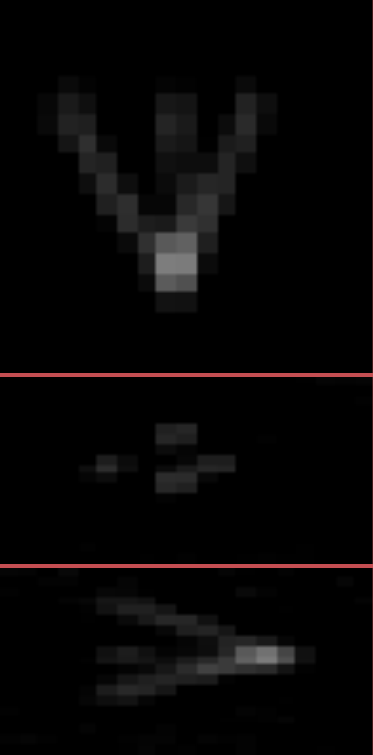}} &
\raisebox{-\totalheight/2}{\includegraphics[width=0.06\linewidth]{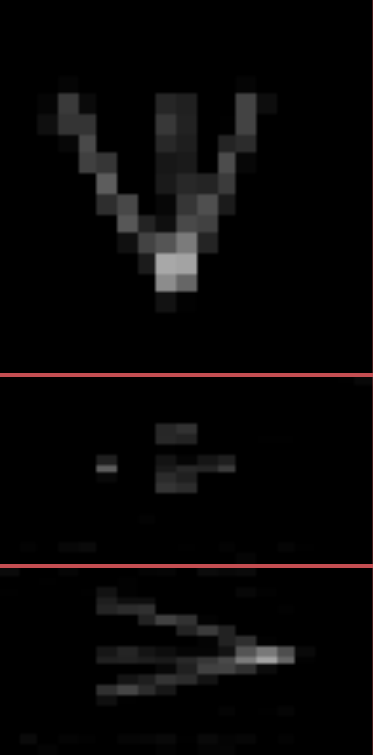}} &
\raisebox{-\totalheight/2}{\includegraphics[width=0.06\linewidth]{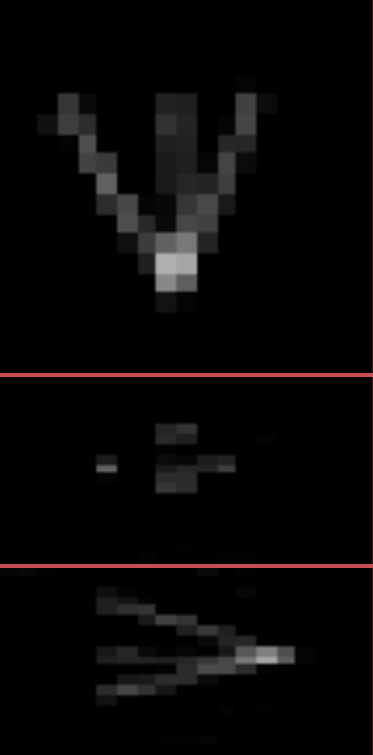}} &
\raisebox{-\totalheight/2}{\includegraphics[width=0.06\linewidth]{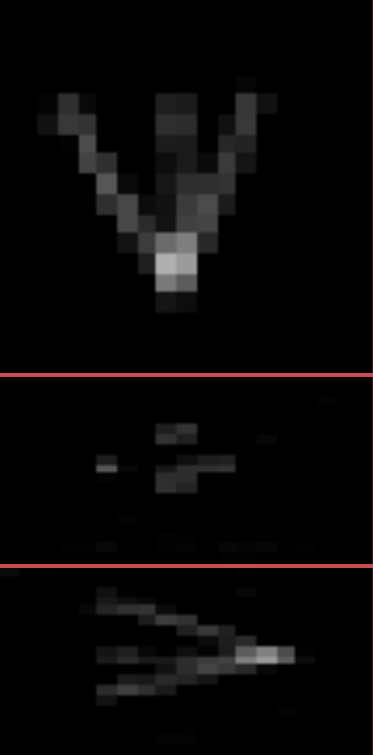}} &
\raisebox{-\totalheight/2}{\includegraphics[width=0.06\linewidth]{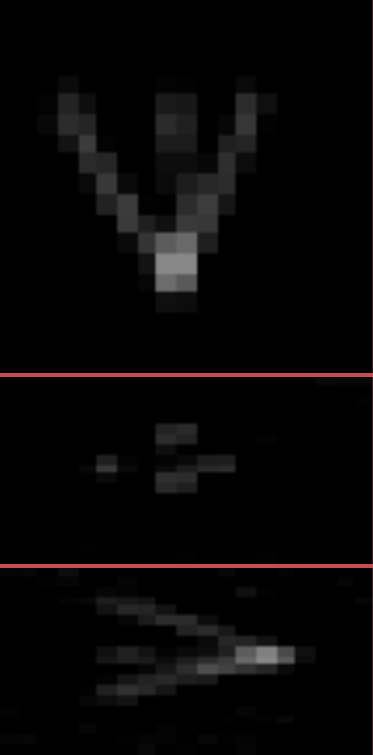}} &
\raisebox{-\totalheight/2}{\includegraphics[width=0.06\linewidth]{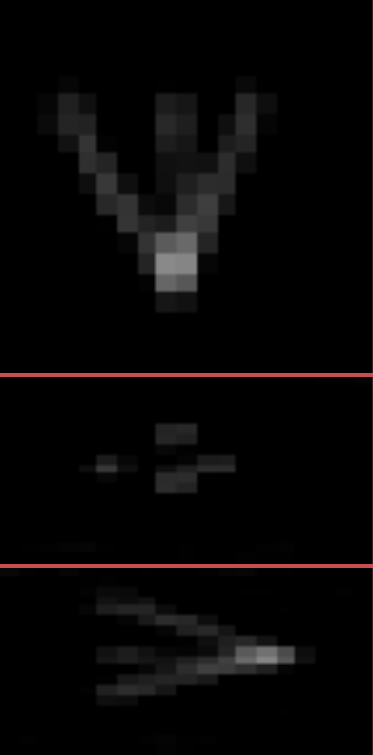}} &
\raisebox{-\totalheight/2}{\includegraphics[width=0.06\linewidth]{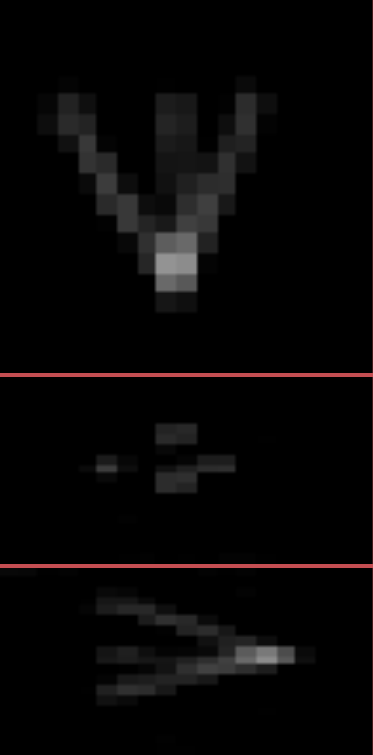}} &
\raisebox{-\totalheight/2}{\includegraphics[width=0.06\linewidth]{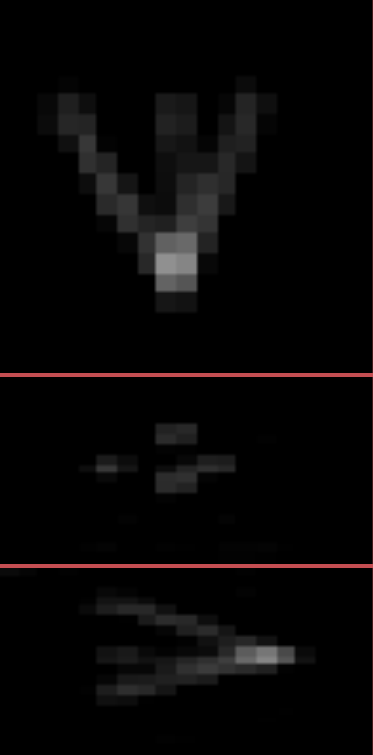}}\\

\vspace{.5mm} KACZ D$\ell^2$ + P$\ell^1$ &
\raisebox{-\totalheight/2}{\includegraphics[width=0.06\linewidth]{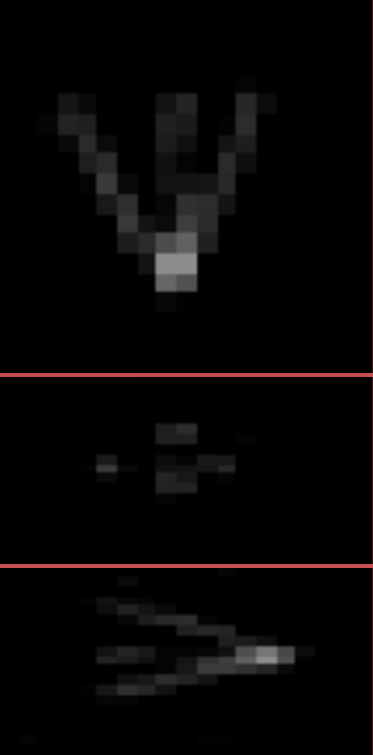}} &
\raisebox{-\totalheight/2}{\includegraphics[width=0.06\linewidth]{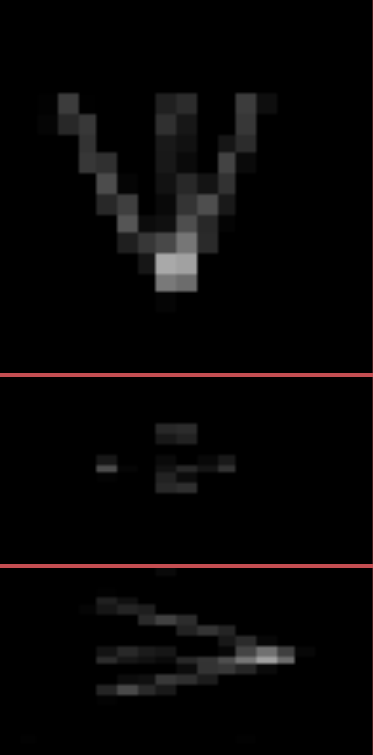}} &
\raisebox{-\totalheight/2}{\includegraphics[width=0.06\linewidth]{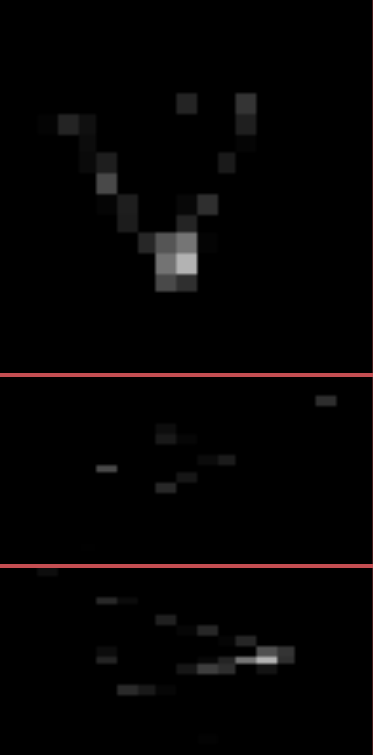}} &
\raisebox{-\totalheight/2}{\includegraphics[width=0.06\linewidth]{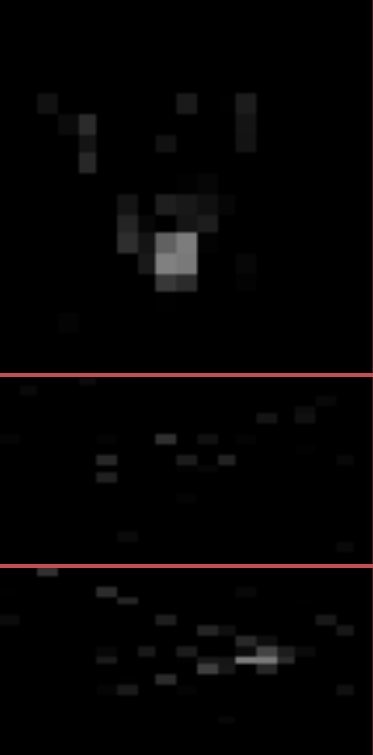}} &
\raisebox{-\totalheight/2}{\includegraphics[width=0.06\linewidth]{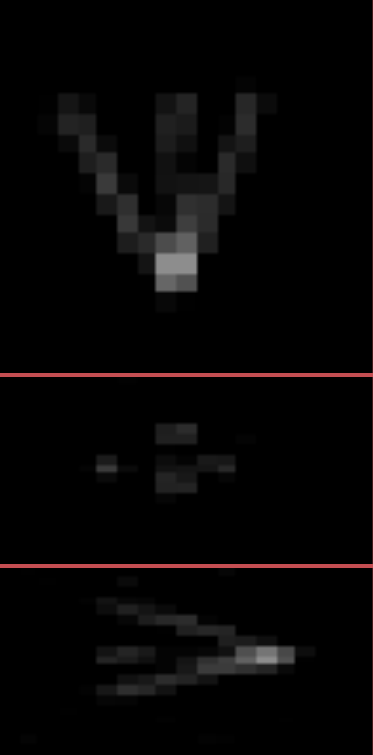}} &
\raisebox{-\totalheight/2}{\includegraphics[width=0.06\linewidth]{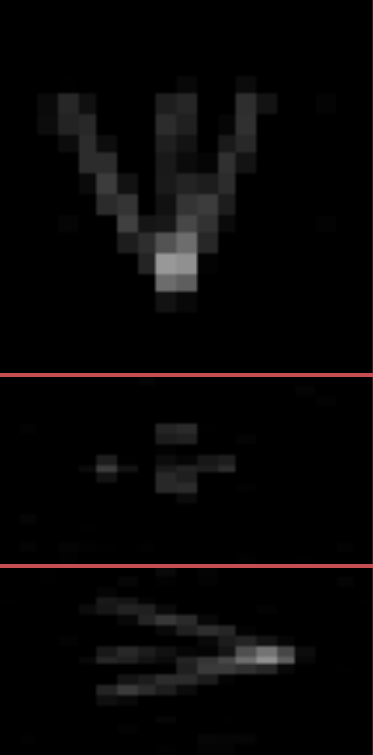}} &
\raisebox{-\totalheight/2}{\includegraphics[width=0.06\linewidth]{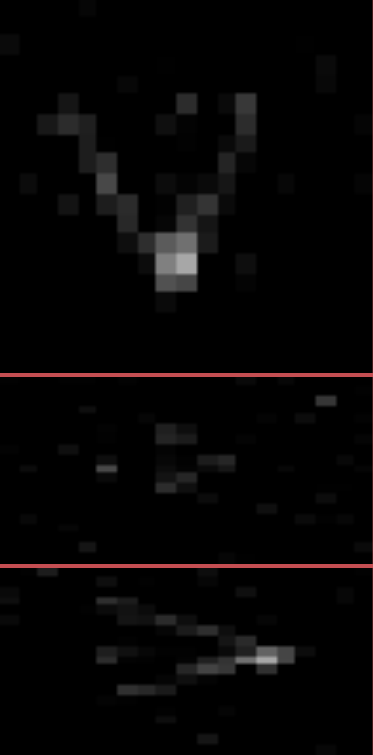}} &
\raisebox{-\totalheight/2}{\includegraphics[width=0.06\linewidth]{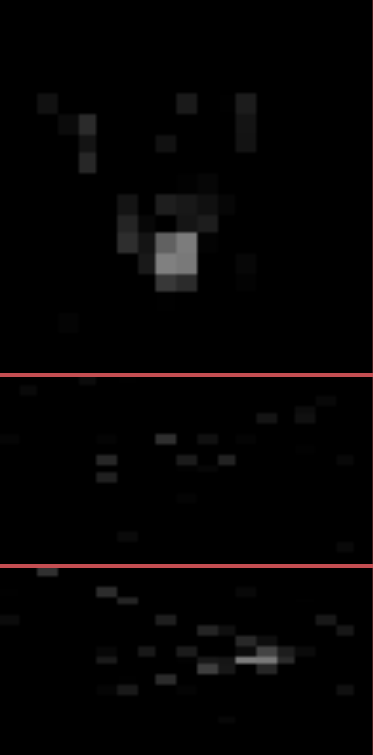}}\\

\vspace{.5mm} KACZ+TSVD D$\ell^2$ + P$\ell^1$ &
\raisebox{-\totalheight/2}{\includegraphics[width=0.06\linewidth]{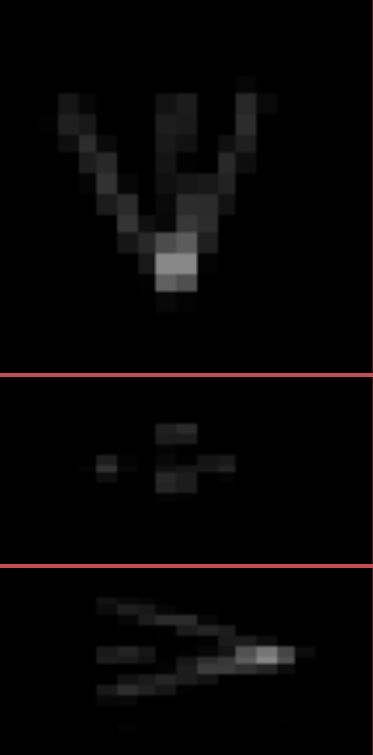}} &
\raisebox{-\totalheight/2}{\includegraphics[width=0.06\linewidth]{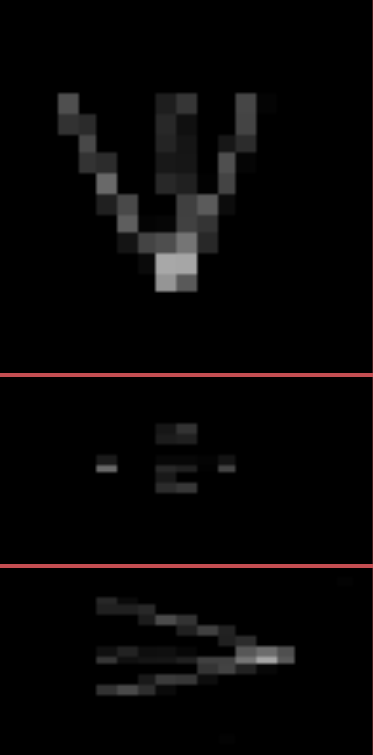}} &
\raisebox{-\totalheight/2}{\includegraphics[width=0.06\linewidth]{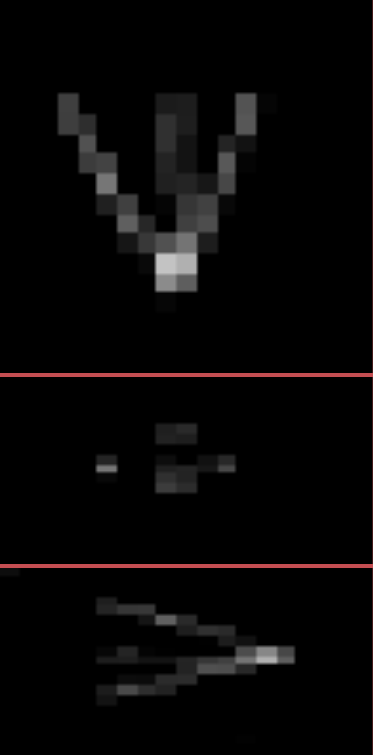}} &
\raisebox{-\totalheight/2}{\includegraphics[width=0.06\linewidth]{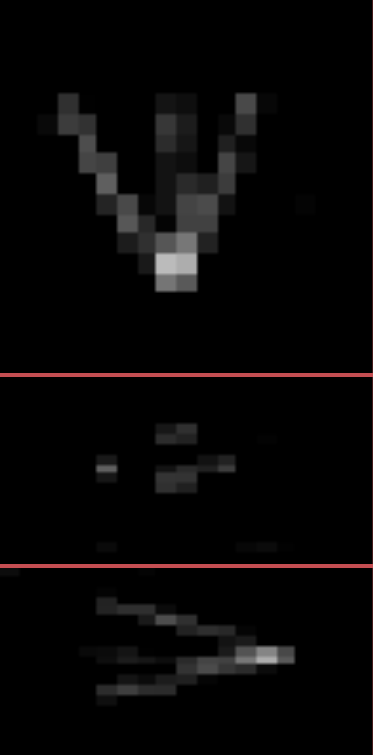}} &
\raisebox{-\totalheight/2}{\includegraphics[width=0.06\linewidth]{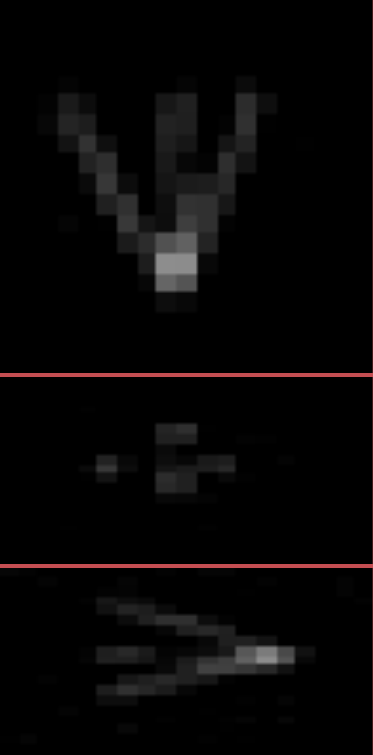}} &
\raisebox{-\totalheight/2}{\includegraphics[width=0.06\linewidth]{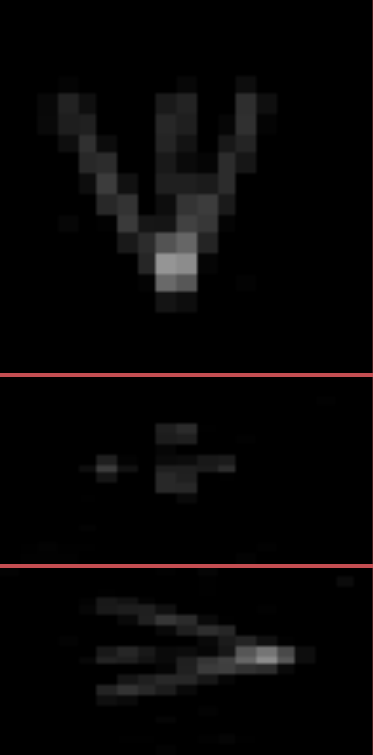}} &
\raisebox{-\totalheight/2}{\includegraphics[width=0.06\linewidth]{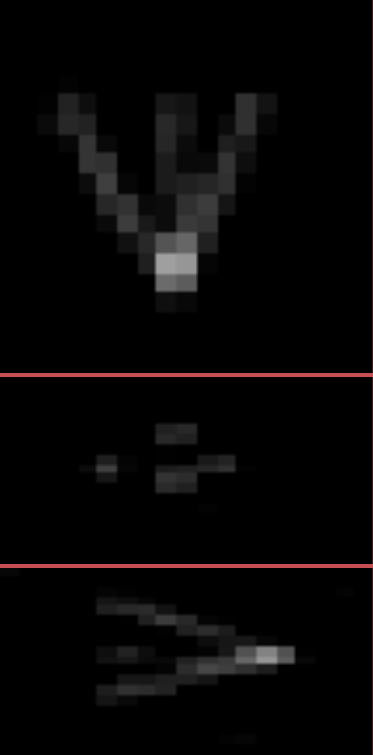}} &
\raisebox{-\totalheight/2}{\includegraphics[width=0.06\linewidth]{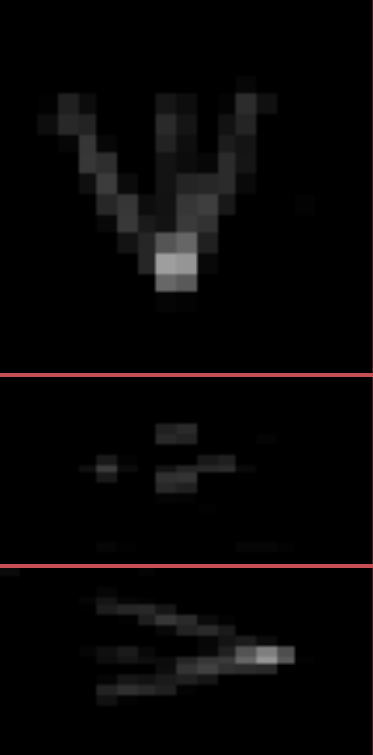}}\\

\vspace{.5mm} KACZ D$\ell^2$ + P$(\ell^1$+$\ell^2)$ &
\raisebox{-\totalheight/2}{\includegraphics[width=0.06\linewidth]{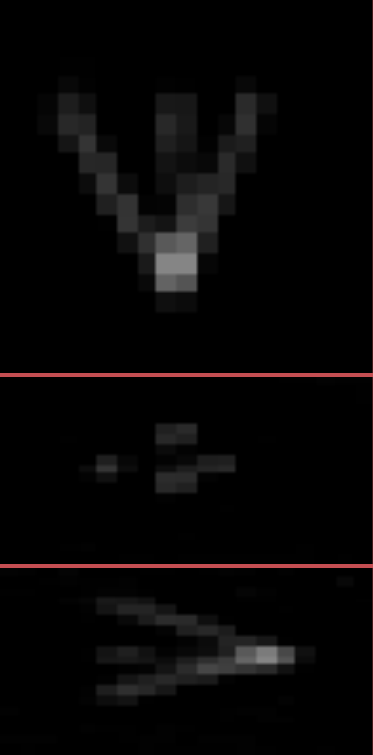}} &
\raisebox{-\totalheight/2}{\includegraphics[width=0.06\linewidth]{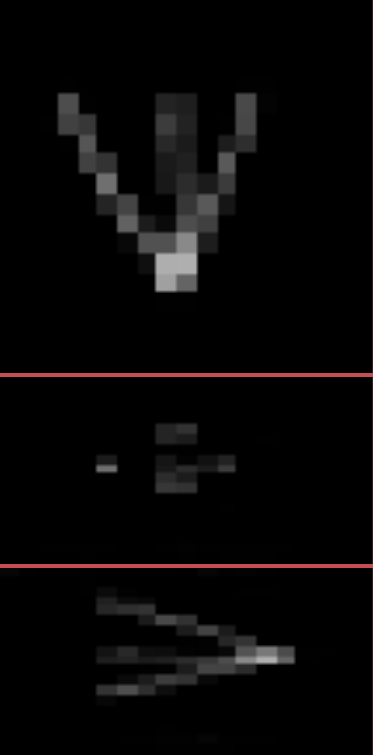}} &
\raisebox{-\totalheight/2}{\includegraphics[width=0.06\linewidth]{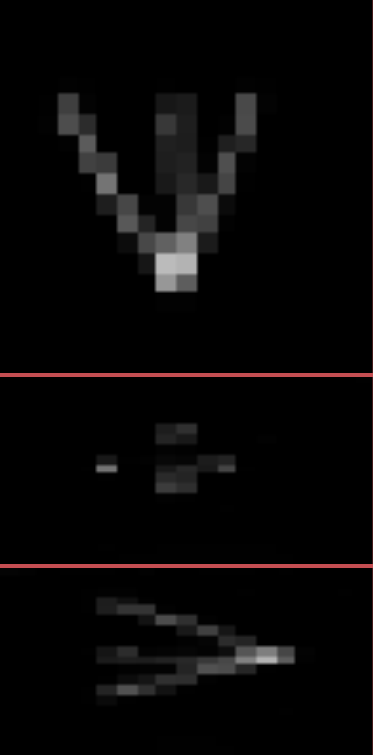}} &
\raisebox{-\totalheight/2}{\includegraphics[width=0.06\linewidth]{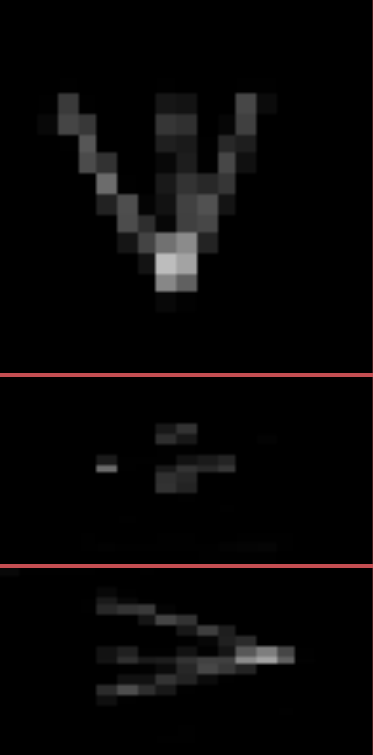}} &
\raisebox{-\totalheight/2}{\includegraphics[width=0.06\linewidth]{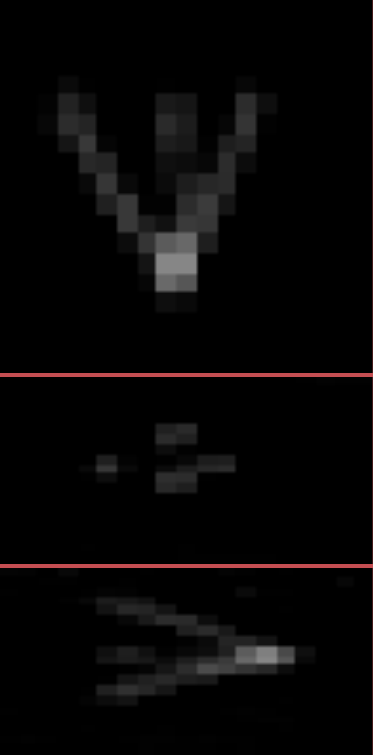}} &
\raisebox{-\totalheight/2}{\includegraphics[width=0.06\linewidth]{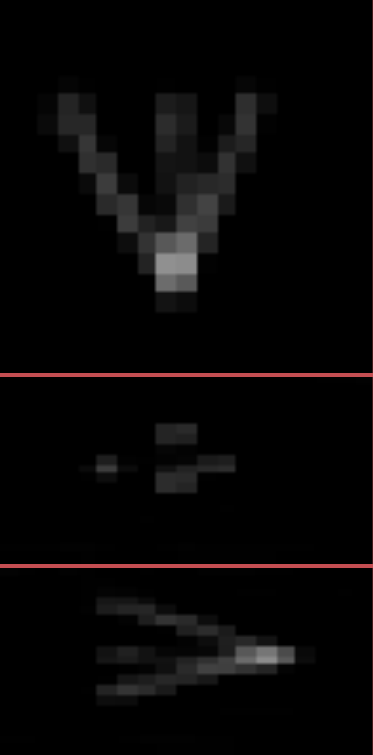}} &
\raisebox{-\totalheight/2}{\includegraphics[width=0.06\linewidth]{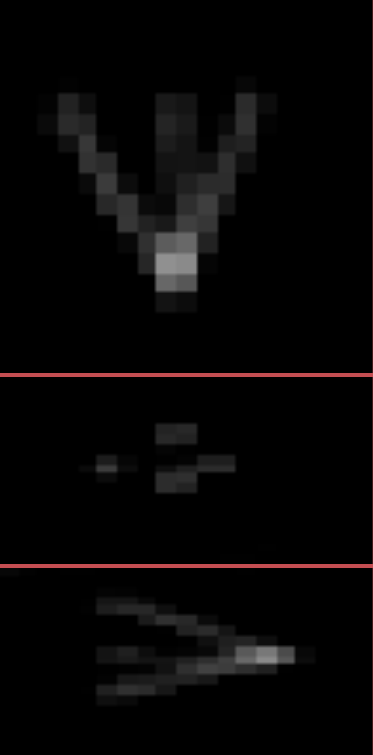}} &
\raisebox{-\totalheight/2}{\includegraphics[width=0.06\linewidth]{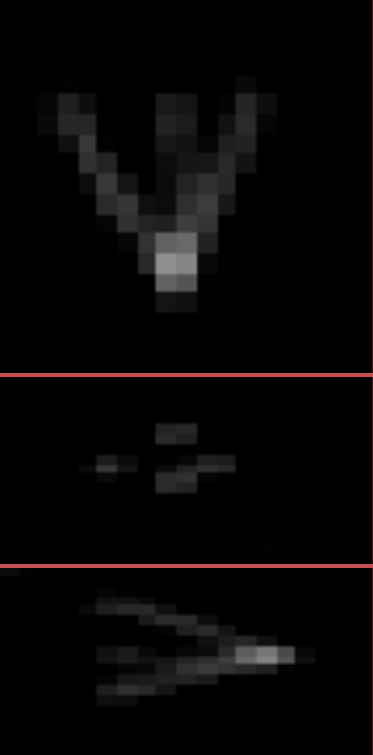}}\\

\vspace{.5mm} VAR D$\ell^1$ + P$\ell^1$  &
\raisebox{-\totalheight/2}{\includegraphics[width=0.06\linewidth]{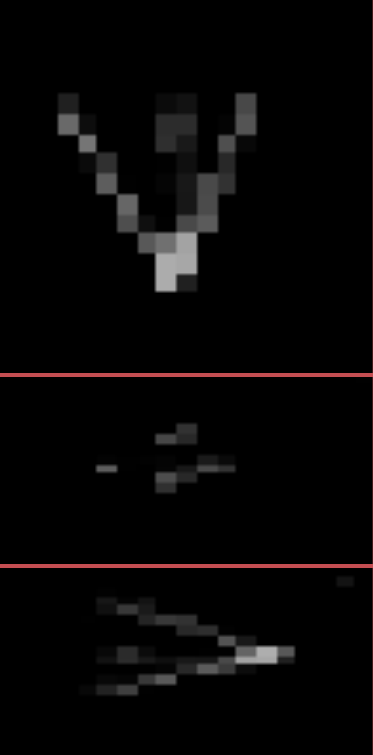}} &
\raisebox{-\totalheight/2}{\includegraphics[width=0.06\linewidth]{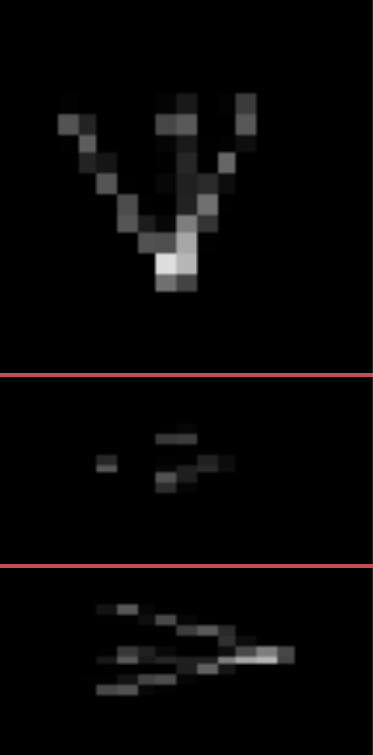}} &
\raisebox{-\totalheight/2}{\includegraphics[width=0.06\linewidth]{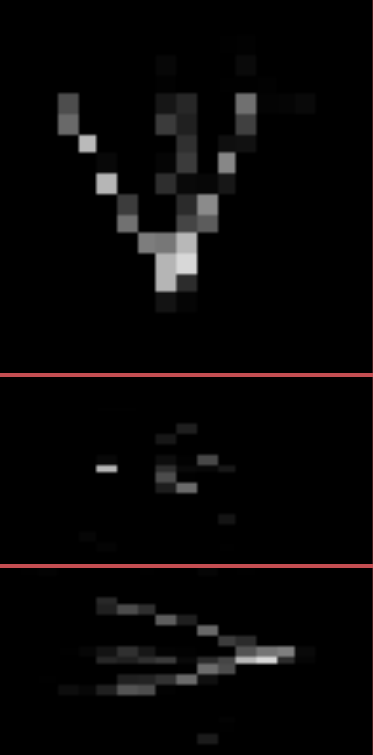}} &
\raisebox{-\totalheight/2}{\includegraphics[width=0.06\linewidth]{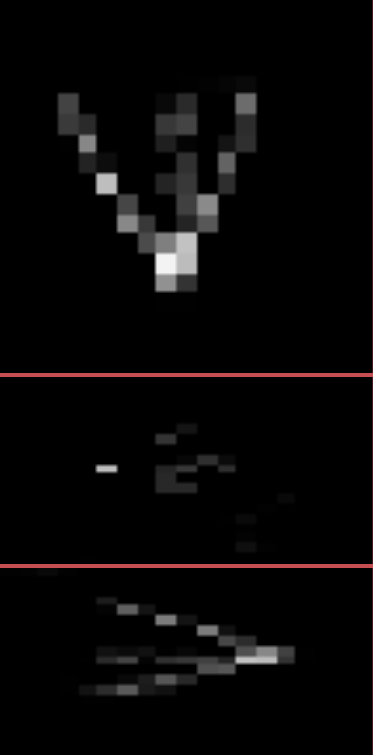}} &
\raisebox{-\totalheight/2}{\includegraphics[width=0.06\linewidth]{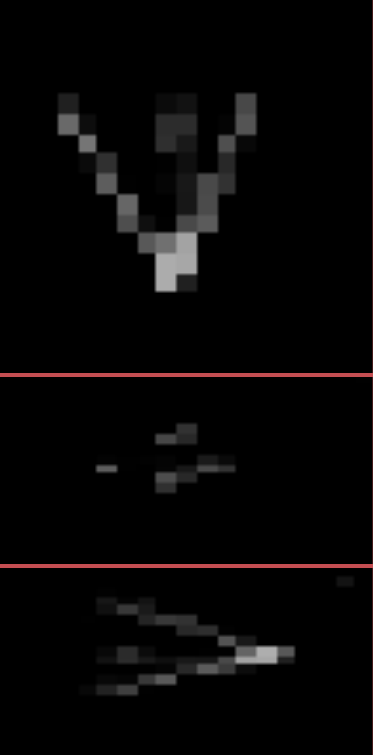}} &
\raisebox{-\totalheight/2}{\includegraphics[width=0.06\linewidth]{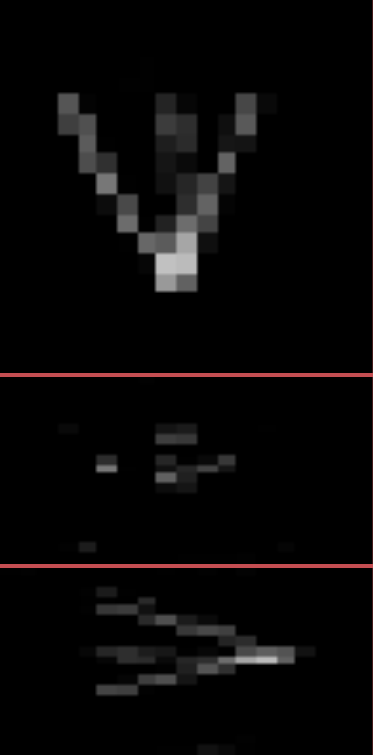}} &
\raisebox{-\totalheight/2}{\includegraphics[width=0.06\linewidth]{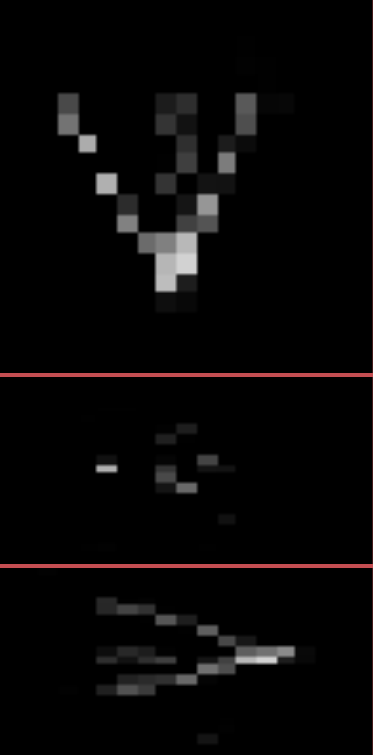}} &
\raisebox{-\totalheight/2}{\includegraphics[width=0.06\linewidth]{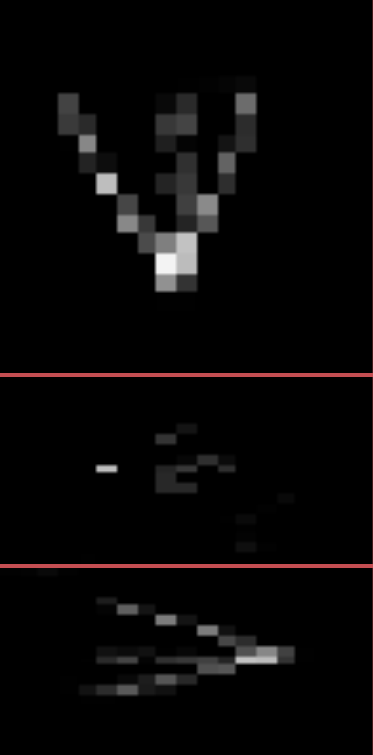}}\\

\vspace{.5mm} VAR D$\ell^1$ + P$\ell^2$ &
\raisebox{-\totalheight/2}{\includegraphics[width=0.06\linewidth]{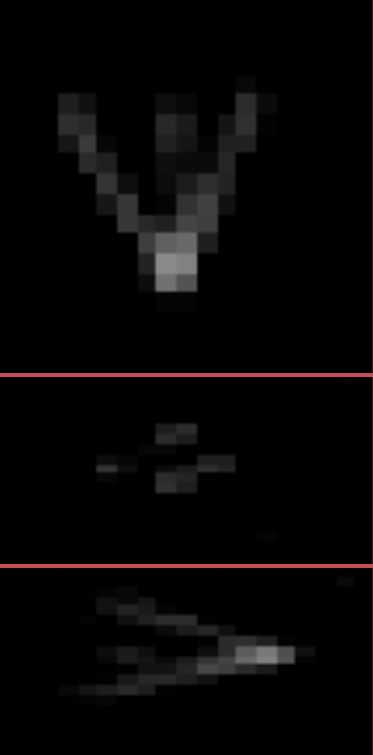}} &
\raisebox{-\totalheight/2}{\includegraphics[width=0.06\linewidth]{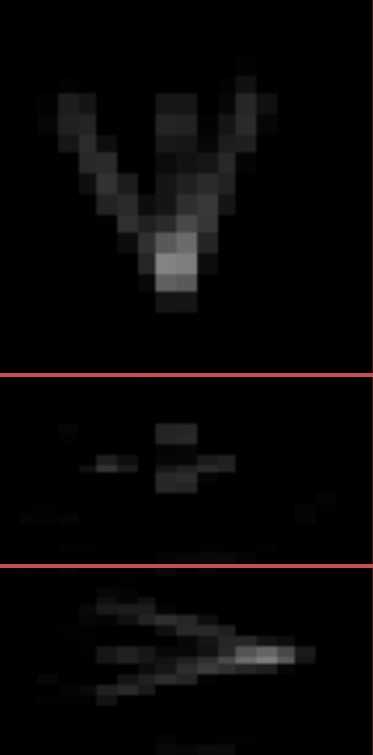}} &
\raisebox{-\totalheight/2}{\includegraphics[width=0.06\linewidth]{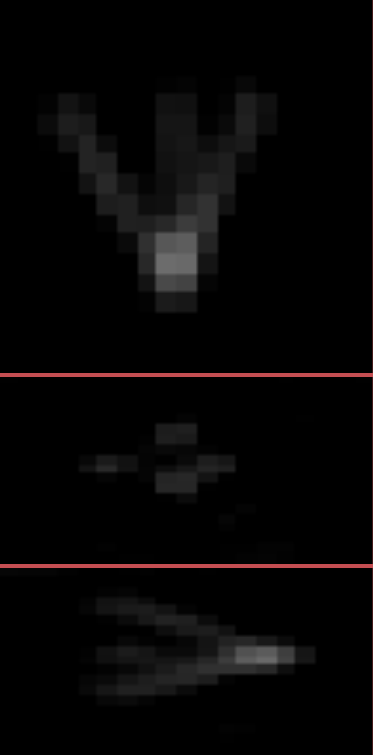}} &
\raisebox{-\totalheight/2}{\includegraphics[width=0.06\linewidth]{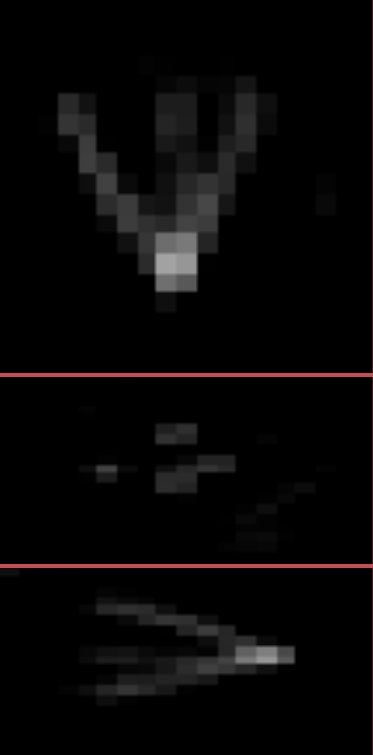}} &
\raisebox{-\totalheight/2}{\includegraphics[width=0.06\linewidth]{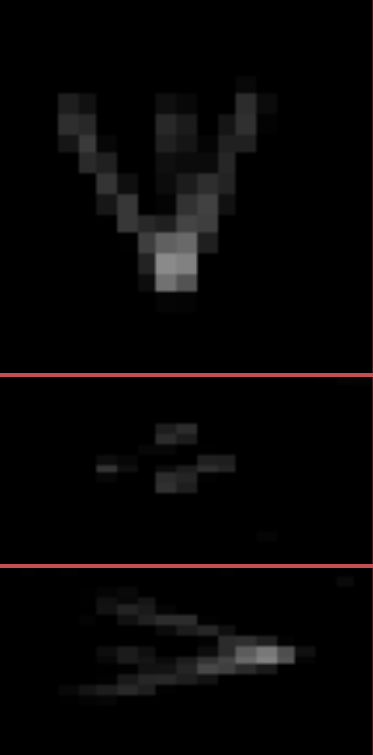}} &
\raisebox{-\totalheight/2}{\includegraphics[width=0.06\linewidth]{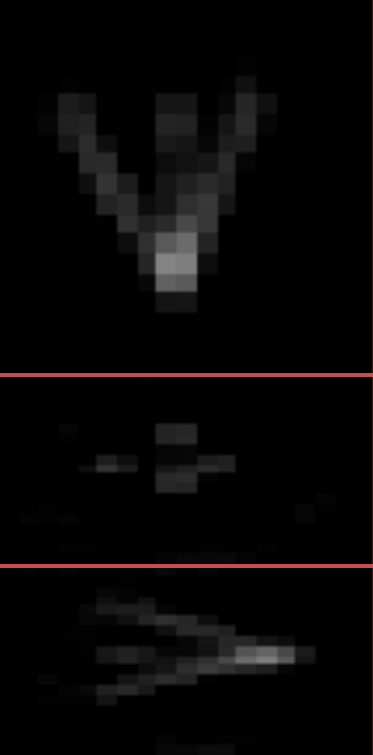}} &
\raisebox{-\totalheight/2}{\includegraphics[width=0.06\linewidth]{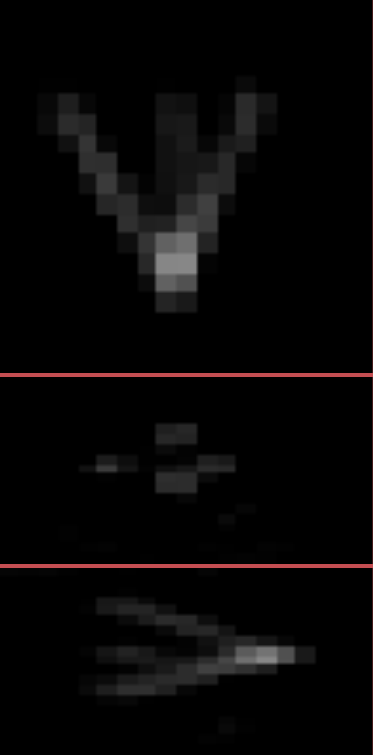}} &
\raisebox{-\totalheight/2}{\includegraphics[width=0.06\linewidth]{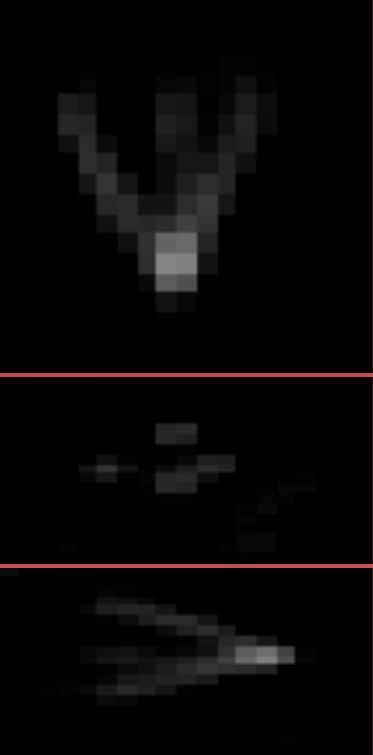}}\\

\vspace{.5mm} VAR D$\ell^1$ + P TV & 
\raisebox{-\totalheight/2}{\includegraphics[width=0.06\linewidth]{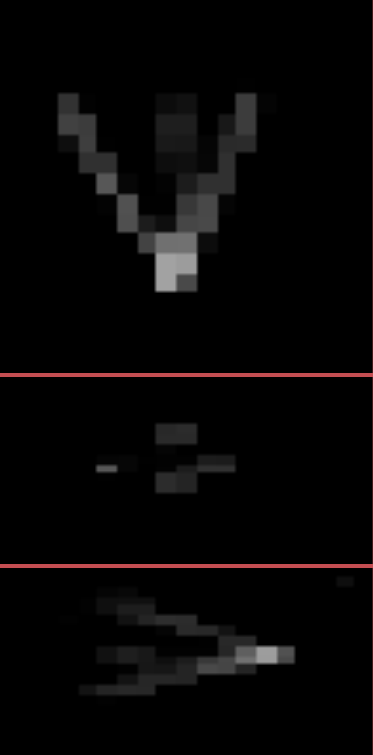}} &
\raisebox{-\totalheight/2}{\includegraphics[width=0.06\linewidth]{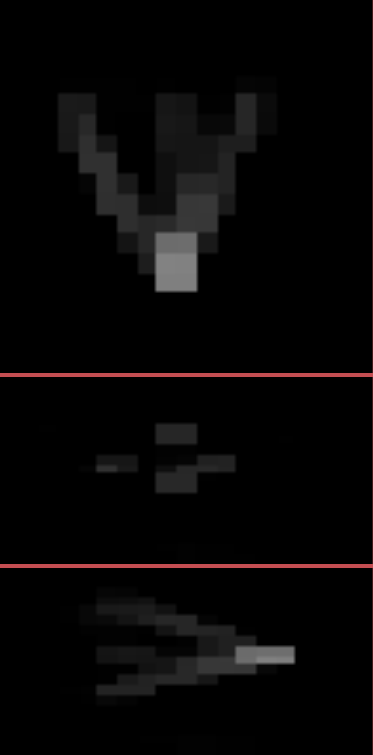}} &
\raisebox{-\totalheight/2}{\includegraphics[width=0.06\linewidth]{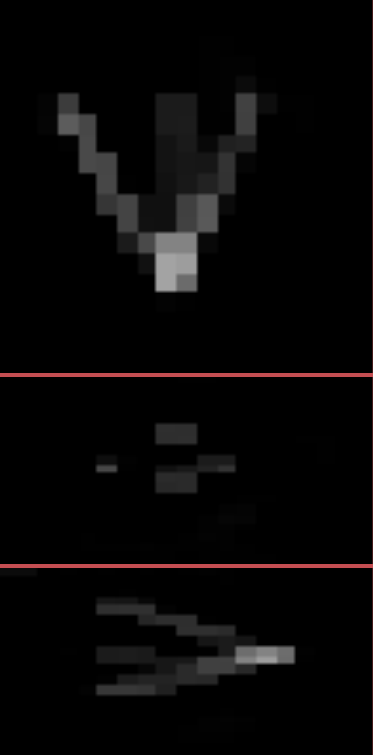}} &
\raisebox{-\totalheight/2}{\includegraphics[width=0.06\linewidth]{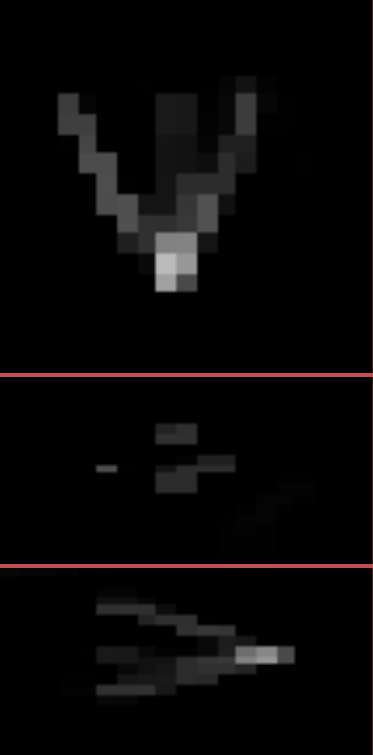}} &
\raisebox{-\totalheight/2}{\includegraphics[width=0.06\linewidth]{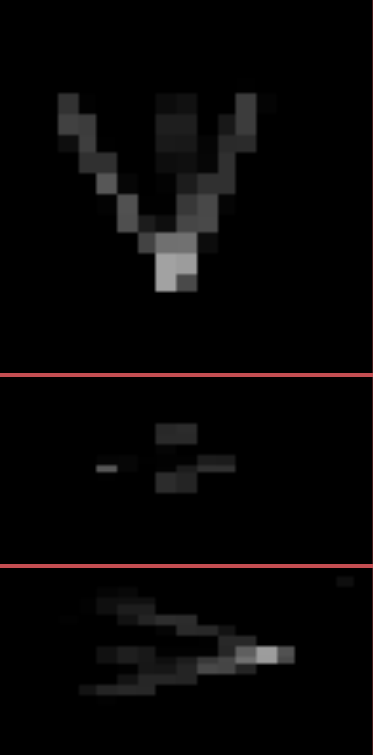}} &
\raisebox{-\totalheight/2}{\includegraphics[width=0.06\linewidth]{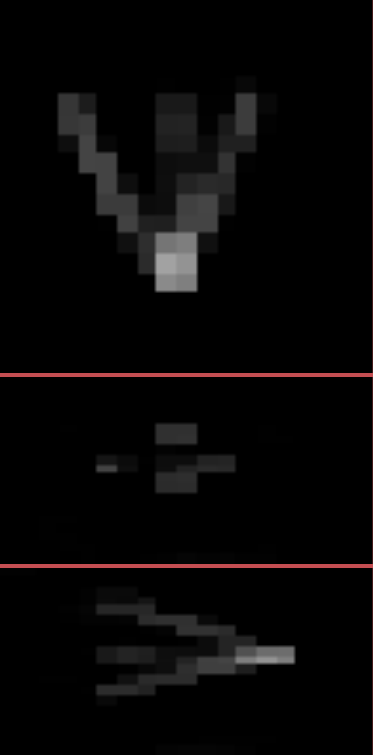}} &
\raisebox{-\totalheight/2}{\includegraphics[width=0.06\linewidth]{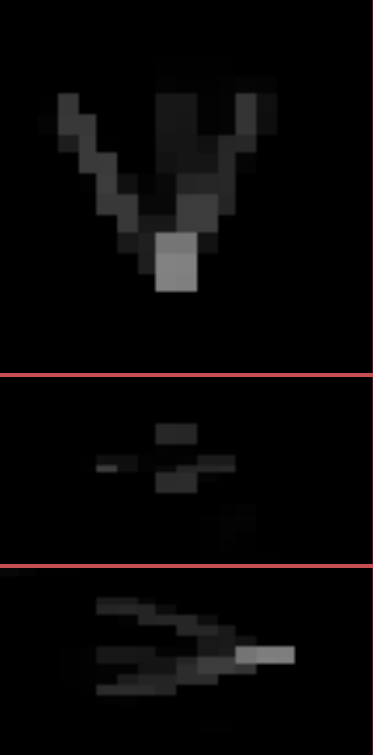}} &
\raisebox{-\totalheight/2}{\includegraphics[width=0.06\linewidth]{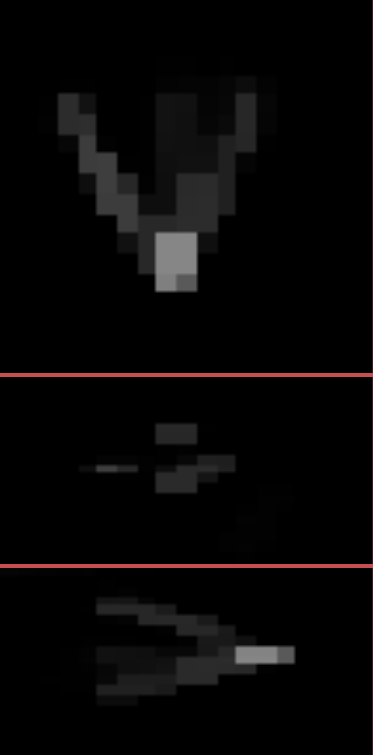}}\\

 \hline
\end{tabular}
\caption{The best reconstructions for the resolution phantom, corresponding to the values in Tables~\ref{tab:whitened_resolution_metrics} and~\ref{tab:whitened_resolution_parameters}. The color scale goes from $0$ (black) to $40$ (white). Each image is separated into three vertically stacked parts, which are separated by red lines. Each of these parts represents one of the three central slices/planes of the three-dimensional reconstruction (x-y-, x-z-, and y-z-plane from top to bottom).}
\label{tab:whitened_resolution_images}
\end{figure*}

Qualitatively we present the results in Figures~\ref{tab:3dTable}, \ref{tab:whitened_shape_images}, and~\ref{tab:whitened_resolution_images}. Figure~\ref{tab:3dTable} utilizes a three-dimensional visualization to compare the best PSNR reconstruction with the ground truth phantom for the shape and resolution phantom respectively. Figures~\ref{tab:whitened_shape_images} and~\ref{tab:whitened_resolution_images} display two-dimensional cross sections of the reconstructions corresponding to the values of Tables~\ref{tab:whitened_shape_metrics} and~\ref{tab:whitened_resolution_metrics} -- except for VAR D$\ell^2$ since these methods result in the worst reconstructions.
The main observations are as follows:
\begin{enumerate}
    \item The proposed DIP approach achieves the overall highest PSNR and SSIM values for both the shape and the resolution phantom. 
    Even for threshold settings ($\tau$'s/columns), in which the DIP produces inferior peak results, they are usually comparable to the best method.
    \item Quantitatively, the VAR D$\ell^1$ reconstructions are superior to the VAR D$\ell^2$ reconstructions as the Tables~\ref{tab:whitened_shape_metrics}, and~\ref{tab:whitened_resolution_metrics} show.
Although in the present work, we project the data as well as the noise onto the subspace spanned by the singular values corresponding to the largest singular values, this finding is in line with the observation in~\cite{KluthJin2019b_preprint}.
    \item Within the KACZ methods, we can observe that the combination of $\ell^2$- and $\ell^1$-term is superior to the standard KACZ method solely using the $\ell^2$-term. 
    \item The KACZ methods, which utilize early-stopping (see Tables \ref{tab:whitened_shape_parameters} and \ref{tab:whitened_resolution_parameters}, produce better results than the purely variational methods (VAR).
    \item The KACZ D$\ell^2$ + P$\ell^1$ method can suffer from instability issues if it has to deal with matrix rows of small norm. However, further reducing the rank using TSVD has a beneficial influence on the reconstruction quality, but in almost all cases, it cannot reach the performance of KACZ D$\ell^2$ + P$(\ell^1+\ell^2)$ which is an alternative approach to stabilize the reconstruction.
    \item Particularly for the resolution phantom, we can observe that in almost all cases, larger SNR thresholds result in worse reconstructions.
\end{enumerate}

\section{Discussion and conclusion}
\label{sec:discussion}
In summary, we have multiple main findings illustrated by the quantitative results in the present work. First, the proposed deep image prior approach based on an autoencoder architecture outperforms all other methods considered in this study. The superior behavior of the DIP is particularly found for smaller numbers of training steps (early stopping) when determining the network parameters. Second, the results for the KACZ class, which is also characterized by limited numbers of iterations (early stopping), illustrate the success of the algebraic reconstruction technique with sole $\ell^2$-penalty in MPI \cite{knopp2017magnetic}. The data and noise structure in MPI seem to be beneficial for the iterative nature of DIP and KACZ methods. Within the KACZ methods, it turned out that the combination of $\ell^2$-term and $\ell^1$-term can result in improved reconstructions compared to the commonly used reconstruction method in MPI. One drawback of the superior reachable image quality in this method is the additional regularization parameter, which needs to be appropriately chosen. Third, this study emphasizes the need for a careful discussion and distinction between different regularization techniques. Variational approaches (VAR) strongly rely on a proper choice of the data fidelity term, i.e., prior knowledge on the data space and noise distribution. A standard $\ell^2$-data fidelity term is not well suited for MPI reconstructions. Similar to the findings in \cite{KluthJin2019b_preprint}, an $\ell^1$-data fidelity term is better suited for MPI reconstruction, not only in the DIP method but also in the VAR methods. This holds even if the data space is projected onto a subspace spanned by singular vectors. The latter finding also emphasizes the need for future research on the identification and proper treatment of the MPI noise structure for the purpose of image reconstruction.

\bibliographystyle{IEEEtran}
\bibliography{IEEEabrv,literature}

\FloatBarrier
\clearpage

\begin{center}
    \begin{tabular}{c}
        \centering\large
        \specialcell{
        \vspace{8cm}\ 
        \\
        \textbf{Supplementary material of}
        \\
        \textbf{Deep image prior for 3D magnetic particle imaging:}
        \\
        \textbf{A quantitative comparison of regularization techniques on \texttt{Open MPI dataset}}
        }
    \end{tabular}
\end{center}
\clearpage

\begin{table*}
\centering
\rowcolors{1}{}{lightgray}
\begin{tabular}{ |l|||c|c|c|c||c|c|c|c|  }
 \hline
 Methods & \multicolumn{4}{c||}{PSNRs} & \multicolumn{4}{c|}{SSIMs} \\
 \hline
 \rowcolor{white}
 \diagbox[dir=NE]{}{} & $\tau=0$ & $\tau=1$ & $\tau=3$ & $\tau=5$ & $\tau=0$ & $\tau=1$ & $\tau=3$ & $\tau=5$\\
 \hline \hline
DIP D$\ell^1$ + P-                    &     26.70 &     \textbf{29.59} &     \textbf{29.79} &     \underline{\textbf{29.81}} &     0.926 &     \underline{\textbf{0.966}} &     \textbf{0.962} &     \textbf{0.965} \\
KACZ D$\ell^2$ + P$\ell^2$            &     27.50 &     27.69 &     27.78 &     27.78 &     0.925 &     0.935 &     0.943 &     0.939 \\
KACZ D$\ell^2$ + P$\ell^1$            &     24.06 &     24.03 &     16.71 &     14.69 &     0.794 &     0.786 &     0.458 &     0.209 \\
KACZ+TSVD D$\ell^2$ + P$\ell^1$       &     27.14 &     27.23 &     27.39 &     27.54 &     0.919 &     0.916 &     0.937 &     0.939 \\
KACZ D$\ell^2$ + P$(\ell^1$+$\ell^2)$ &     \textbf{28.03} &     28.28 &     28.36 &     28.20 &     \textbf{0.939} &     0.943 &     0.947 &     0.944 \\
VAR D$\ell^1$ + P$\ell^1$             &     18.25 &     19.93 &     21.47 &     21.83 &     0.749 &     0.799 &     0.809 &     0.842 \\
VAR D$\ell^1$ + P$\ell^2$             &     23.78 &     25.84 &     24.88 &     25.44 &     0.835 &     0.834 &     0.792 &     0.798 \\
VAR D$\ell^1$ + P TV                  &     24.37 &     26.21 &     24.87 &     26.42 &     0.848 &     0.916 &     0.866 &     0.900 \\
VAR D$\ell^2$ + P$\ell^1$             &     17.04 &     19.01 &     23.62 &     24.96 &     0.596 &     0.684 &     0.887 &     0.896 \\
VAR D$\ell^2$ + P$\ell^2$             &     19.85 &     21.32 &     25.12 &     25.43 &     0.638 &     0.700 &     0.887 &     0.883 \\
VAR D$\ell^2$ + P TV                  &     19.52 &     20.69 &     24.70 &     24.78 &     0.613 &     0.685 &     0.887 &     0.891 \\
\hline
\end{tabular}
\caption{PSNR and SSIM values as achieved by the different SNR thresholds $\tau$ settings for the shape phantom \underline{without} whitening. Largest values for each column are in bold font. The overall largest value is underlined.}
\label{tab:non_whitened_shape_metrics}
\end{table*}

\begin{table*}
\centering
\rowcolors{1}{}{lightgray}
\begin{tabular}{ |l|||c|c|c|c||c|c|c|c|  }
 \hline
 Methods & \multicolumn{4}{c||}{PSNRs} & \multicolumn{4}{c|}{SSIMs} \\
 \hline
 \rowcolor{white}
 \diagbox[dir=NE]{}{} & $\tau=0$ & $\tau=1$ & $\tau=3$ & $\tau=5$ & $\tau=0$ & $\tau=1$ & $\tau=3$ & $\tau=5$\\
 \hline \hline
DIP D$\ell^1$ + P-                    &     \underline{\textbf{32.02}} &     \textbf{31.97} &     \textbf{31.66} &     \textbf{31.63} &     0.912 &     \textbf{0.945} &     0.939 &     0.943 \\
KACZ D$\ell^2$ + P$\ell^2$            &     31.16 &     31.38 &     31.45 &     31.27 &     0.938 &     0.941 &     0.946 &     0.944 \\
KACZ D$\ell^2$ + P$\ell^1$            &     31.66 &     31.58 &     29.34 &     28.99 &     0.938 &     0.937 &     0.828 &     0.761 \\
KACZ+TSVD D$\ell^2$ + P$\ell^1$       &     31.68 &     31.71 &     31.22 &     30.86 &     0.939 &     0.941 &     0.936 &     0.934 \\
KACZ D$\ell^2$ + P$(\ell^1$+$\ell^2)$ &     31.56 &     31.69 &     31.65 &     31.46 &     \textbf{0.942} &     0.944 &     \underline{\textbf{0.947}} &     \textbf{0.945} \\
VAR D$\ell^1$ + P$\ell^1$             &     29.85 &     30.57 &     29.80 &     29.72 &     0.896 &     0.904 &     0.905 &     0.889 \\
VAR D$\ell^1$ + P$\ell^2$             &     30.87 &     31.34 &     31.00 &     30.93 &     0.921 &     0.940 &     0.936 &     0.934 \\
VAR D$\ell^1$ + P TV                  &     31.02 &     31.29 &     30.61 &     30.35 &     0.916 &     0.938 &     0.929 &     0.927 \\
VAR D$\ell^2$ + P$\ell^1$             &     28.87 &     29.17 &     29.14 &     29.82 &     0.782 &     0.829 &     0.882 &     0.898 \\
VAR D$\ell^2$ + P$\ell^2$             &     29.49 &     29.53 &     30.42 &     30.42 &     0.811 &     0.847 &     0.906 &     0.911 \\
VAR D$\ell^2$ + P TV                  &     29.20 &     29.39 &     29.87 &     29.85 &     0.808 &     0.845 &     0.899 &     0.908 \\
 \hline
\end{tabular}
\caption{PSNR and SSIM values as achieved by the different SNR thresholds $\tau$ settings for the resolution phantom \underline{without} whitening. Largest values for each column are in bold font. The overall largest value is underlined.}
\label{tab:non_whitened_resolution_metrics}
\end{table*}

\begin{figure*}
\centering
\rowcolors{1}{}{lightgray}
\begin{tabular}{c   c c c c  c c c c}
 \hline
 Methods & \multicolumn{4}{c  }{PSNRs} & \multicolumn{4}{c }{SSIMs} \\
 \hline
 \rowcolor{white}
  & $\tau=0$ & $\tau=1$ & $\tau=3$ & $\tau=5$ & $\tau=0$ & $\tau=1$ & $\tau=3$ & $\tau=5$\\
 \hline\\[-2.0ex]

\vspace{.5mm} DIP D$\ell^1$ + P- &
\raisebox{-\totalheight/2}{\includegraphics[width=0.06\linewidth]{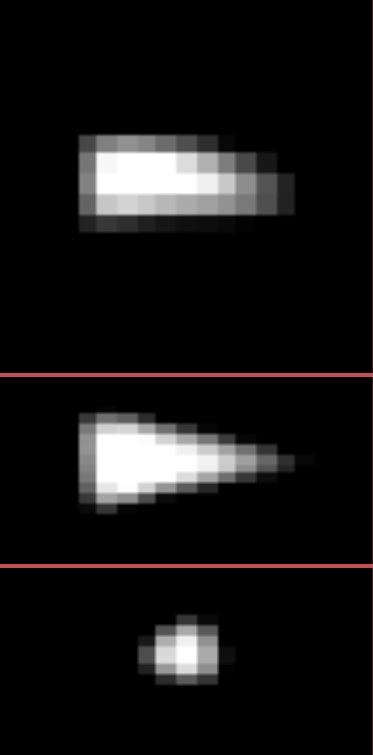}} &
\raisebox{-\totalheight/2}{\includegraphics[width=0.06\linewidth]{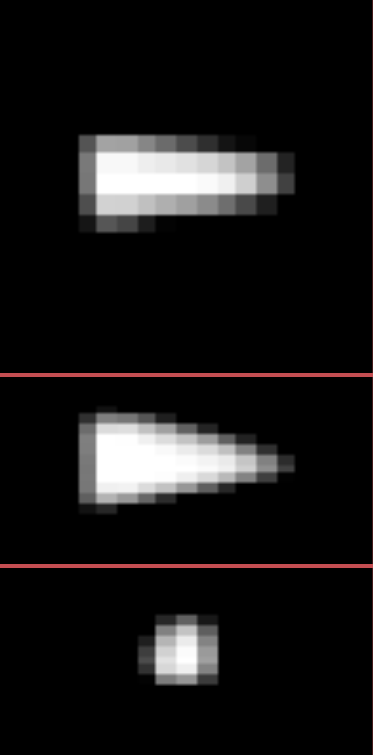}} &
\raisebox{-\totalheight/2}{\includegraphics[width=0.06\linewidth]{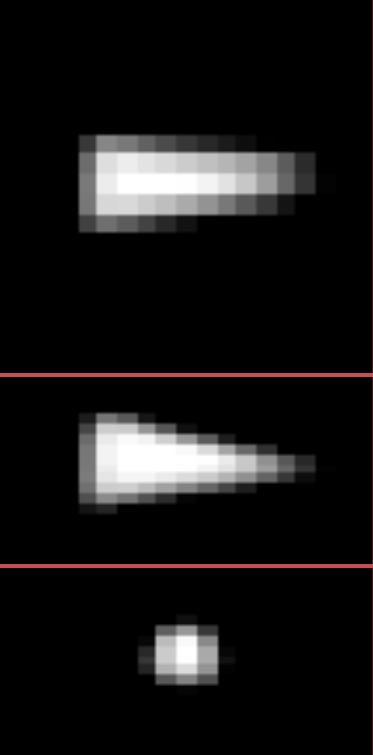}} &
\raisebox{-\totalheight/2}{\includegraphics[width=0.06\linewidth]{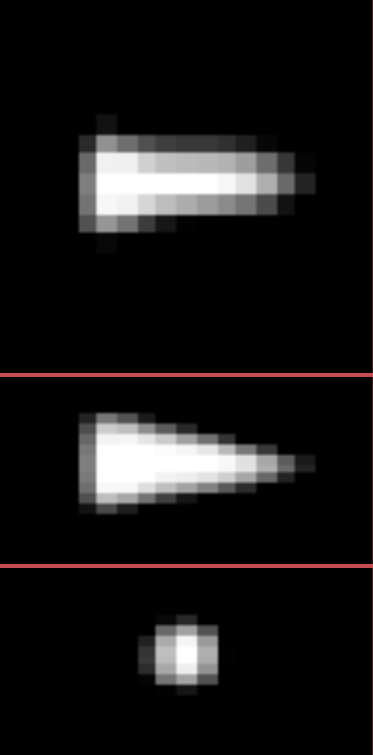}} &
\raisebox{-\totalheight/2}{\includegraphics[width=0.06\linewidth]{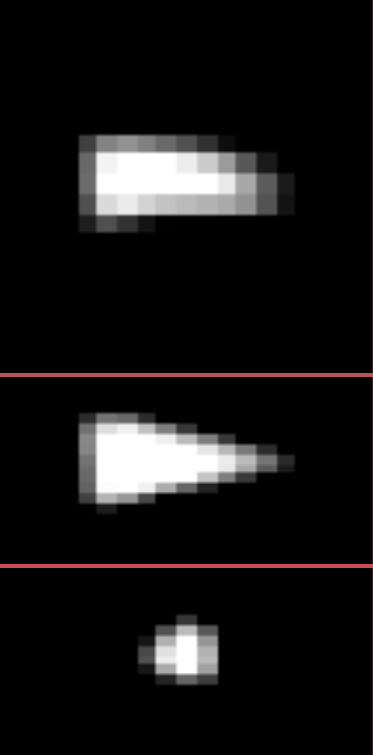}} &
\raisebox{-\totalheight/2}{\includegraphics[width=0.06\linewidth]{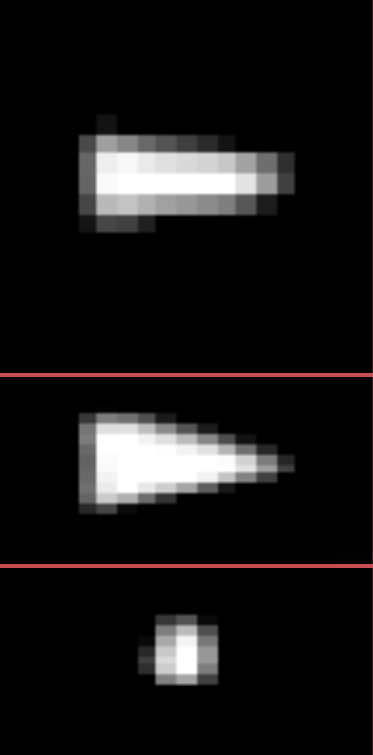}} &
\raisebox{-\totalheight/2}{\includegraphics[width=0.06\linewidth]{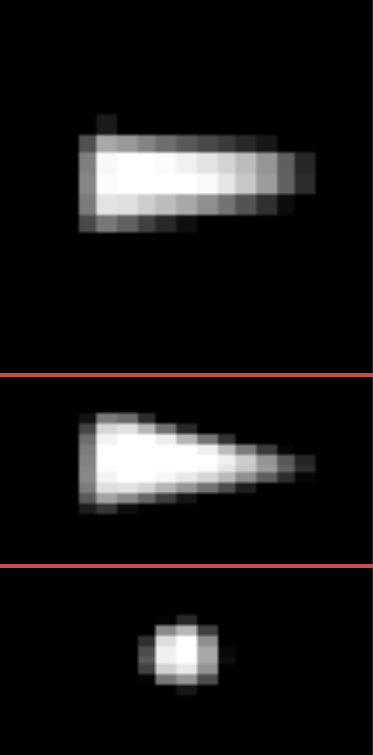}} &
\raisebox{-\totalheight/2}{\includegraphics[width=0.06\linewidth]{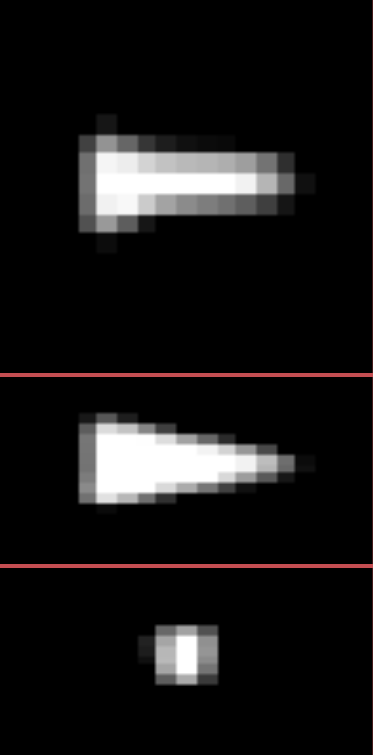}}\\

\vspace{.5mm} KACZ D$\ell^2$ + P$\ell^2$ &
\raisebox{-\totalheight/2}{\includegraphics[width=0.06\linewidth]{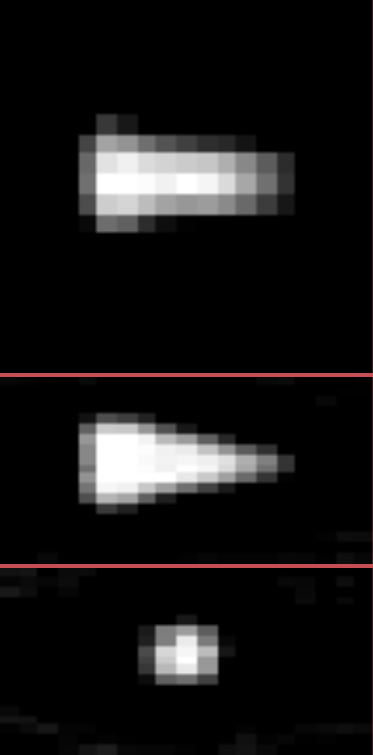}} &
\raisebox{-\totalheight/2}{\includegraphics[width=0.06\linewidth]{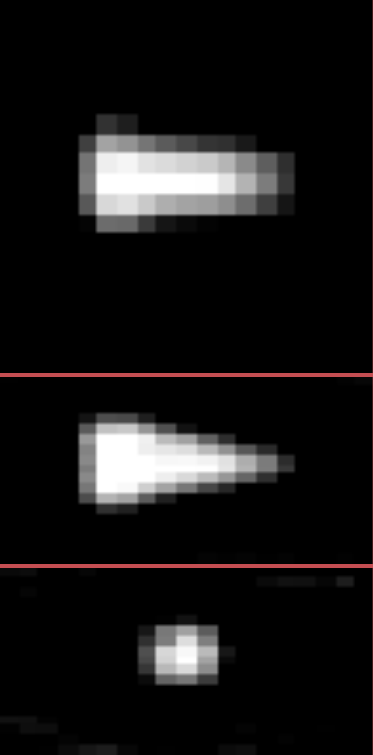}} &
\raisebox{-\totalheight/2}{\includegraphics[width=0.06\linewidth]{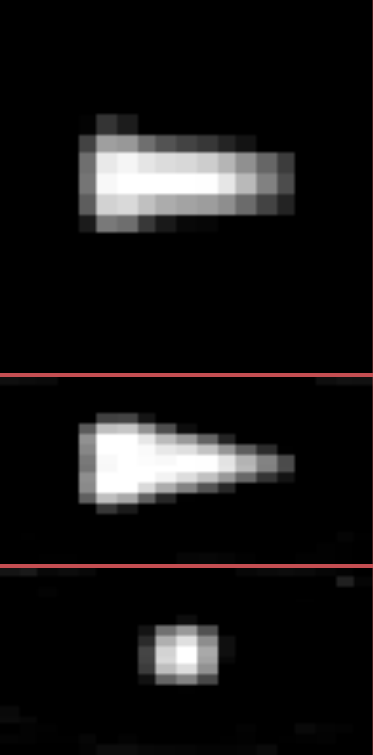}} &
\raisebox{-\totalheight/2}{\includegraphics[width=0.06\linewidth]{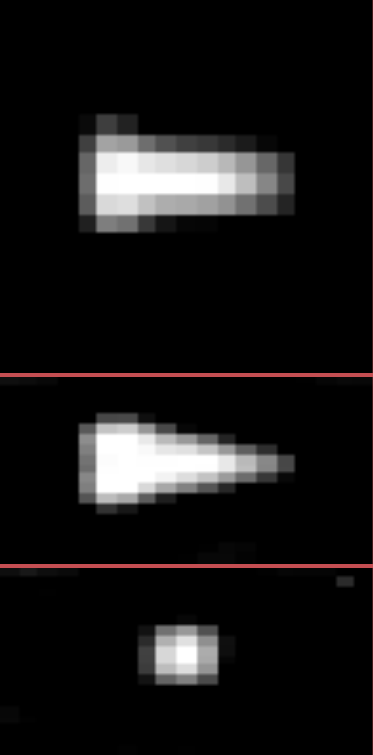}} &
\raisebox{-\totalheight/2}{\includegraphics[width=0.06\linewidth]{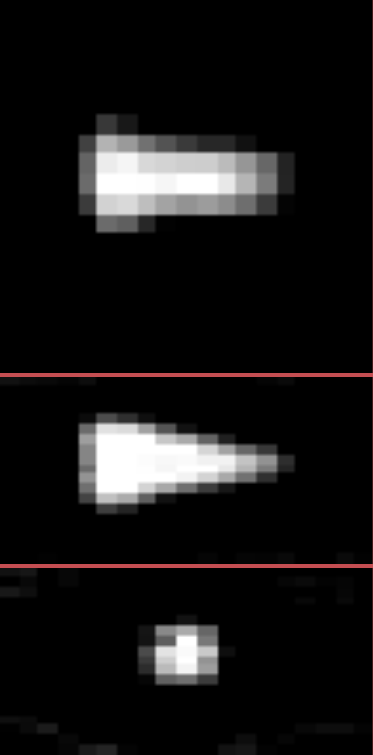}} &
\raisebox{-\totalheight/2}{\includegraphics[width=0.06\linewidth]{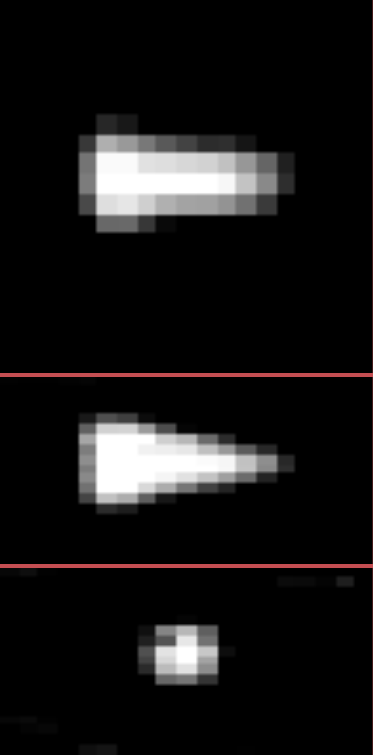}} &
\raisebox{-\totalheight/2}{\includegraphics[width=0.06\linewidth]{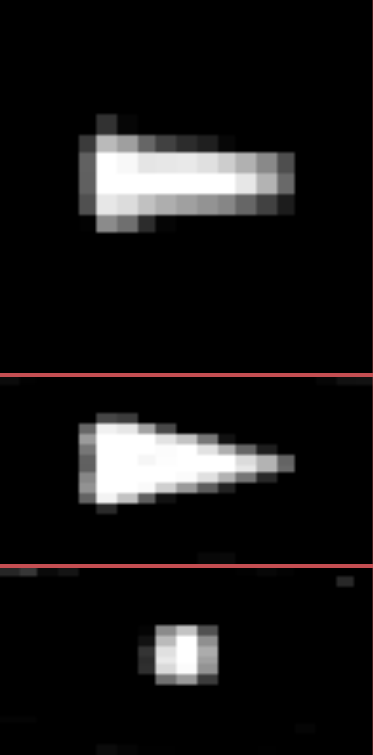}} &
\raisebox{-\totalheight/2}{\includegraphics[width=0.06\linewidth]{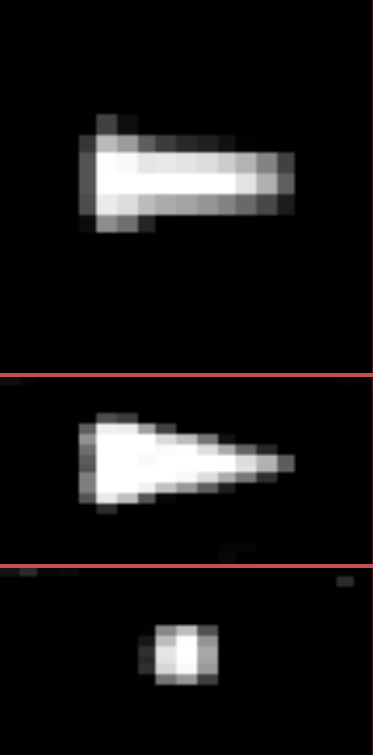}}\\

\vspace{.5mm} KACZ D$\ell^2$ + P$\ell^1$ &
\raisebox{-\totalheight/2}{\includegraphics[width=0.06\linewidth]{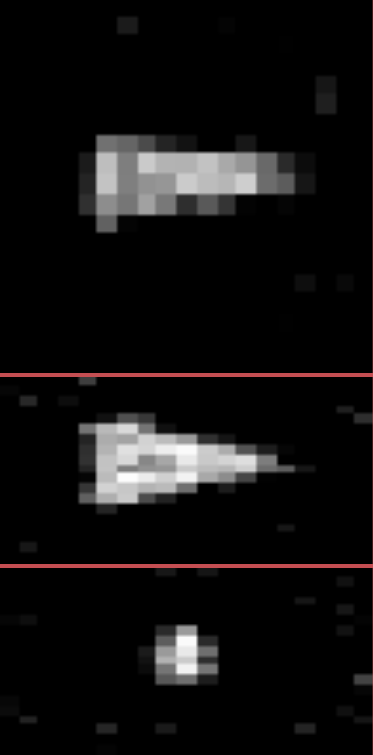}} &
\raisebox{-\totalheight/2}{\includegraphics[width=0.06\linewidth]{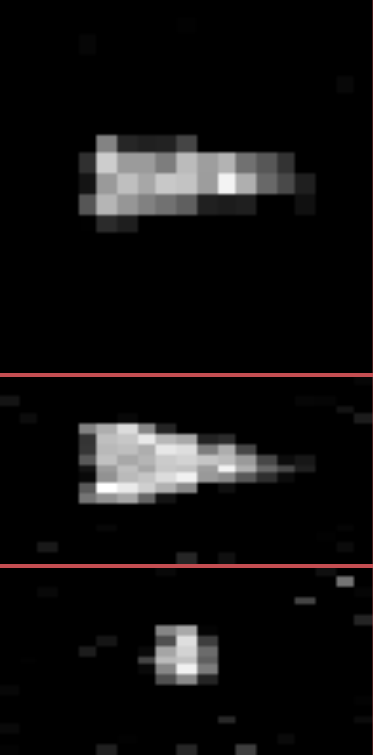}} &
\raisebox{-\totalheight/2}{\includegraphics[width=0.06\linewidth]{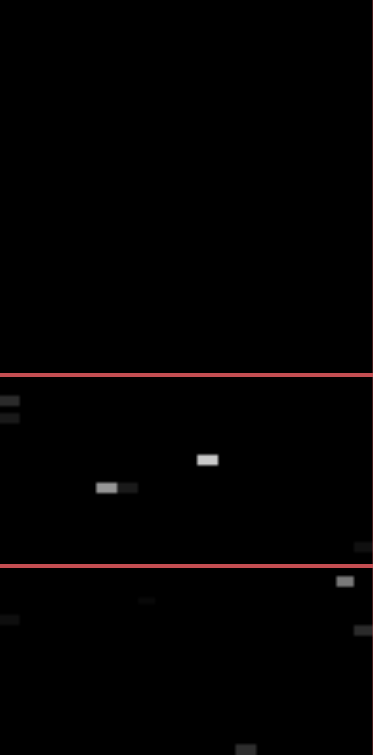}} &
\raisebox{-\totalheight/2}{\includegraphics[width=0.06\linewidth]{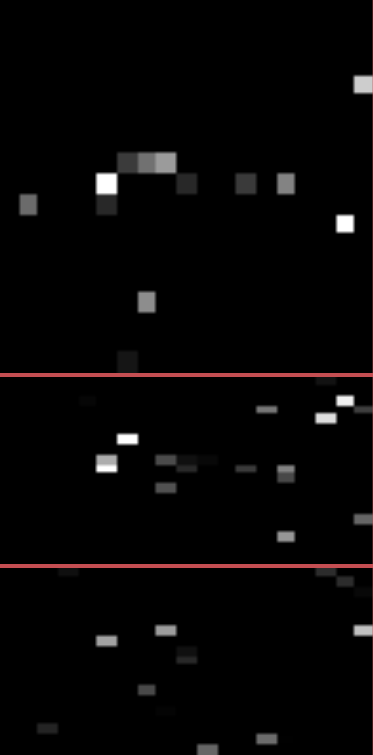}} &
\raisebox{-\totalheight/2}{\includegraphics[width=0.06\linewidth]{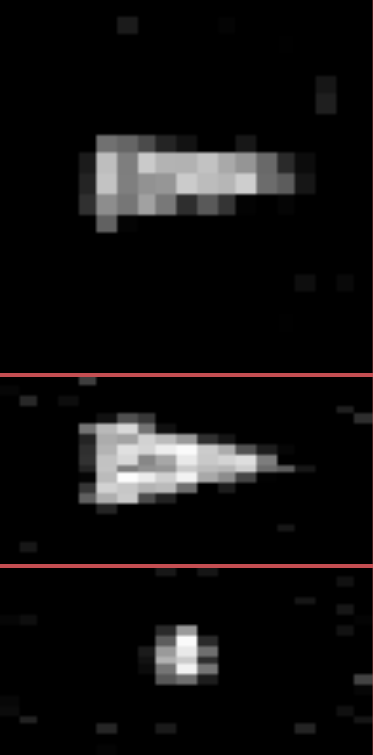}} &
\raisebox{-\totalheight/2}{\includegraphics[width=0.06\linewidth]{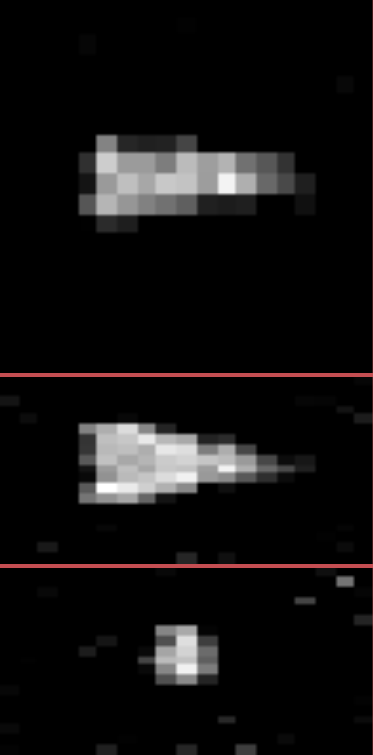}} &
\raisebox{-\totalheight/2}{\includegraphics[width=0.06\linewidth]{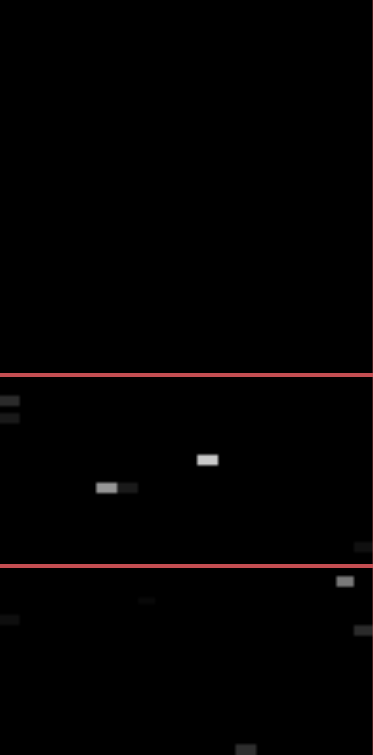}} &
\raisebox{-\totalheight/2}{\includegraphics[width=0.06\linewidth]{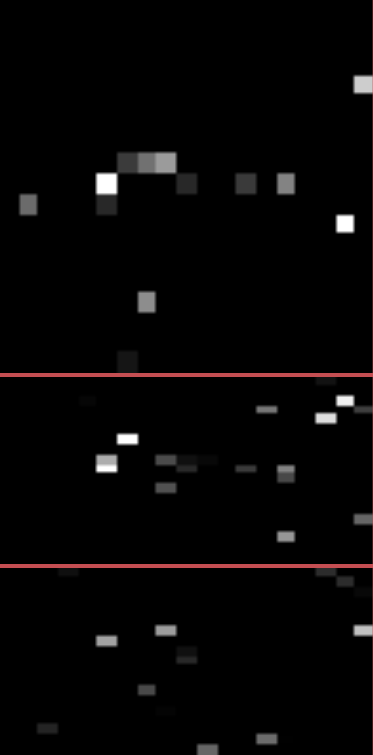}}\\

\vspace{.5mm} KACZ+TSVD D$\ell^2$ + P$\ell^1$ &
\raisebox{-\totalheight/2}{\includegraphics[width=0.06\linewidth]{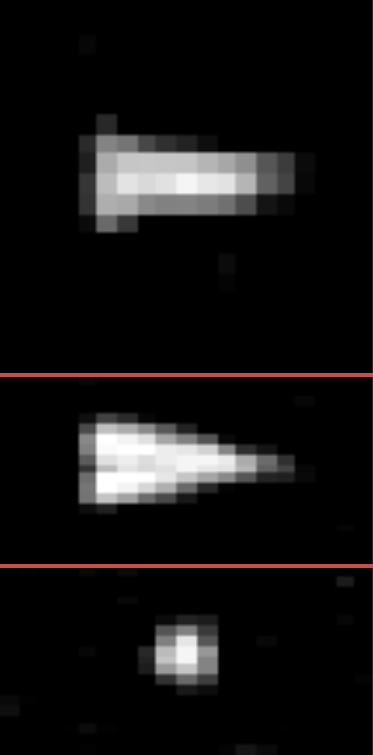}} &
\raisebox{-\totalheight/2}{\includegraphics[width=0.06\linewidth]{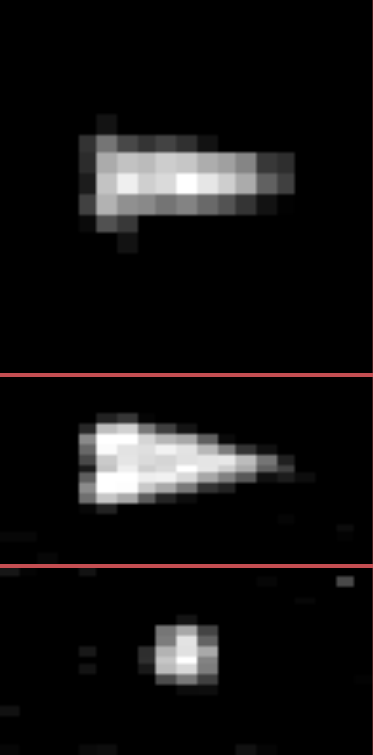}} &
\raisebox{-\totalheight/2}{\includegraphics[width=0.06\linewidth]{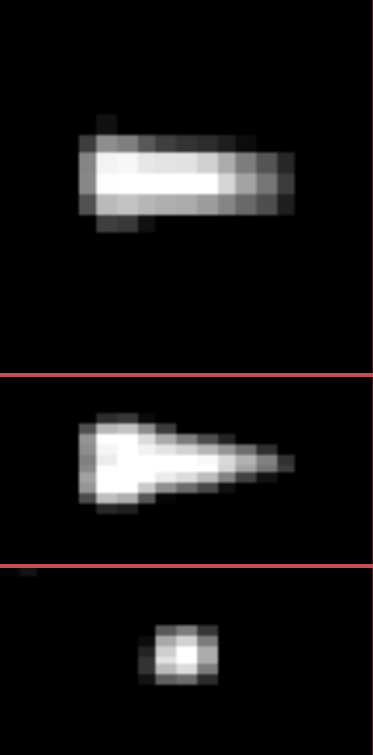}} &
\raisebox{-\totalheight/2}{\includegraphics[width=0.06\linewidth]{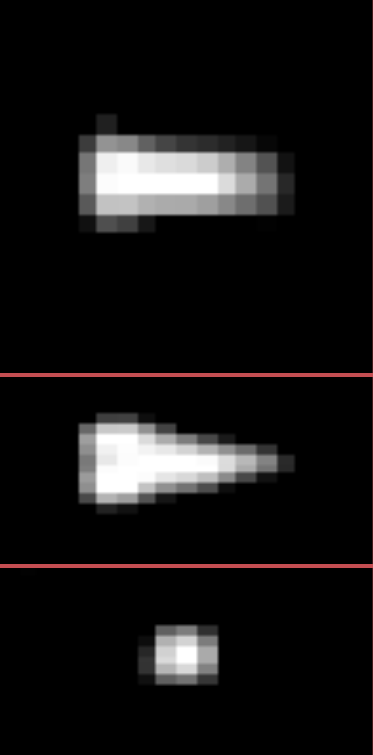}} &
\raisebox{-\totalheight/2}{\includegraphics[width=0.06\linewidth]{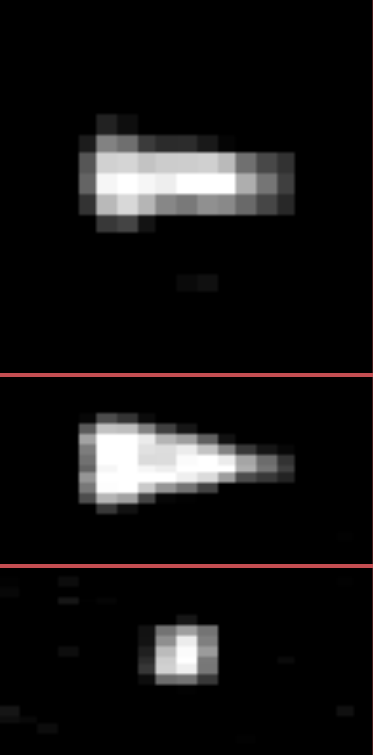}} &
\raisebox{-\totalheight/2}{\includegraphics[width=0.06\linewidth]{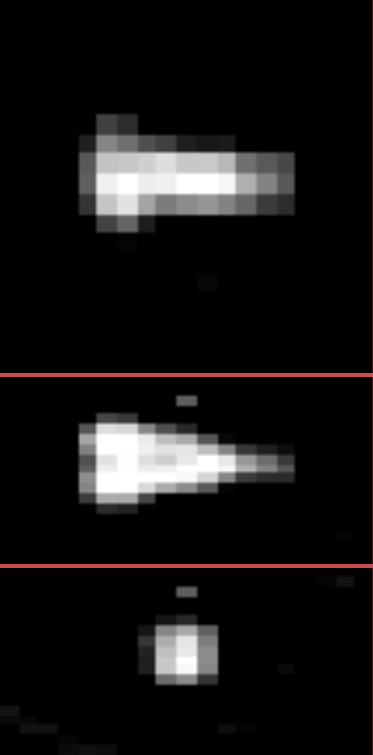}} &
\raisebox{-\totalheight/2}{\includegraphics[width=0.06\linewidth]{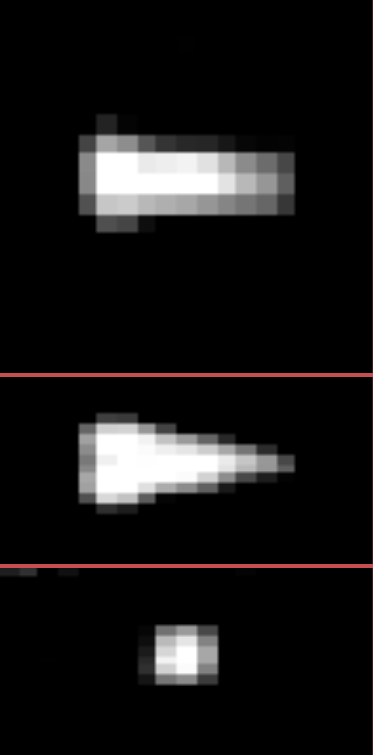}} &
\raisebox{-\totalheight/2}{\includegraphics[width=0.06\linewidth]{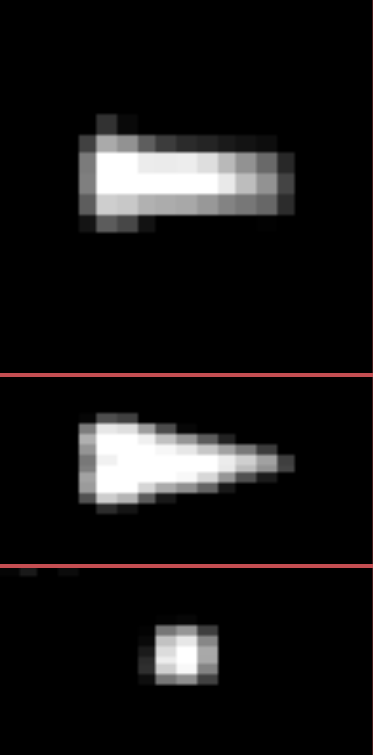}}\\

\vspace{.5mm} KACZ D$\ell^2$ + P$(\ell^1$+$\ell^2)$ &
\raisebox{-\totalheight/2}{\includegraphics[width=0.06\linewidth]{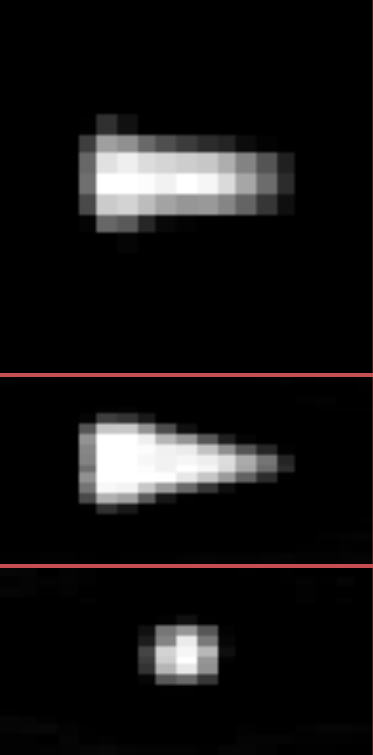}} &
\raisebox{-\totalheight/2}{\includegraphics[width=0.06\linewidth]{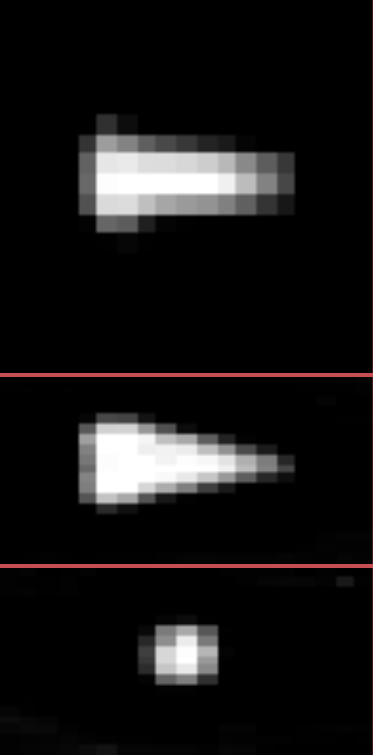}} &
\raisebox{-\totalheight/2}{\includegraphics[width=0.06\linewidth]{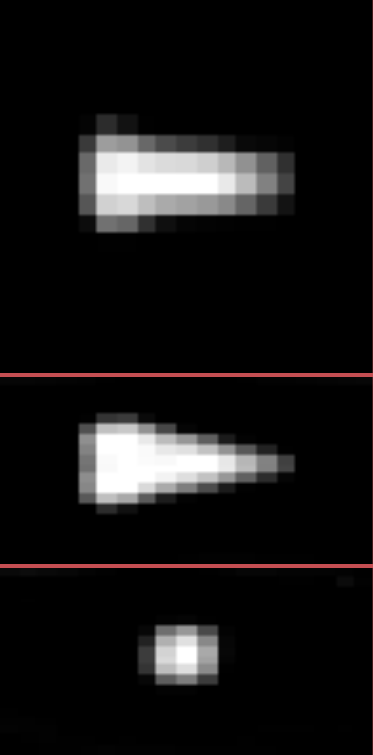}} &
\raisebox{-\totalheight/2}{\includegraphics[width=0.06\linewidth]{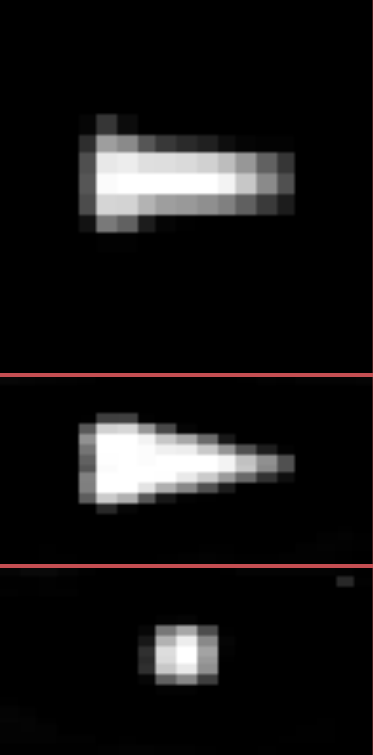}} &
\raisebox{-\totalheight/2}{\includegraphics[width=0.06\linewidth]{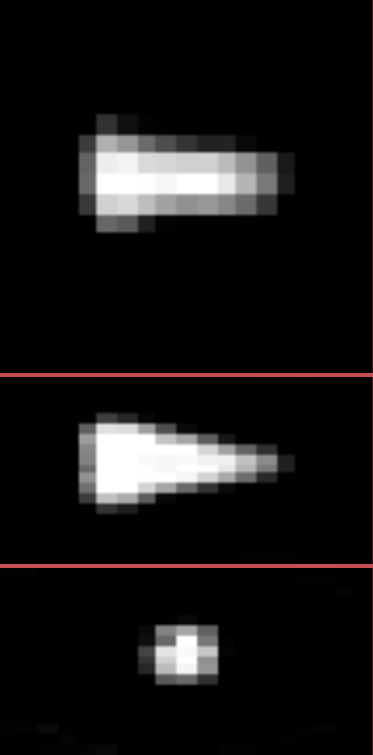}} &
\raisebox{-\totalheight/2}{\includegraphics[width=0.06\linewidth]{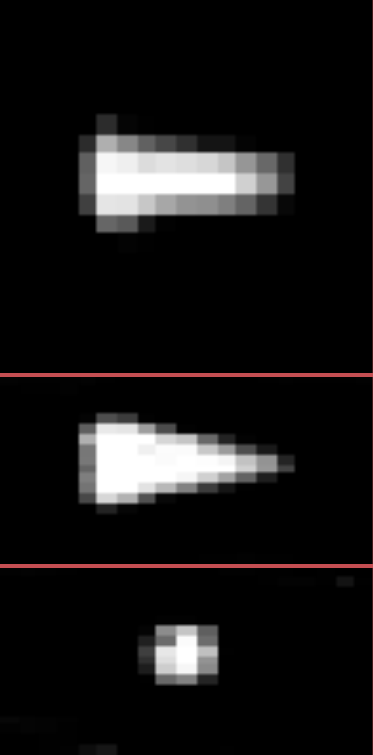}} &
\raisebox{-\totalheight/2}{\includegraphics[width=0.06\linewidth]{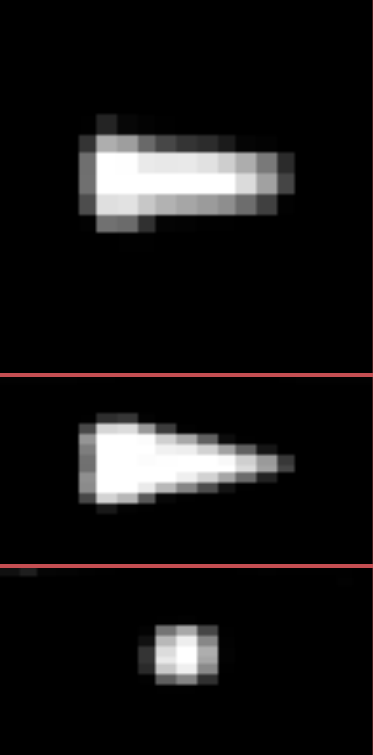}} &
\raisebox{-\totalheight/2}{\includegraphics[width=0.06\linewidth]{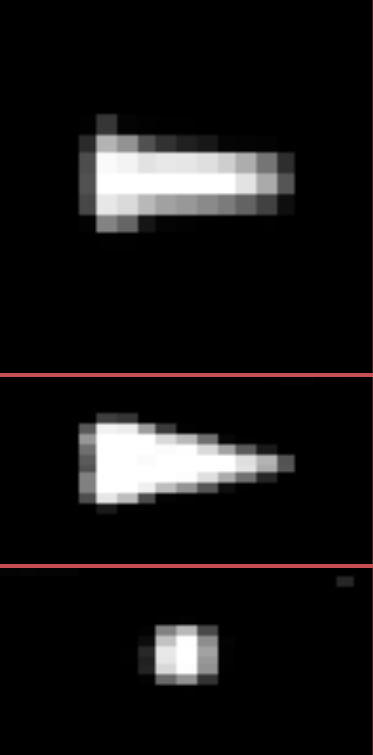}}\\

\vspace{.5mm} VAR D$\ell^1$ + P$\ell^1$  &
\raisebox{-\totalheight/2}{\includegraphics[width=0.06\linewidth]{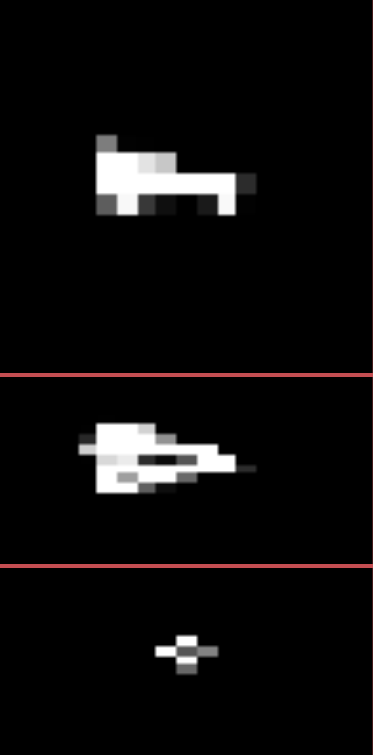}} &
\raisebox{-\totalheight/2}{\includegraphics[width=0.06\linewidth]{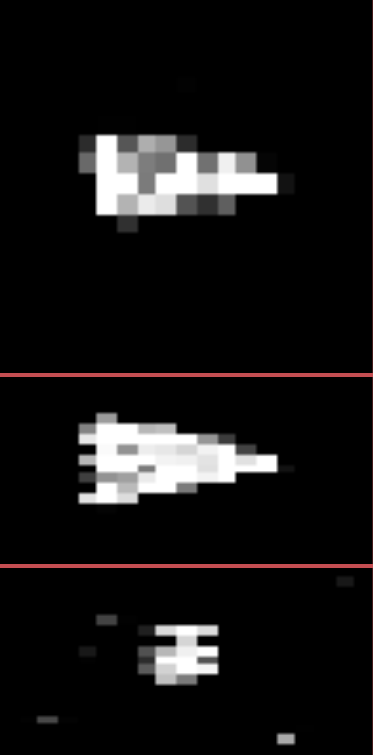}} &
\raisebox{-\totalheight/2}{\includegraphics[width=0.06\linewidth]{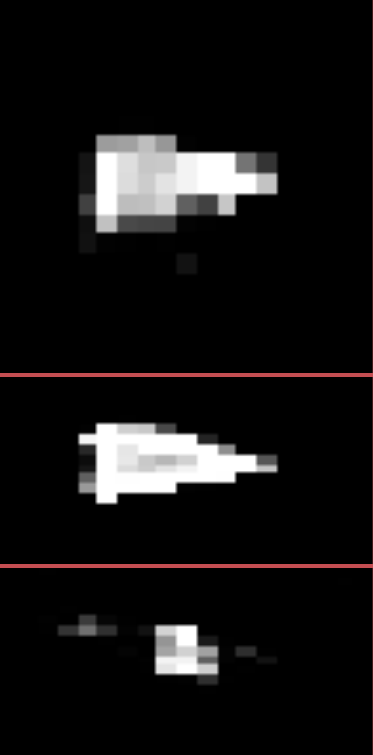}} &
\raisebox{-\totalheight/2}{\includegraphics[width=0.06\linewidth]{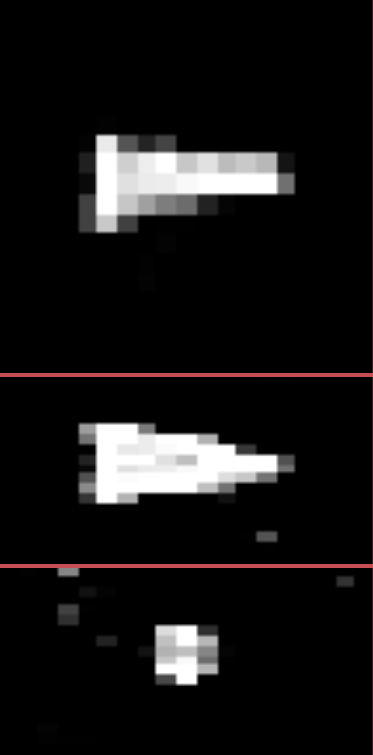}} &
\raisebox{-\totalheight/2}{\includegraphics[width=0.06\linewidth]{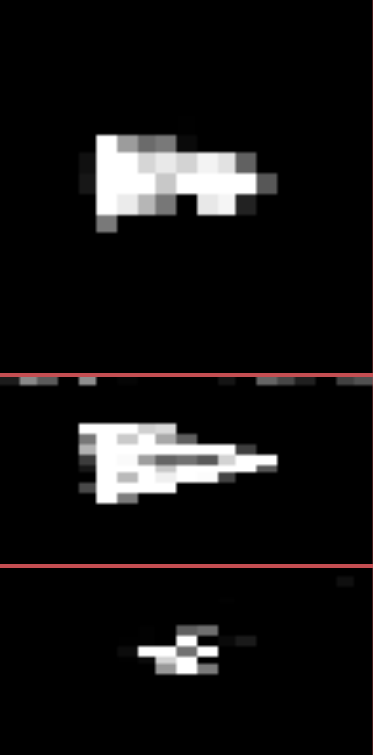}} &
\raisebox{-\totalheight/2}{\includegraphics[width=0.06\linewidth]{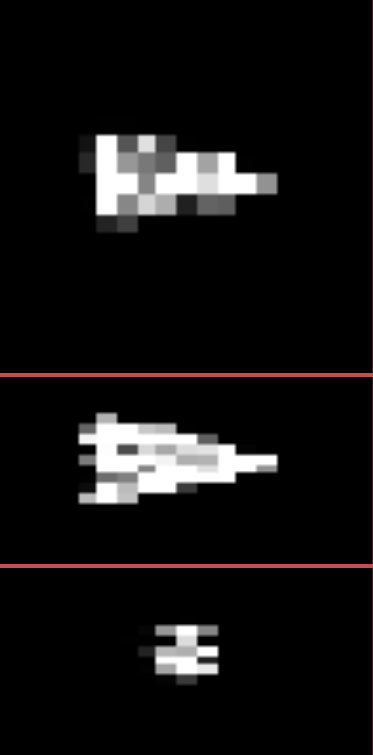}} &
\raisebox{-\totalheight/2}{\includegraphics[width=0.06\linewidth]{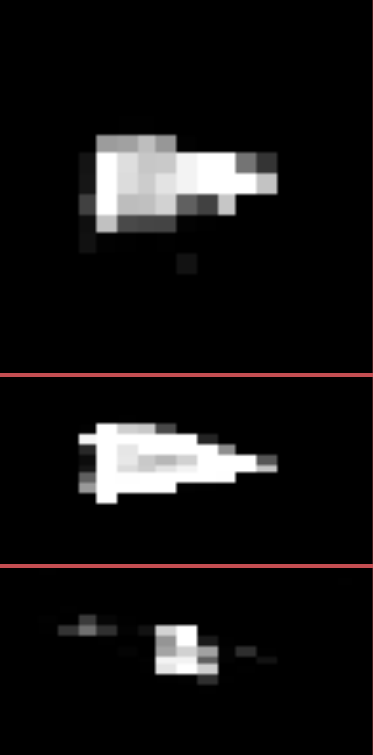}} &
\raisebox{-\totalheight/2}{\includegraphics[width=0.06\linewidth]{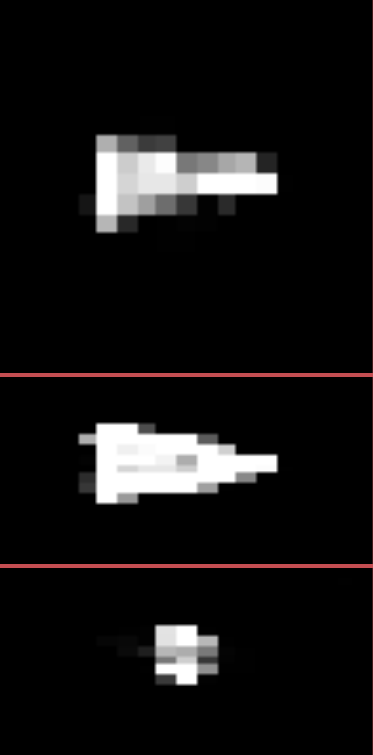}}\\

\vspace{.5mm} VAR D$\ell^1$ + P$\ell^2$ &
\raisebox{-\totalheight/2}{\includegraphics[width=0.06\linewidth]{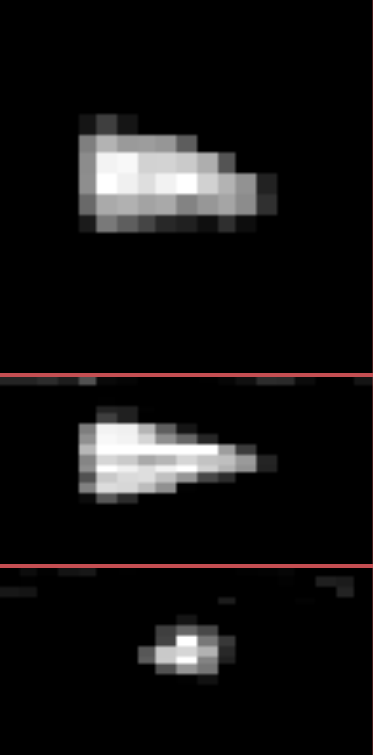}} &
\raisebox{-\totalheight/2}{\includegraphics[width=0.06\linewidth]{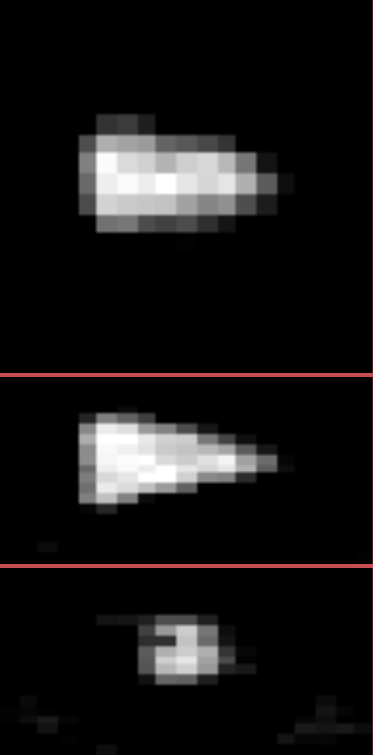}} &
\raisebox{-\totalheight/2}{\includegraphics[width=0.06\linewidth]{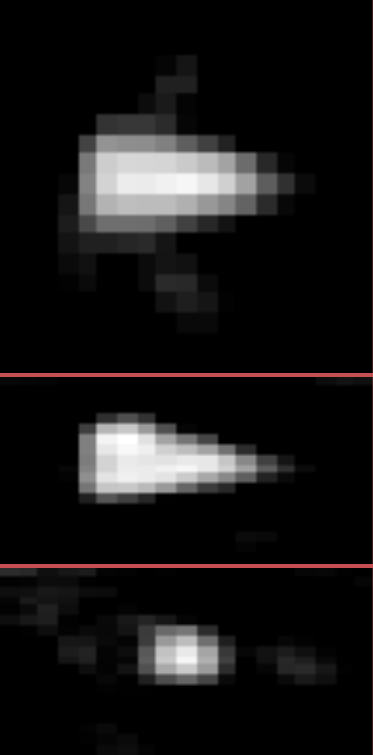}} &
\raisebox{-\totalheight/2}{\includegraphics[width=0.06\linewidth]{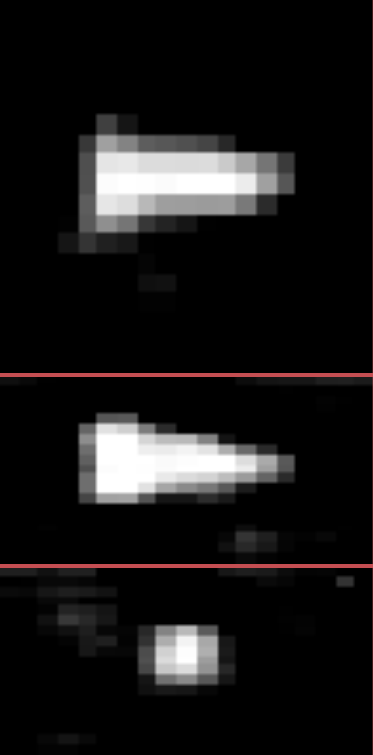}} &
\raisebox{-\totalheight/2}{\includegraphics[width=0.06\linewidth]{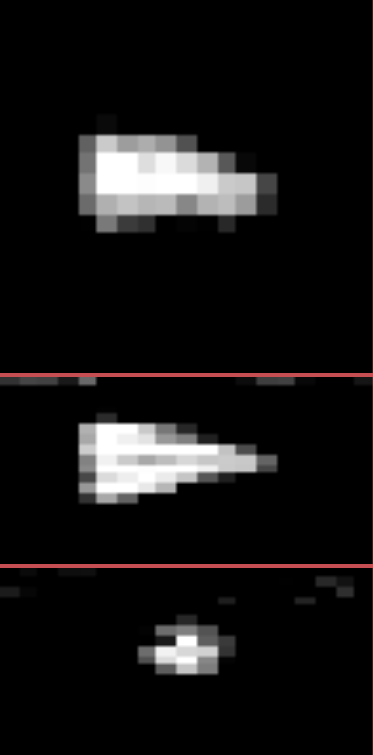}} &
\raisebox{-\totalheight/2}{\includegraphics[width=0.06\linewidth]{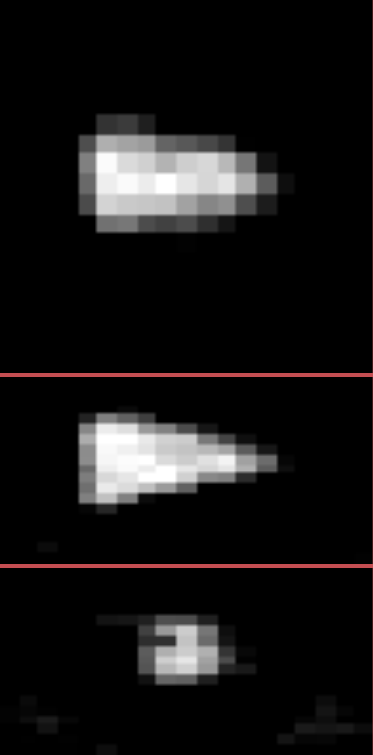}} &
\raisebox{-\totalheight/2}{\includegraphics[width=0.06\linewidth]{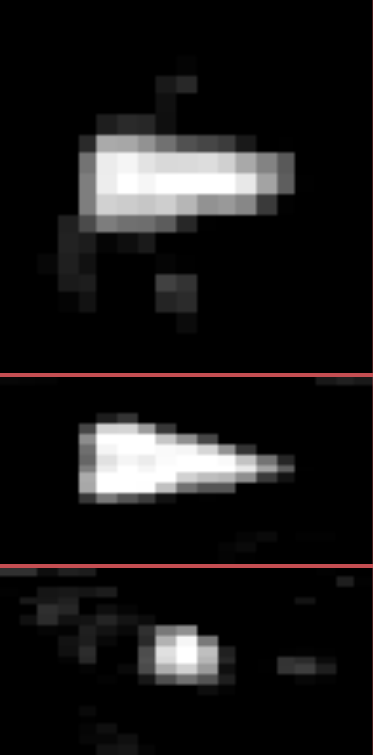}} &
\raisebox{-\totalheight/2}{\includegraphics[width=0.06\linewidth]{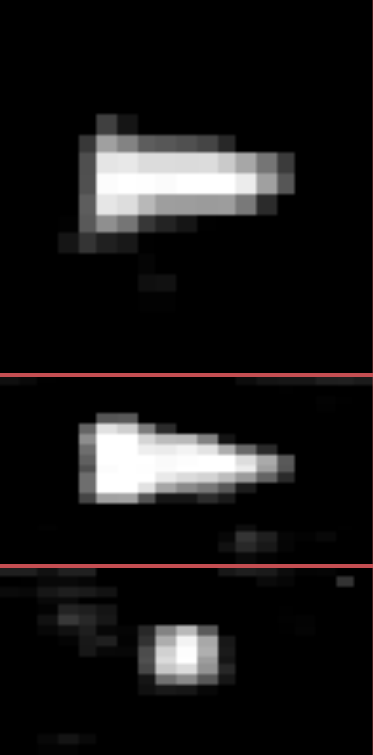}}\\

\vspace{.5mm} VAR D$\ell^1$ + P TV & 
\raisebox{-\totalheight/2}{\includegraphics[width=0.06\linewidth]{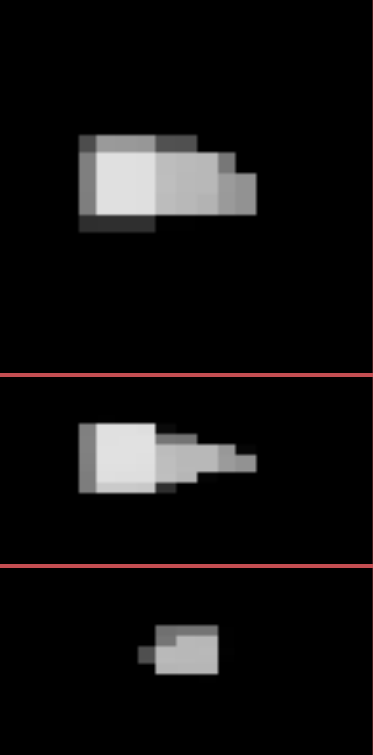}} &
\raisebox{-\totalheight/2}{\includegraphics[width=0.06\linewidth]{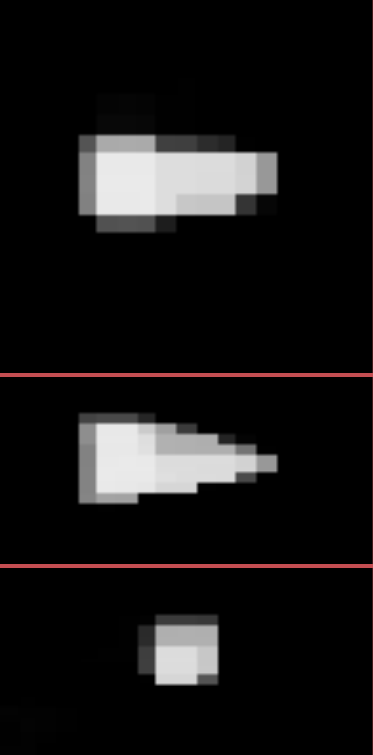}} &
\raisebox{-\totalheight/2}{\includegraphics[width=0.06\linewidth]{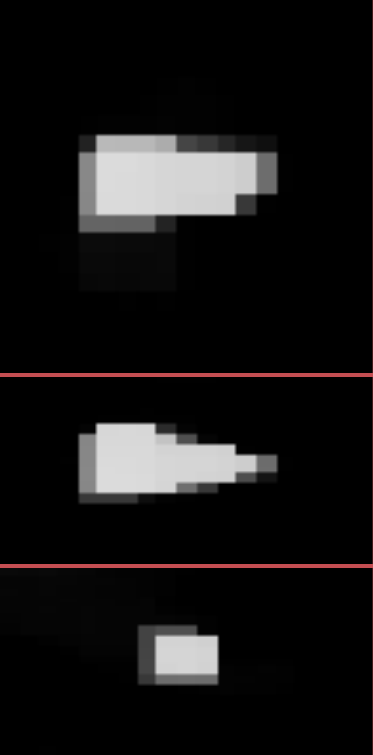}} &
\raisebox{-\totalheight/2}{\includegraphics[width=0.06\linewidth]{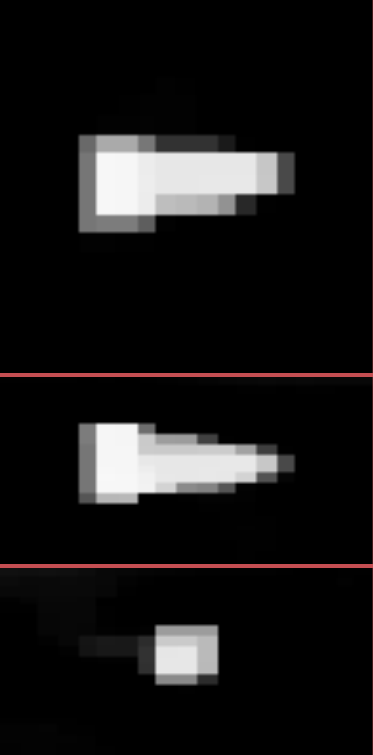}} &
\raisebox{-\totalheight/2}{\includegraphics[width=0.06\linewidth]{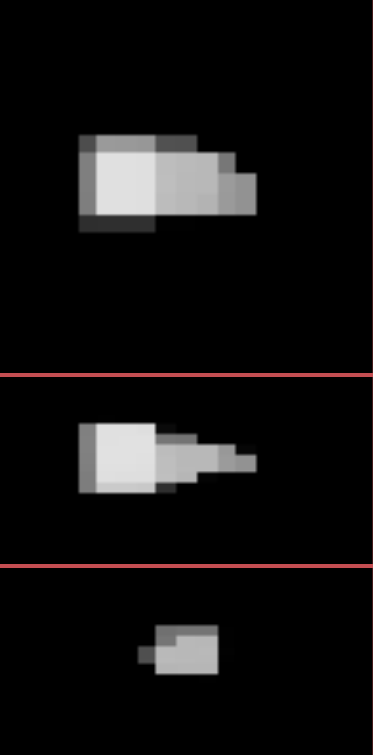}} &
\raisebox{-\totalheight/2}{\includegraphics[width=0.06\linewidth]{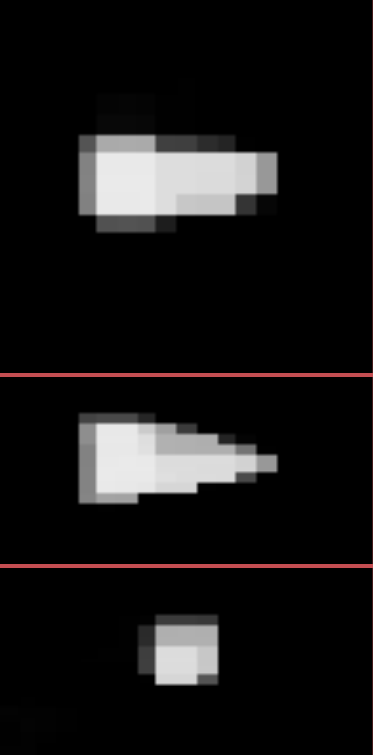}} &
\raisebox{-\totalheight/2}{\includegraphics[width=0.06\linewidth]{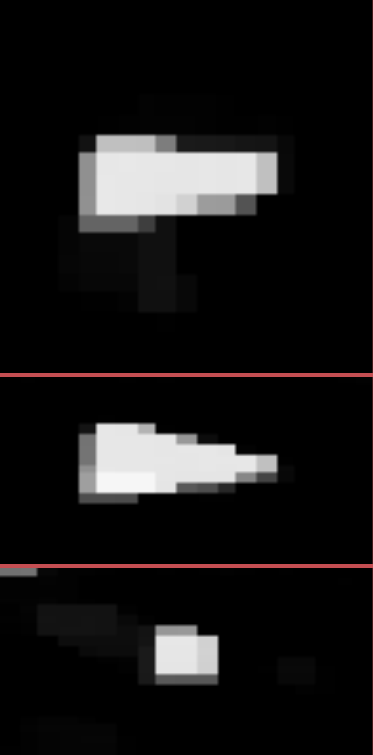}} &
\raisebox{-\totalheight/2}{\includegraphics[width=0.06\linewidth]{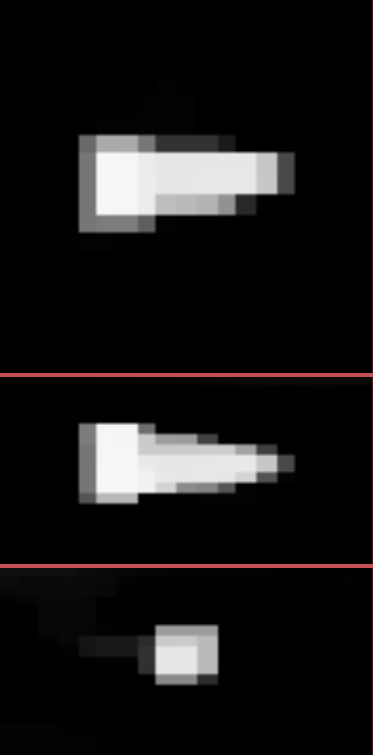}}\\

 \hline
\end{tabular}
\caption{The best reconstructions for the shape phantom without whitening, corresponding to the values in Tables~\ref{tab:non_whitened_shape_metrics} and~\ref{tab:non_whitened_shape_parameters}. The color scale goes from $0$ (black) to $60$ (white). Each image is separated into three vertically stacked parts, which are separated by red lines. Each of these parts represents one of the three central slices/planes of the three-dimensional reconstruction (x-y-, x-z-, and y-z-plane from top to bottom).}
\label{tab:non_whitened_shape_images}
\end{figure*}

\begin{figure*}
\centering
\rowcolors{1}{}{lightgray}
\begin{tabular}{c   c c c c  c c c c}
 \hline
 Methods & \multicolumn{4}{c  }{PSNRs} & \multicolumn{4}{c }{SSIMs} \\
 \hline
 \rowcolor{white}
  & $\tau=0$ & $\tau=1$ & $\tau=3$ & $\tau=5$ & $\tau=0$ & $\tau=1$ & $\tau=3$ & $\tau=5$\\
 \hline\\[-2.0ex]

\vspace{.5mm} DIP D$\ell^1$ + P- &
\raisebox{-\totalheight/2}{\includegraphics[width=0.06\linewidth]{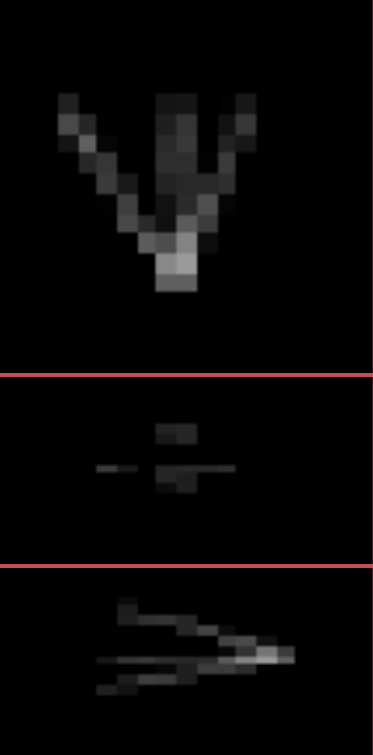}} &
\raisebox{-\totalheight/2}{\includegraphics[width=0.06\linewidth]{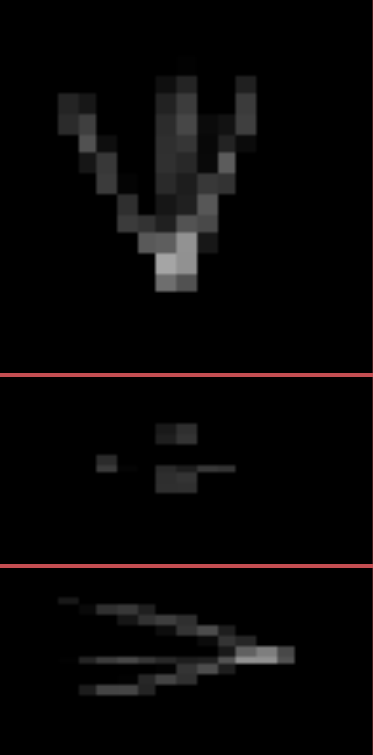}} &
\raisebox{-\totalheight/2}{\includegraphics[width=0.06\linewidth]{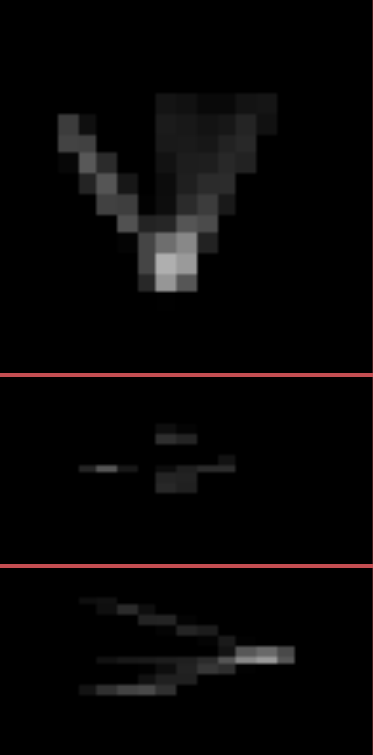}} &
\raisebox{-\totalheight/2}{\includegraphics[width=0.06\linewidth]{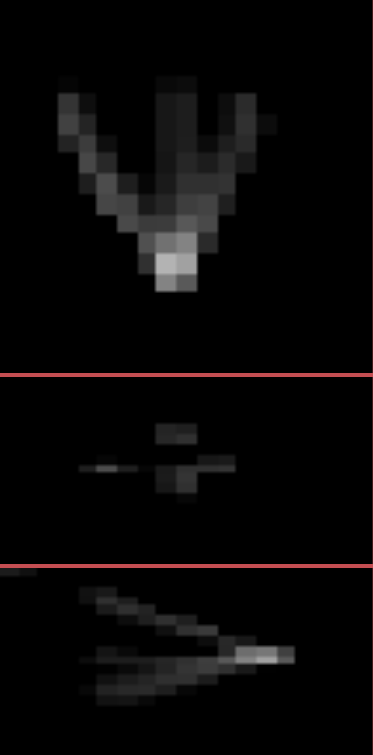}} &
\raisebox{-\totalheight/2}{\includegraphics[width=0.06\linewidth]{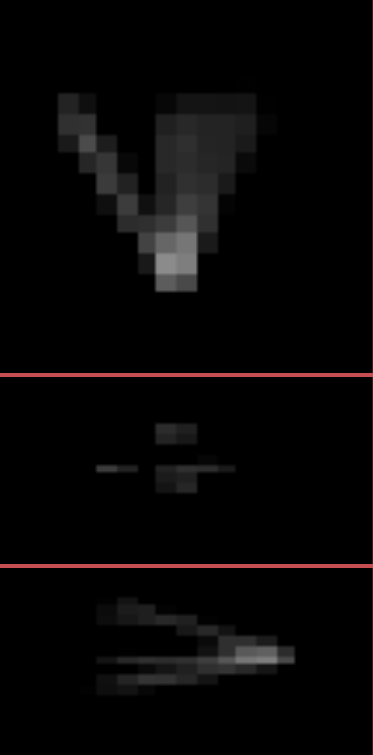}} &
\raisebox{-\totalheight/2}{\includegraphics[width=0.06\linewidth]{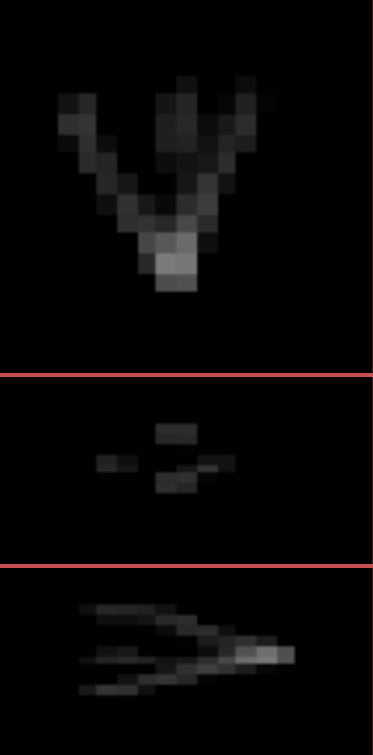}} &
\raisebox{-\totalheight/2}{\includegraphics[width=0.06\linewidth]{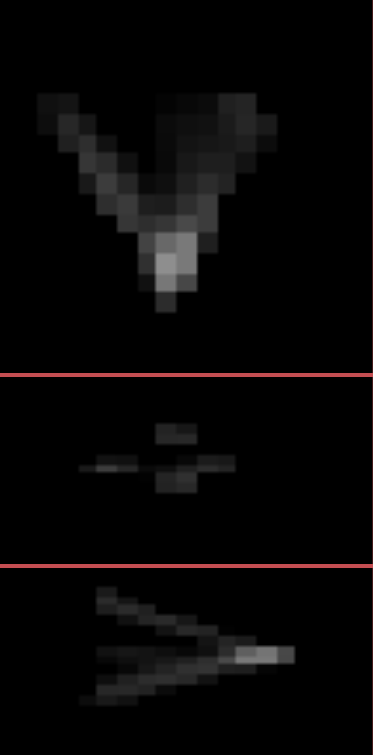}} &
\raisebox{-\totalheight/2}{\includegraphics[width=0.06\linewidth]{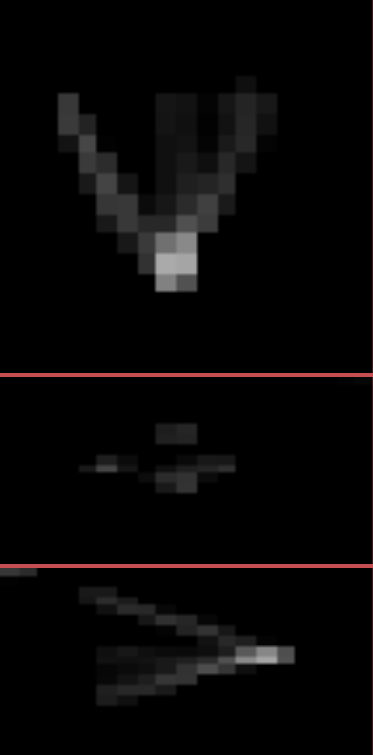}}\\

\vspace{.5mm} KACZ D$\ell^2$ + P$\ell^2$ &
\raisebox{-\totalheight/2}{\includegraphics[width=0.06\linewidth]{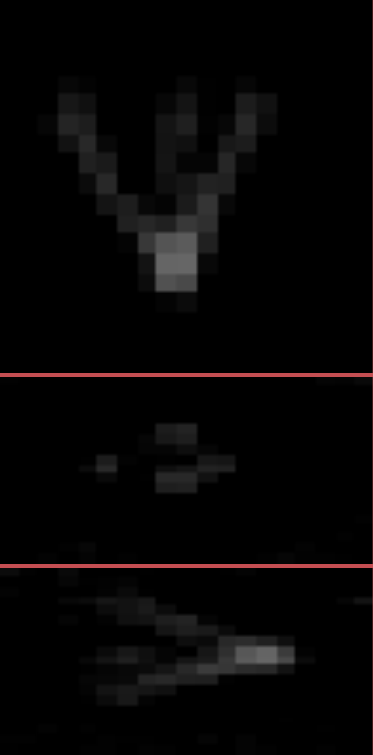}} &
\raisebox{-\totalheight/2}{\includegraphics[width=0.06\linewidth]{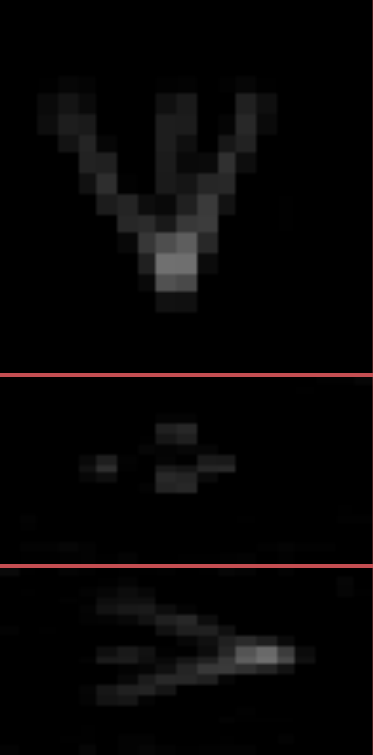}} &
\raisebox{-\totalheight/2}{\includegraphics[width=0.06\linewidth]{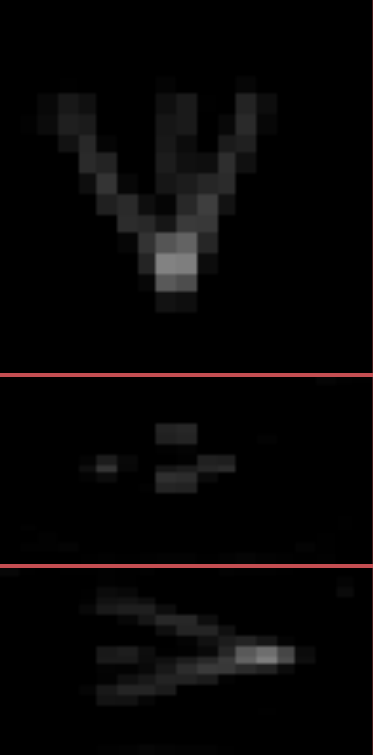}} &
\raisebox{-\totalheight/2}{\includegraphics[width=0.06\linewidth]{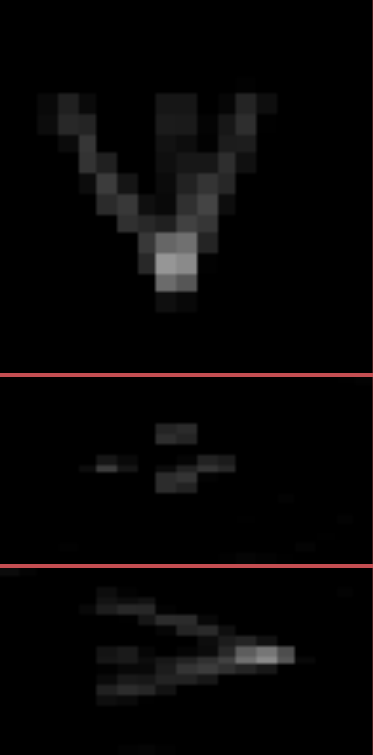}} &
\raisebox{-\totalheight/2}{\includegraphics[width=0.06\linewidth]{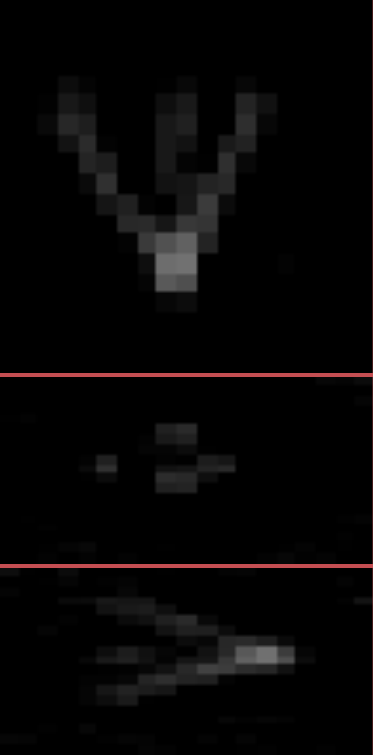}} &
\raisebox{-\totalheight/2}{\includegraphics[width=0.06\linewidth]{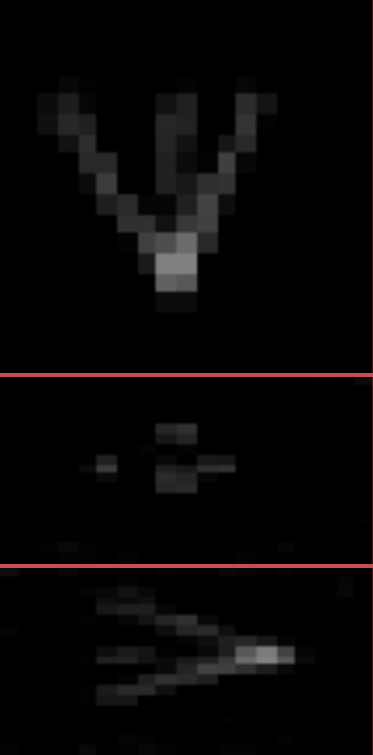}} &
\raisebox{-\totalheight/2}{\includegraphics[width=0.06\linewidth]{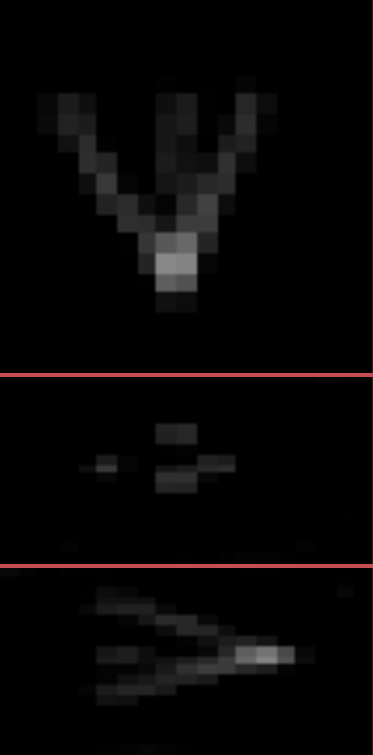}} &
\raisebox{-\totalheight/2}{\includegraphics[width=0.06\linewidth]{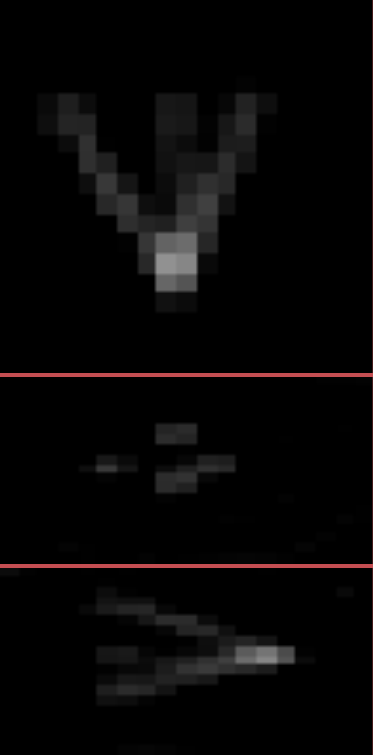}}\\

\vspace{.5mm} KACZ D$\ell^2$ + P$\ell^1$ &
\raisebox{-\totalheight/2}{\includegraphics[width=0.06\linewidth]{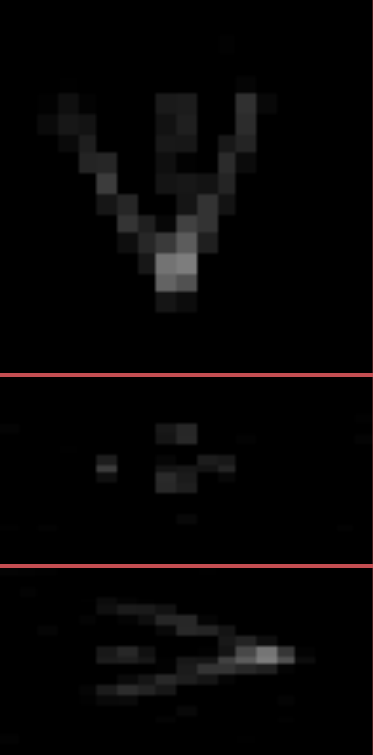}} &
\raisebox{-\totalheight/2}{\includegraphics[width=0.06\linewidth]{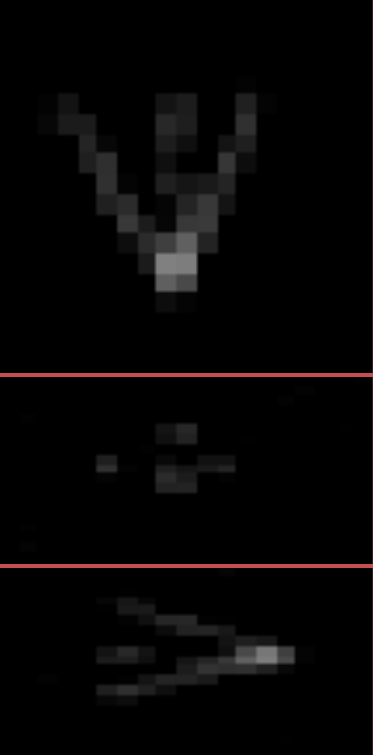}} &
\raisebox{-\totalheight/2}{\includegraphics[width=0.06\linewidth]{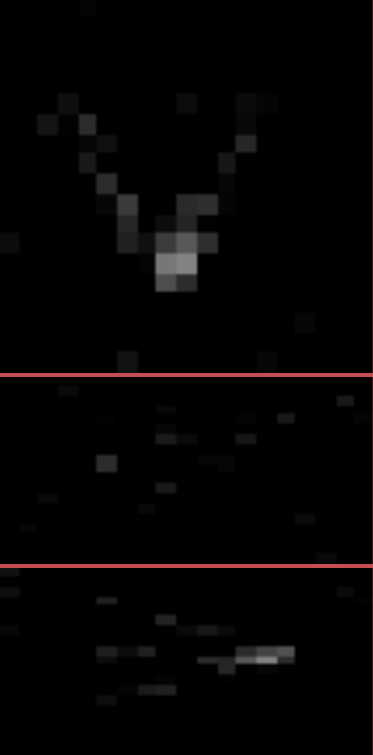}} &
\raisebox{-\totalheight/2}{\includegraphics[width=0.06\linewidth]{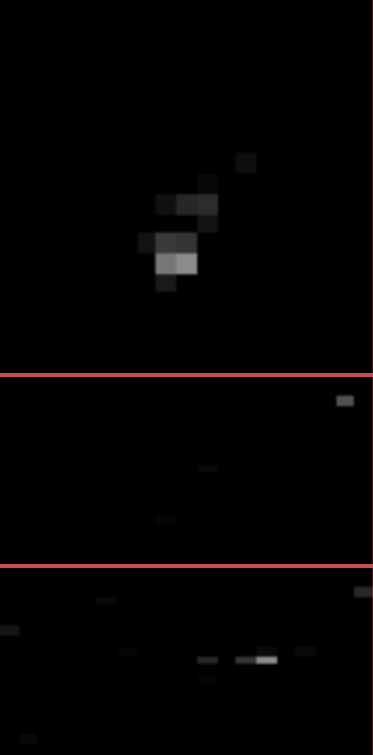}} &
\raisebox{-\totalheight/2}{\includegraphics[width=0.06\linewidth]{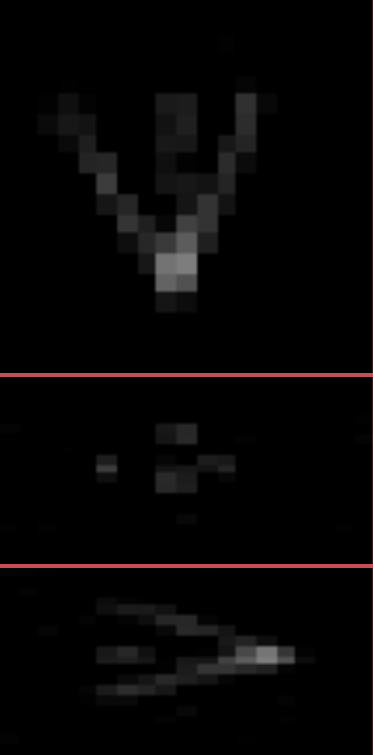}} &
\raisebox{-\totalheight/2}{\includegraphics[width=0.06\linewidth]{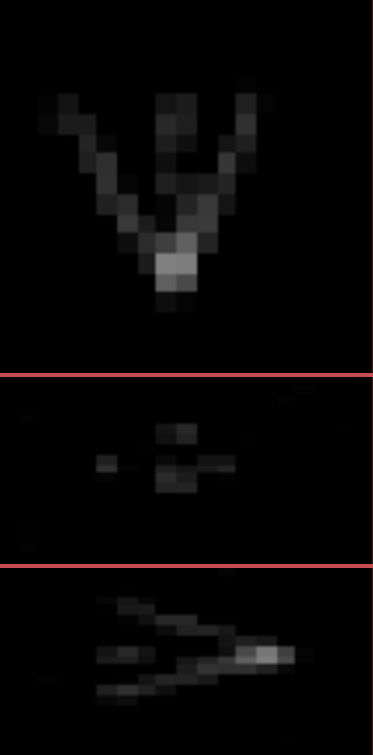}} &
\raisebox{-\totalheight/2}{\includegraphics[width=0.06\linewidth]{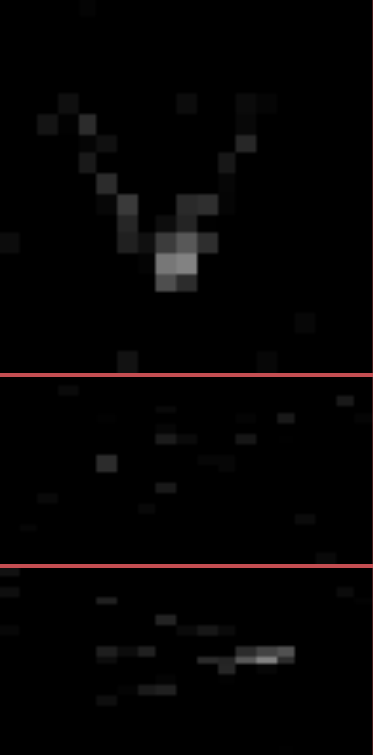}} &
\raisebox{-\totalheight/2}{\includegraphics[width=0.06\linewidth]{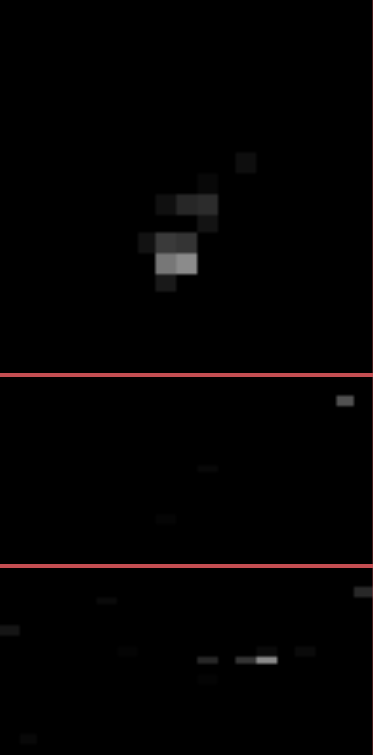}}\\

\vspace{.5mm} KACZ+TSVD D$\ell^2$ + P$\ell^1$ &
\raisebox{-\totalheight/2}{\includegraphics[width=0.06\linewidth]{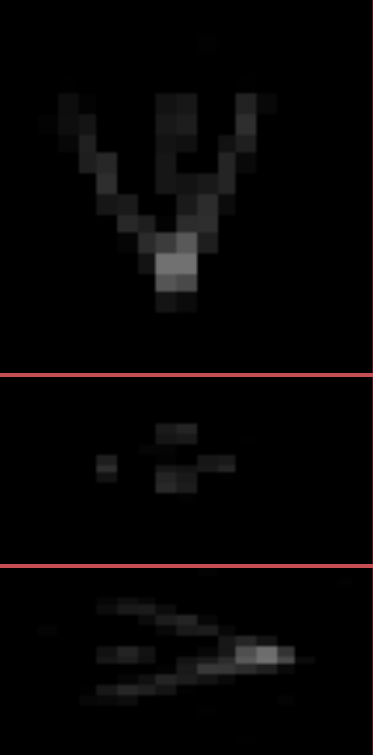}} &
\raisebox{-\totalheight/2}{\includegraphics[width=0.06\linewidth]{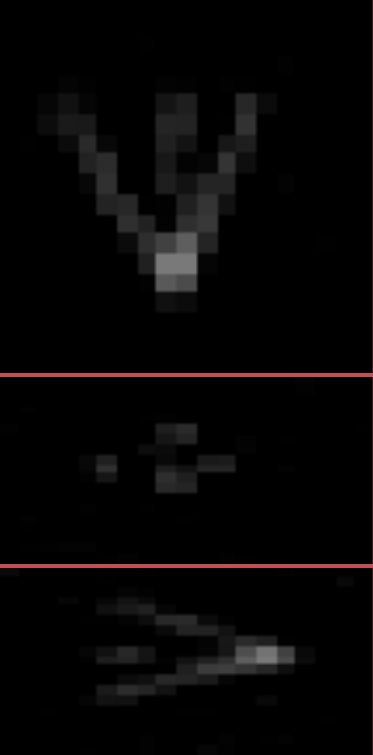}} &
\raisebox{-\totalheight/2}{\includegraphics[width=0.06\linewidth]{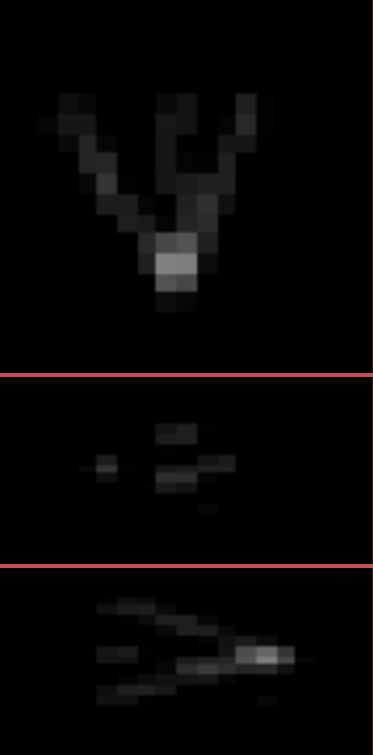}} &
\raisebox{-\totalheight/2}{\includegraphics[width=0.06\linewidth]{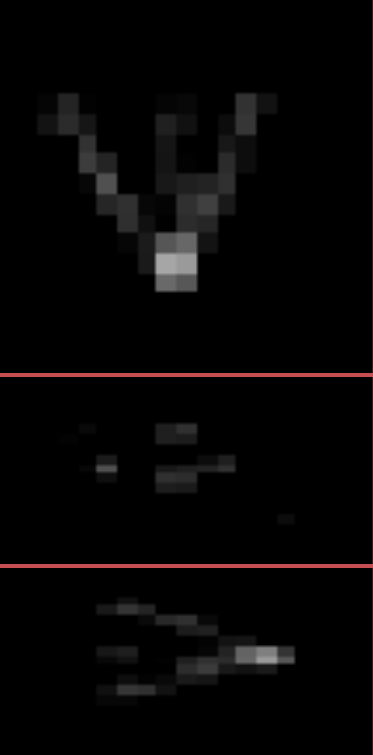}} &
\raisebox{-\totalheight/2}{\includegraphics[width=0.06\linewidth]{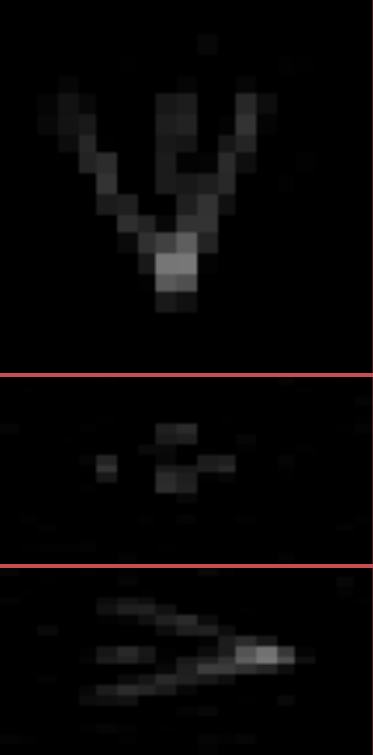}} &
\raisebox{-\totalheight/2}{\includegraphics[width=0.06\linewidth]{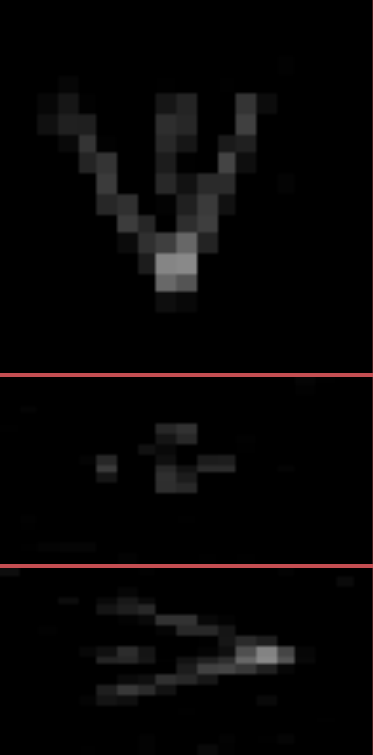}} &
\raisebox{-\totalheight/2}{\includegraphics[width=0.06\linewidth]{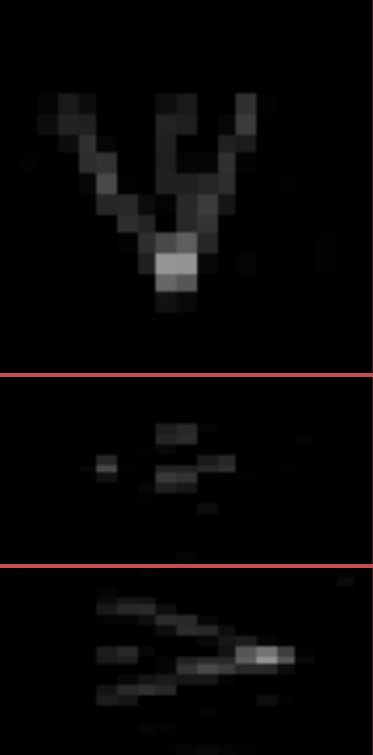}} &
\raisebox{-\totalheight/2}{\includegraphics[width=0.06\linewidth]{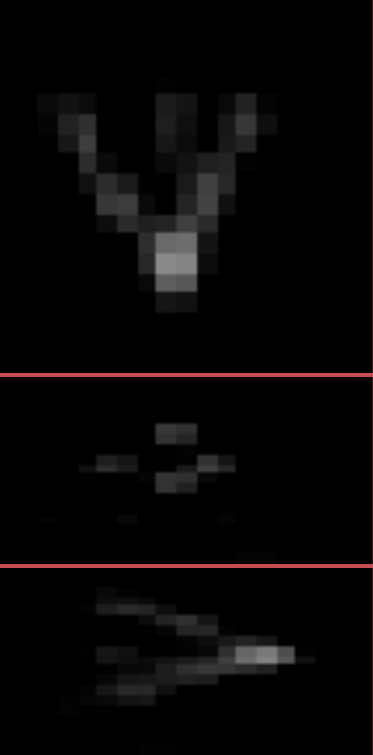}}\\

\vspace{.5mm} KACZ D$\ell^2$ + P$(\ell^1$+$\ell^2)$ &
\raisebox{-\totalheight/2}{\includegraphics[width=0.06\linewidth]{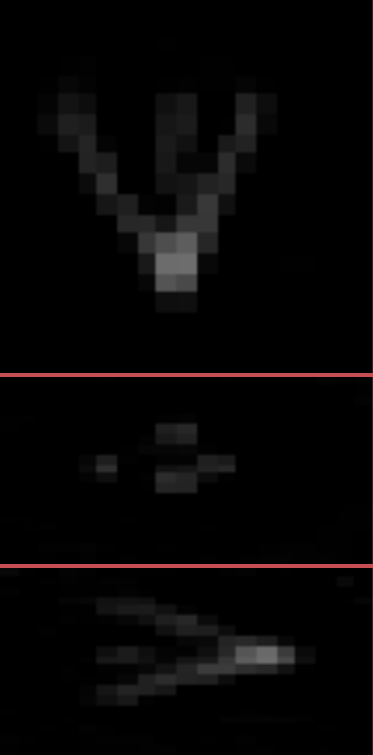}} &
\raisebox{-\totalheight/2}{\includegraphics[width=0.06\linewidth]{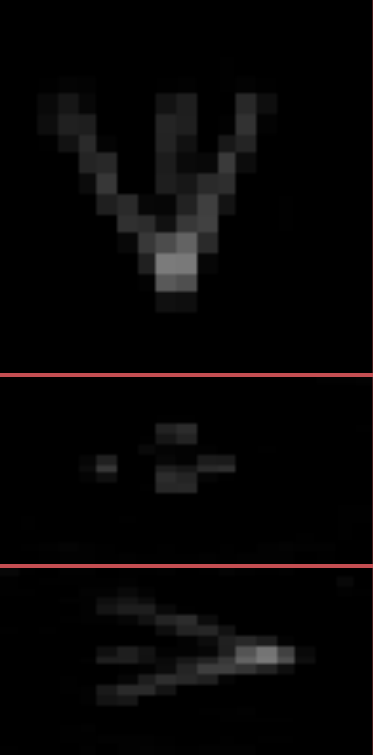}} &
\raisebox{-\totalheight/2}{\includegraphics[width=0.06\linewidth]{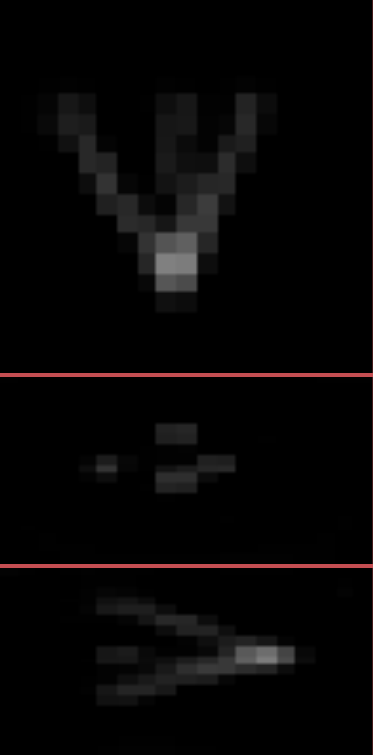}} &
\raisebox{-\totalheight/2}{\includegraphics[width=0.06\linewidth]{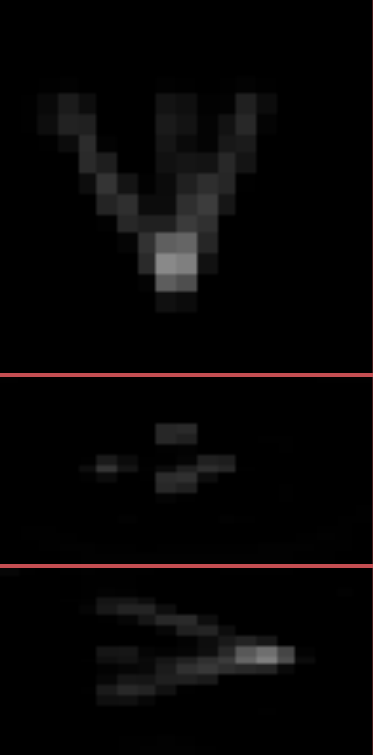}} &
\raisebox{-\totalheight/2}{\includegraphics[width=0.06\linewidth]{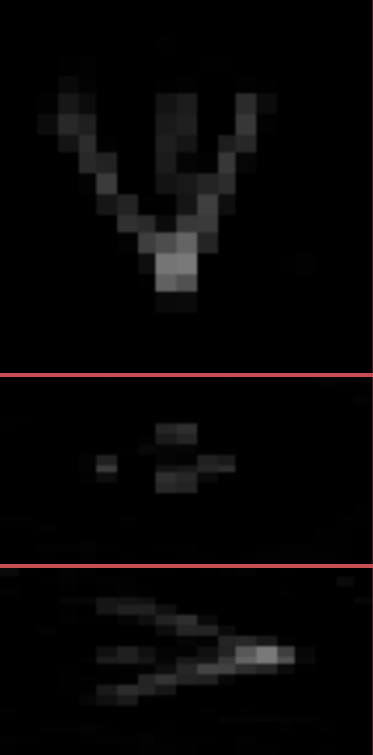}} &
\raisebox{-\totalheight/2}{\includegraphics[width=0.06\linewidth]{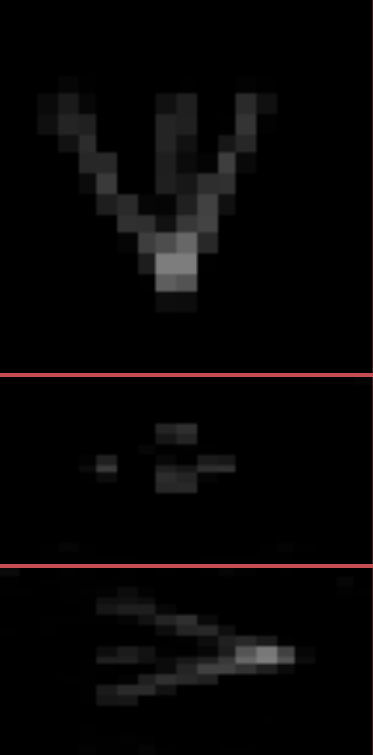}} &
\raisebox{-\totalheight/2}{\includegraphics[width=0.06\linewidth]{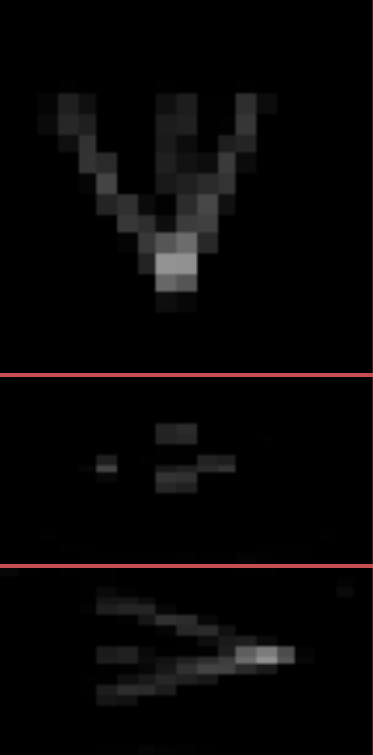}} &
\raisebox{-\totalheight/2}{\includegraphics[width=0.06\linewidth]{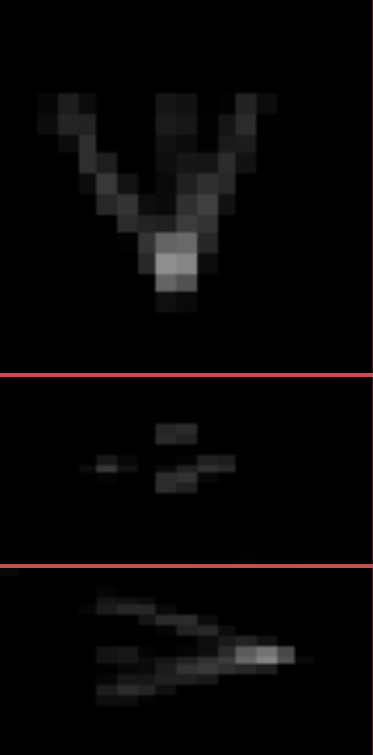}}\\

\vspace{.5mm} VAR D$\ell^1$ + P$\ell^1$  &
\raisebox{-\totalheight/2}{\includegraphics[width=0.06\linewidth]{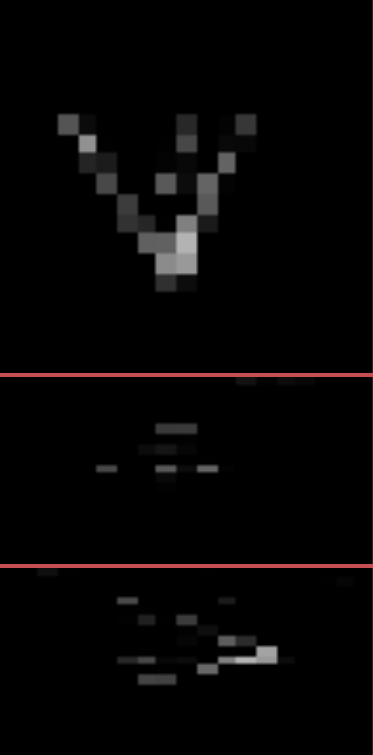}} &
\raisebox{-\totalheight/2}{\includegraphics[width=0.06\linewidth]{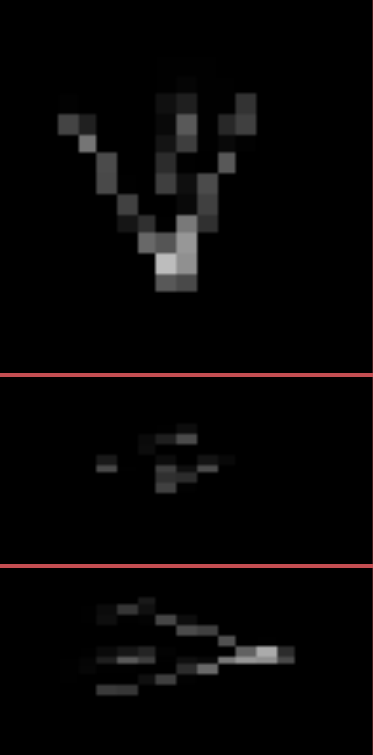}} &
\raisebox{-\totalheight/2}{\includegraphics[width=0.06\linewidth]{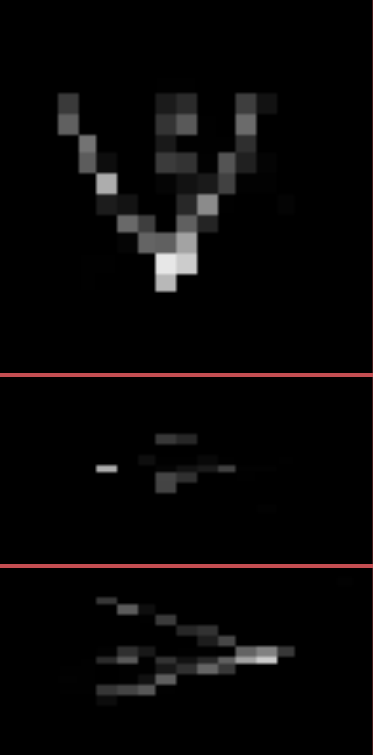}} &
\raisebox{-\totalheight/2}{\includegraphics[width=0.06\linewidth]{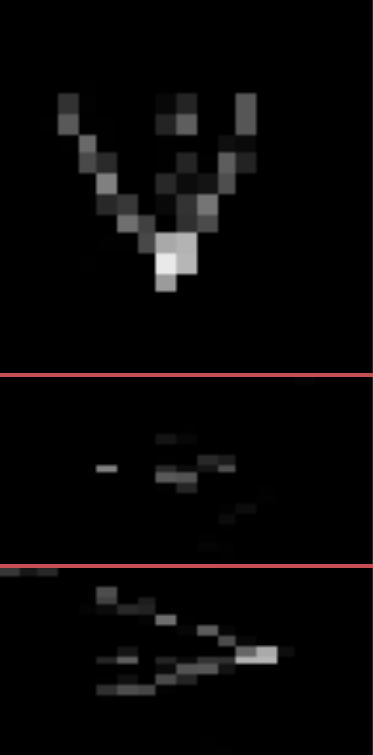}} &
\raisebox{-\totalheight/2}{\includegraphics[width=0.06\linewidth]{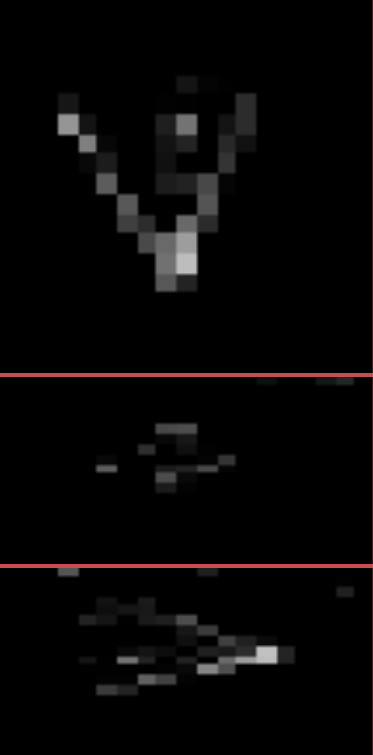}} &
\raisebox{-\totalheight/2}{\includegraphics[width=0.06\linewidth]{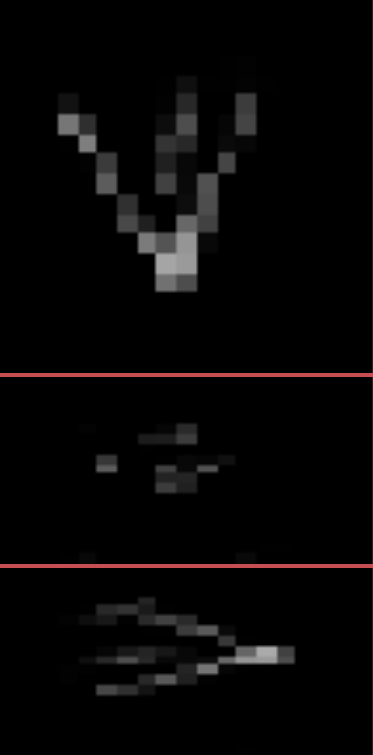}} &
\raisebox{-\totalheight/2}{\includegraphics[width=0.06\linewidth]{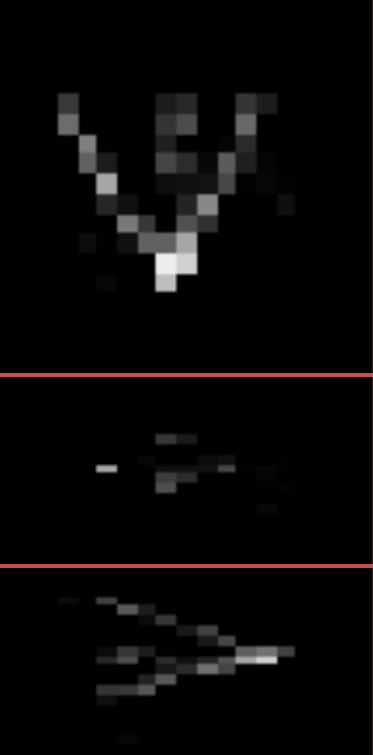}} &
\raisebox{-\totalheight/2}{\includegraphics[width=0.06\linewidth]{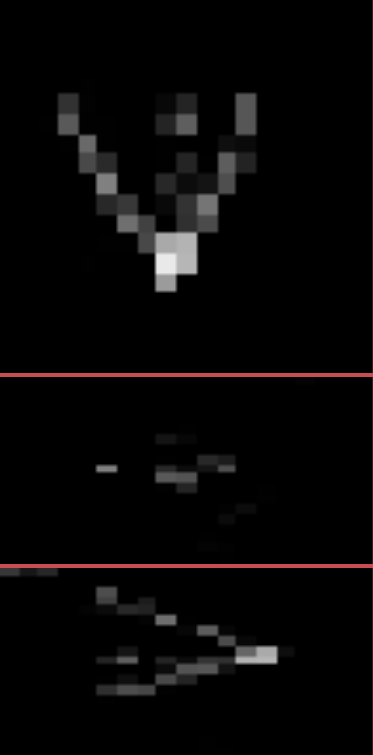}}\\

\vspace{.5mm} VAR D$\ell^1$ + P$\ell^2$ &
\raisebox{-\totalheight/2}{\includegraphics[width=0.06\linewidth]{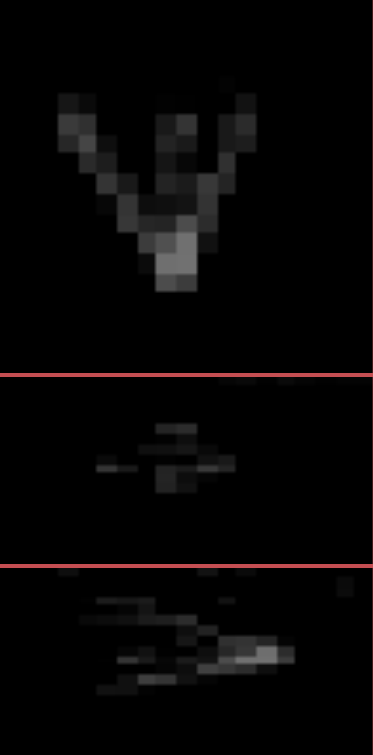}} &
\raisebox{-\totalheight/2}{\includegraphics[width=0.06\linewidth]{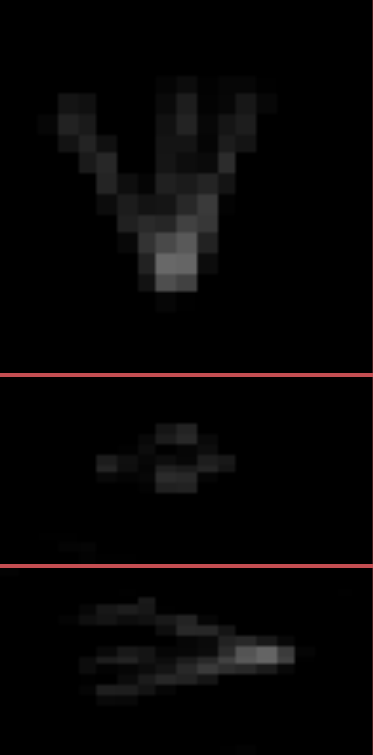}} &
\raisebox{-\totalheight/2}{\includegraphics[width=0.06\linewidth]{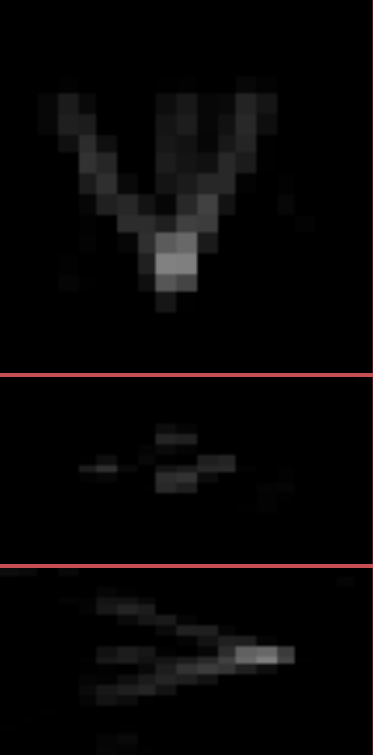}} &
\raisebox{-\totalheight/2}{\includegraphics[width=0.06\linewidth]{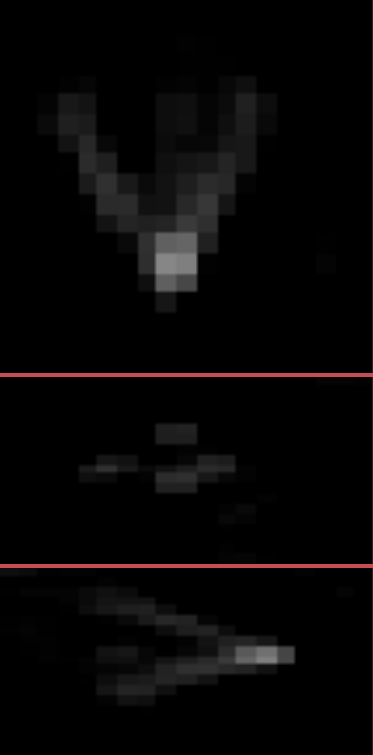}} &
\raisebox{-\totalheight/2}{\includegraphics[width=0.06\linewidth]{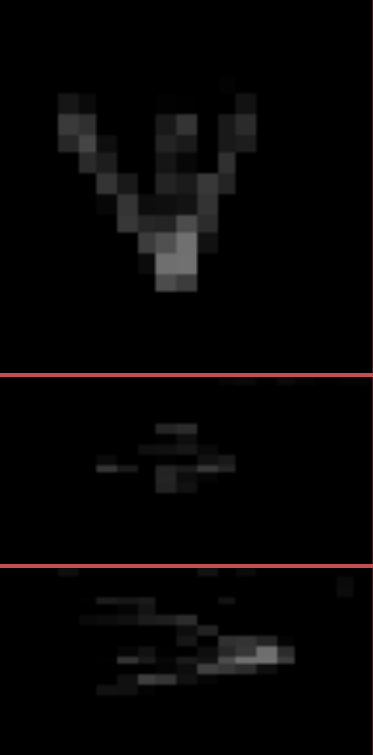}} &
\raisebox{-\totalheight/2}{\includegraphics[width=0.06\linewidth]{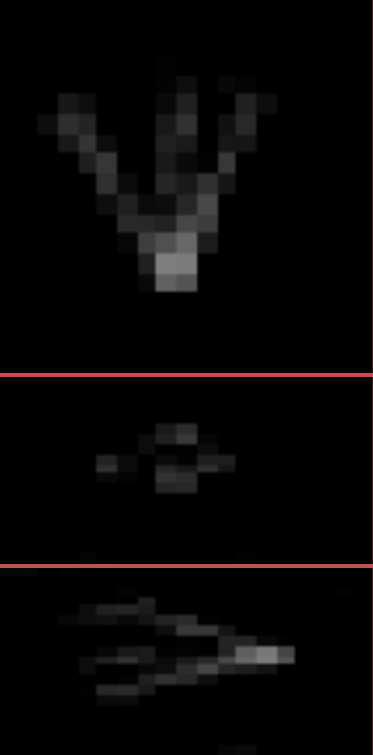}} &
\raisebox{-\totalheight/2}{\includegraphics[width=0.06\linewidth]{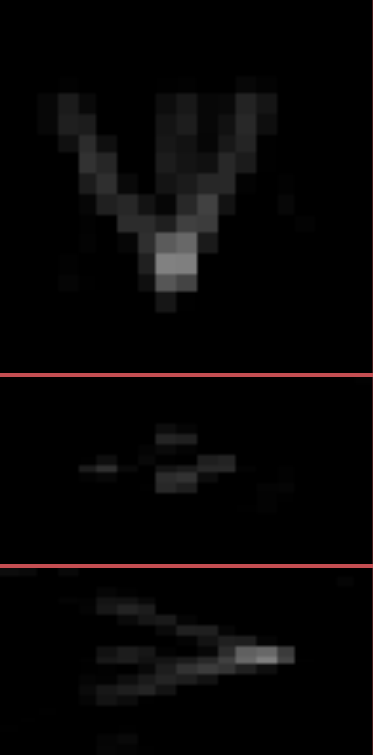}} &
\raisebox{-\totalheight/2}{\includegraphics[width=0.06\linewidth]{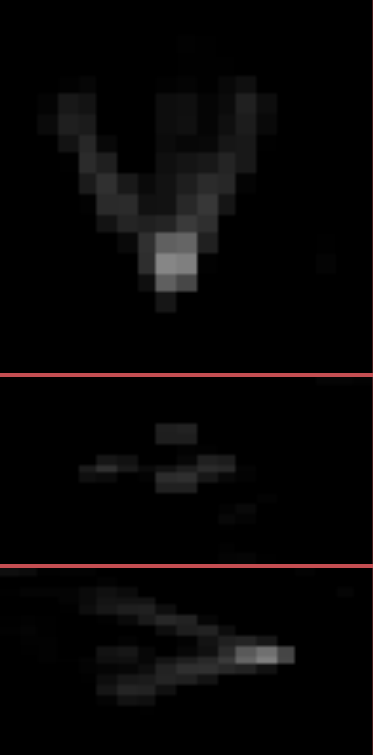}}\\

\vspace{.5mm} VAR D$\ell^1$ + P TV & 
\raisebox{-\totalheight/2}{\includegraphics[width=0.06\linewidth]{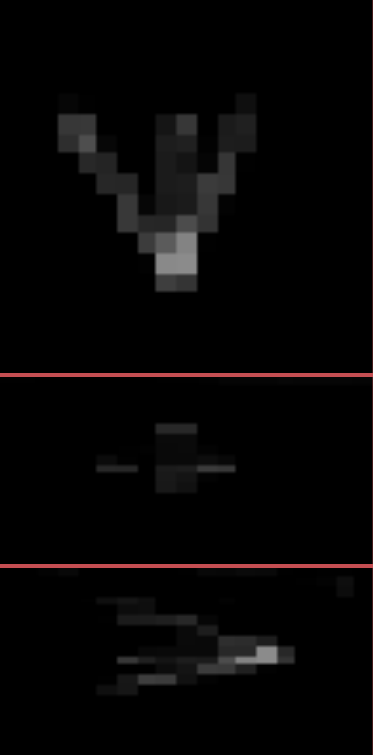}} &
\raisebox{-\totalheight/2}{\includegraphics[width=0.06\linewidth]{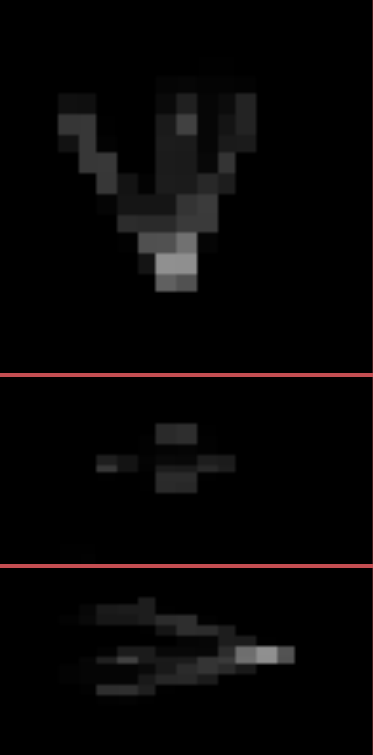}} &
\raisebox{-\totalheight/2}{\includegraphics[width=0.06\linewidth]{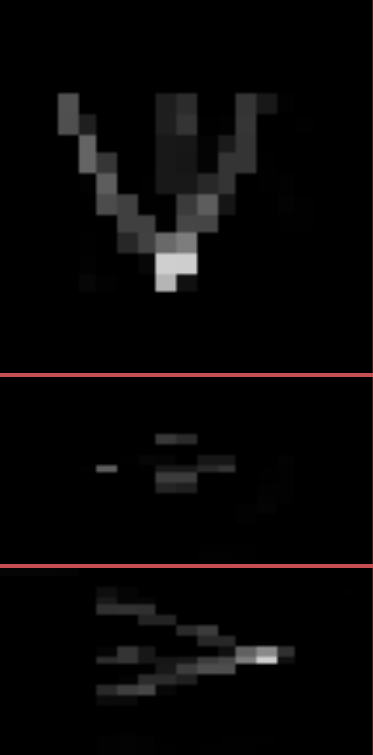}} &
\raisebox{-\totalheight/2}{\includegraphics[width=0.06\linewidth]{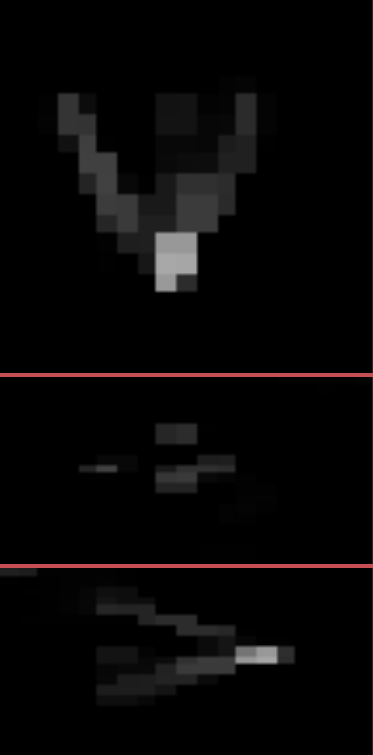}} &
\raisebox{-\totalheight/2}{\includegraphics[width=0.06\linewidth]{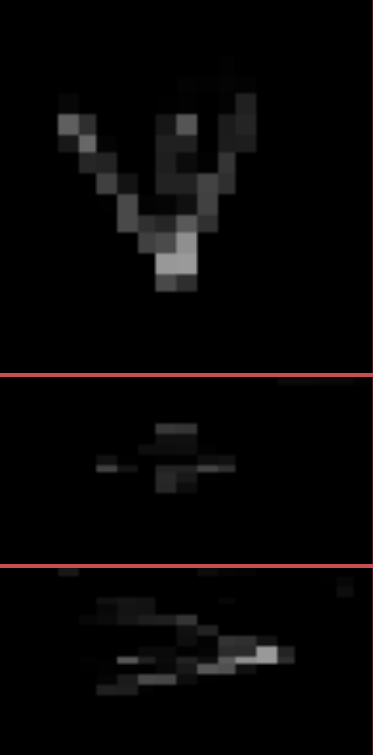}} &
\raisebox{-\totalheight/2}{\includegraphics[width=0.06\linewidth]{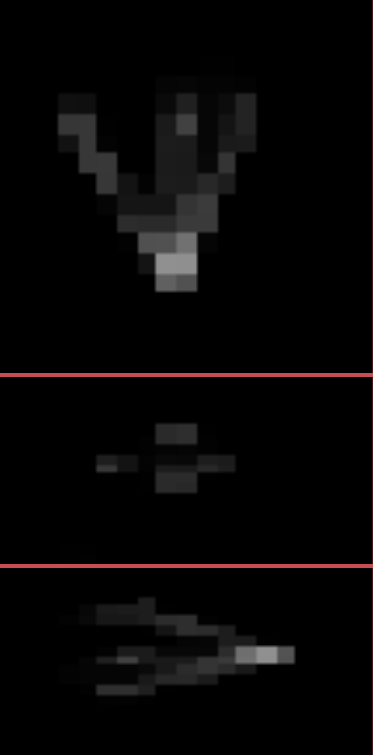}} &
\raisebox{-\totalheight/2}{\includegraphics[width=0.06\linewidth]{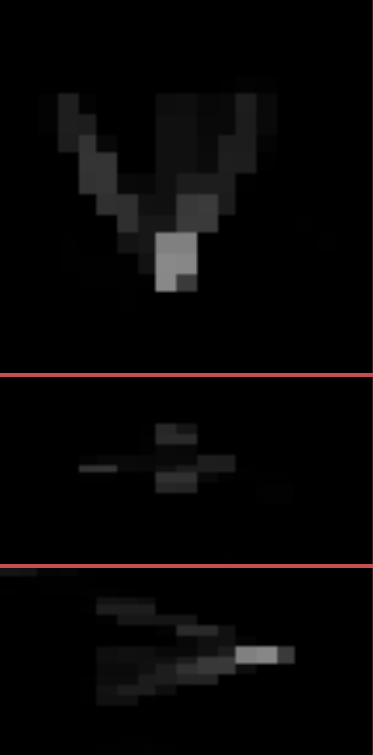}} &
\raisebox{-\totalheight/2}{\includegraphics[width=0.06\linewidth]{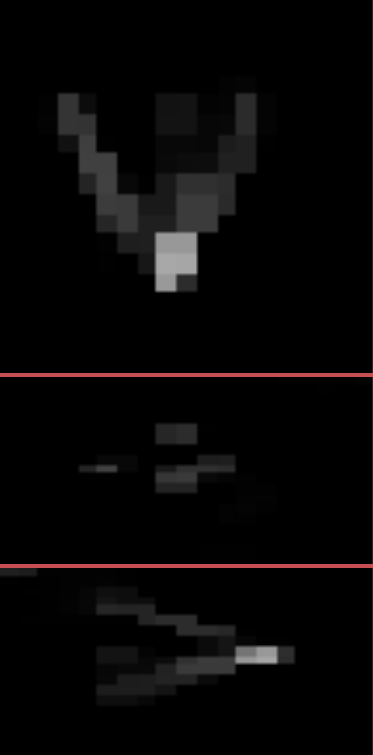}}\\

 \hline
\end{tabular}
\caption{The best reconstructions for the resolution phantom without whitening, corresponding to the values in Tables~\ref{tab:non_whitened_resolution_metrics} and~\ref{tab:non_whitened_resolution_parameters}. The color scale goes from $0$ (black) to $40$ (white). Each image is separated into three vertically stacked parts, which are separated by red lines. Each of these parts represents one of the three central slices/planes of the three-dimensional reconstruction.}
\label{tab:non_whitened_resolution_images}
\end{figure*}

\begin{table*}
\scriptsize
\centering
\rowcolors{1}{}{lightgray}
\begin{tabular}{ |l||c|c|c|c|  }
 \hline
 & $\tau=0$ & $\tau=1$ & $\tau=3$ & $\tau=5$\\
 \hline
 \rowcolor{white}
  & \multicolumn{4}{c|}{PSNRs} \\

DIP D$\ell^1$ + P-                    &$      \alpha_3$, $s=200 $&$      \alpha_4$, $s=150 $&$      \alpha_4$, $s=225 $&$      \alpha_3$, $s=350 $\\
KACZ D$\ell^2$ + P$\ell^2$            &$        \rho_{15}$, $s=5 $&$        \rho_{12}$, $s=4 $&$        \rho_{12}$, $s=4 $&$        \rho_{12}$, $s=5 $\\
KACZ D$\ell^2$ + P$\ell^1$            &$         \lambda_5$, $s=1 $&$         \lambda_5$, $s=1 $&$         \lambda_3$, $s=1 $&$         \lambda_1$, $s=1 $\\
KACZ+TSVD D$\ell^2$ + P$\ell^1$       &$  d=512$, $\lambda_6$, $s=1 $&$  d=512$, $\lambda_6$, $s=1 $&$  d=512$, $\lambda_5$, $s=1 $&$  d=512$, $\lambda_5$, $s=1 $\\
KACZ D$\ell^2$ + P$(\ell^1$+$\ell^2)$ &$   \rho_{16}$, $\lambda_6$, $s=2 $&$   \rho_{14}$, $\lambda_6$, $s=2 $&$   \rho_{13}$, $\lambda_6$, $s=2 $&$   \rho_{13}$, $\lambda_6$, $s=2 $\\
VAR D$\ell^1$ + P$\ell^1$             &$             \lambda_16 $&$              \lambda_8 $&$              \lambda_6 $&$              \lambda_7 $\\
VAR D$\ell^1$ + P$\ell^2$             &$              \rho_6 $&$              \rho_4 $&$              \rho_4 $&$              \rho_4 $\\
VAR D$\ell^1$ + P TV                  &$              \mu_8 $&$              \mu_5 $&$              \mu_7 $&$              \mu_6 $\\
VAR D$\ell^2$ + P$\ell^1$             &$             \lambda_50 $&$             \lambda_50 $&$             \lambda_11 $&$             \lambda_12 $\\
VAR D$\ell^2$ + P$\ell^2$             &$              \rho_9 $&$              \rho_8 $&$              \rho_9 $&$              \rho_9 $\\
VAR D$\ell^2$ + P TV                  &$             \mu_10 $&$              \mu_8 $&$             \mu_12 $&$             \mu_13 $\\

  & \multicolumn{4}{c|}{SSIMs} \\

DIP D$\ell^1$ + P-                    &$      \alpha_3$, $s=400 $&$      \alpha_4$, $s=250 $&$      \alpha_3$, $s=375 $&$      \alpha_3$, $s=250 $\\
KACZ D$\ell^2$ + P$\ell^2$            &$       \rho_{15}$, $s=10 $&$       \rho_{12}$, $s=11 $&$       \rho_{11}$, $s=14 $&$       \rho_{12}$, $s=12 $\\
KACZ D$\ell^2$ + P$\ell^1$            &$         \lambda_4$, $s=1 $&$         \lambda_4$, $s=1 $&$         \lambda_1$, $s=1 $&$         \lambda_1$, $s=1 $\\
KACZ+TSVD D$\ell^2$ + P$\ell^1$       &$  d=256$, $\lambda_6$, $s=2 $&$  d=256$, $\lambda_6$, $s=2 $&$  d=128$, $\lambda_7$, $s=7 $&$  d=256$, $\lambda_6$, $s=2 $\\
KACZ D$\ell^2$ + P$(\ell^1$+$\ell^2)$ &$   \rho_{16}$, $\lambda_7$, $s=6 $&$   \rho_{13}$, $\lambda_6$, $s=4 $&$   \rho_{13}$, $\lambda_6$, $s=5 $&$   \rho_{13}$, $\lambda_6$, $s=5 $\\
VAR D$\ell^1$ + P$\ell^1$             &$             \lambda_12 $&$              \lambda_6 $&$              \lambda_6 $&$              \lambda_6 $\\
VAR D$\ell^1$ + P$\ell^2$             &$              \rho_7 $&$              \rho_5 $&$              \rho_4 $&$              \rho_5 $\\
VAR D$\ell^1$ + P TV                  &$             \mu_10 $&$              \mu_7 $&$              \mu_7 $&$              \mu_6 $\\
VAR D$\ell^2$ + P$\ell^1$             &$             \lambda_50 $&$             \lambda_50 $&$             \lambda_14 $&$             \lambda_12 $\\
VAR D$\ell^2$ + P$\ell^2$             &$             \rho_10 $&$              \rho_8 $&$             \rho_10 $&$             \rho_10 $\\
VAR D$\ell^2$ + P TV                  &$             \mu_13 $&$             \mu_12 $&$             \mu_12 $&$             \mu_13 $\\
 \hline
\end{tabular}
\caption{Optimal parameters creating the results for the shape phantom with whitening as displayed in Table~\ref{tab:whitened_shape_metrics}. $\rho_i=0.5^{i-1}$, $\lambda_i,\mu_i$ analogously and $\alpha_i=10^{-i}$.}
\label{tab:whitened_shape_parameters}
\end{table*}

\begin{table*}
\scriptsize
\centering
\rowcolors{1}{}{lightgray}
\begin{tabular}{ |l||c|c|c|c|  }
 \hline
 & $\tau=0$ & $\tau=1$ & $\tau=3$ & $\tau=5$\\
 \hline
 \rowcolor{white}
  & \multicolumn{4}{c|}{PSNRs} \\

DIP D$\ell^1$ + P-                    &$      \alpha_4$, $s=1100 $&$      \alpha_3$, $s=5000 $&$     \alpha_3$, $s=11000 $&$      \alpha_4$, $s=4000 $\\
KACZ D$\ell^2$ + P$\ell^2$            &$         \rho_{17}$, $s=5 $&$        \rho_{17}$, $s=25 $&$        \rho_{16}$, $s=25 $&$        \rho_{16}$, $s=27 $\\
KACZ D$\ell^2$ + P$\ell^1$            &$          \lambda_7$, $s=1 $&$          \lambda_7$, $s=4 $&$          \lambda_5$, $s=2 $&$          \lambda_5$, $s=1 $\\
KACZ+TSVD D$\ell^2$ + P$\ell^1$       &$  d=1024$, $\lambda_7$, $s=1 $&$  d=1024$, $\lambda_7$, $s=8 $&$  d=512$, $\lambda_8$, $s=14 $&$  d=512$, $\lambda_8$, $s=10 $\\
KACZ D$\ell^2$ + P$(\ell^1$+$\ell^2)$ &$    \rho_{19}$, $\lambda_8$, $s=2 $&$   \rho_{19}$, $\lambda_9$, $s=35 $&$   \rho_{18}$, $\lambda_8$, $s=18 $&$   \rho_{18}$, $\lambda_9$, $s=24 $\\
VAR D$\ell^1$ + P$\ell^1$             &$               \lambda_8 $&$               \lambda_5 $&$               \lambda_8 $&$               \lambda_7 $\\
VAR D$\ell^1$ + P$\ell^2$             &$               \rho_5 $&$               \rho_3 $&$               \rho_2 $&$               \rho_4 $\\
VAR D$\ell^1$ + P TV                  &$              \mu_{11} $&$               \mu_8 $&$               \mu_9 $&$               \mu_9 $\\
VAR D$\ell^2$ + P$\ell^1$             &$              \lambda_{13} $&$              \lambda_{12} $&$              \lambda_{14} $&$              \lambda_{15} $\\
VAR D$\ell^2$ + P$\ell^2$             &$              \rho_{12} $&$              \rho_{11} $&$              \rho_{12} $&$              \rho_{13} $\\
VAR D$\ell^2$ + P TV                  &$              \mu_{15} $&$              \mu_{13} $&$              \mu_{16} $&$              \mu_{18} $\\

  & \multicolumn{4}{c|}{SSIMs} \\

DIP D$\ell^1$ + P-                    &$      \alpha_4$, $s=3500 $&$      \alpha_4$, $s=1900 $&$     \alpha_3$, $s=3500 $&$      \alpha_4$, $s=900 $\\
KACZ D$\ell^2$ + P$\ell^2$            &$        \rho_{18}$, $s=11 $&$        \rho_{14}$, $s=11 $&$       \rho_{14}$, $s=11 $&$       \rho_{14}$, $s=12 $\\
KACZ D$\ell^2$ + P$\ell^1$            &$          \lambda_7$, $s=1 $&$          \lambda_8$, $s=1 $&$         \lambda_6$, $s=1 $&$         \lambda_5$, $s=1 $\\
KACZ+TSVD D$\ell^2$ + P$\ell^1$       &$  d=1024$, $\lambda_8$, $s=1 $&$  d=1024$, $\lambda_8$, $s=1 $&$  d=512$, $\lambda_8$, $s=3 $&$  d=512$, $\lambda_8$, $s=3 $\\
KACZ D$\ell^2$ + P$(\ell^1$+$\ell^2)$ &$    \rho_{19}$, $\lambda_9$, $s=5 $&$    \rho_{16}$, $\lambda_8$, $s=3 $&$   \rho_{15}$, $\lambda_9$, $s=5 $&$  \rho_{15}$, $\lambda_{10}$, $s=8 $\\
VAR D$\ell^1$ + P$\ell^1$             &$               \lambda_8 $&$               \lambda_8 $&$              \lambda_7 $&$              \lambda_7 $\\
VAR D$\ell^1$ + P$\ell^2$             &$               \rho_5 $&$               \rho_3 $&$              \rho_3 $&$              \rho_3 $\\
VAR D$\ell^1$ + P TV                  &$              \mu_{11} $&$               \mu_9 $&$              \mu_8 $&$              \mu_8 $\\
VAR D$\ell^2$ + P$\ell^1$             &$              \lambda_{40} $&$              \lambda_{17} $&$             \lambda_{40} $&$             \lambda_{15} $\\
VAR D$\ell^2$ + P$\ell^2$             &$              \rho_{12} $&$              \rho_{11} $&$             \rho_{12} $&$             \rho_{12} $\\
VAR D$\ell^2$ + P TV                  &$              \mu_{15} $&$              \mu_{16} $&$             \mu_{17} $&$             \mu_{17} $\\
 \hline
\end{tabular}
\caption{Optimal parameters creating the results for the resolution phantom with whitening as displayed in Table~\ref{tab:whitened_resolution_metrics}. $\rho_i=0.5^{i-1}$, $\lambda_i,\mu_i$ analogously and $\alpha_i=10^{-i}$.}
\label{tab:whitened_resolution_parameters}
\end{table*}

\begin{table*}
\scriptsize
\centering
\rowcolors{1}{}{lightgray}
\begin{tabular}{ |l||c|c|c|c|  }
 \hline
 & $\tau=0$ & $\tau=1$ & $\tau=3$ & $\tau=5$\\
 \hline
 \rowcolor{white}
  & \multicolumn{4}{c|}{PSNRs} \\

DIP D$\ell^1$ + P-                    &$      \alpha_3$, $s=325 $&$      \alpha_4$, $s=350 $&$      \alpha_3$, $s=350 $&$      \alpha_3$, $s=325 $\\
KACZ D$\ell^2$ + P$\ell^2$            &$        \rho_{15}$, $s=6 $&$        \rho_{13}$, $s=7 $&$        \rho_{12}$, $s=5 $&$        \rho_{12}$, $s=7 $\\
KACZ D$\ell^2$ + P$\ell^1$            &$         \lambda_4$, $s=1 $&$         \lambda_4$, $s=1 $&$         \lambda_1$, $s=1 $&$         \lambda_1$, $s=1 $\\
KACZ+TSVD D$\ell^2$ + P$\ell^1$       &$  d=512$, $\lambda_5$, $s=1 $&$  d=512$, $\lambda_5$, $s=1 $&$  d=128$, $\lambda_6$, $s=4 $&$  d=128$, $\lambda_6$, $s=4 $\\
KACZ D$\ell^2$ + P$(\ell^1$+$\ell^2)$ &$   \rho_{16}$, $\lambda_6$, $s=3 $&$   \rho_{15}$, $\lambda_6$, $s=3 $&$   \rho_{13}$, $\lambda_6$, $s=3 $&$   \rho_{14}$, $\lambda_6$, $s=3 $\\
VAR D$\ell^1$ + P$\ell^1$             &$              \lambda_7 $&$              \lambda_8 $&$              \lambda_6 $&$              \lambda_7 $\\
VAR D$\ell^1$ + P$\ell^2$             &$              \rho_7 $&$              \rho_6 $&$              \rho_5 $&$              \rho_6 $\\
VAR D$\ell^1$ + P TV                  &$              \mu_8 $&$              \mu_8 $&$              \mu_7 $&$              \mu_8 $\\
VAR D$\ell^2$ + P$\ell^1$             &$              \lambda_6 $&$             \lambda_{40} $&$             \lambda_{11} $&$             \lambda_{11} $\\
VAR D$\ell^2$ + P$\ell^2$             &$              \rho_9 $&$              \rho_8 $&$             \rho_{10} $&$             \rho_{10} $\\
VAR D$\ell^2$ + P TV                  &$             \mu_{10} $&$              \mu_9 $&$             \mu_{12} $&$             \mu_{12} $\\

  & \multicolumn{4}{c|}{SSIMs} \\

DIP D$\ell^1$ + P-                    &$      \alpha_3$, $s=475 $&$      \alpha_4$, $s=400 $&$      \alpha_3$, $s=425 $&$      \alpha_3$, $s=700 $\\
KACZ D$\ell^2$ + P$\ell^2$            &$       \rho_{15}$, $s=12 $&$       \rho_{13}$, $s=15 $&$       \rho_{13}$, $s=20 $&$       \rho_{13}$, $s=20 $\\
KACZ D$\ell^2$ + P$\ell^1$            &$         \lambda_4$, $s=1 $&$         \lambda_4$, $s=1 $&$         \lambda_1$, $s=1 $&$         \lambda_1$, $s=1 $\\
KACZ+TSVD D$\ell^2$ + P$\ell^1$       &$  d=256$, $\lambda_6$, $s=2 $&$  d=256$, $\lambda_6$, $s=2 $&$  d=128$, $\lambda_7$, $s=7 $&$  d=128$, $\lambda_7$, $s=7 $\\
KACZ D$\ell^2$ + P$(\ell^1$+$\ell^2)$ &$   \rho_{16}$, $\lambda_7$, $s=8 $&$   \rho_{15}$, $\lambda_7$, $s=9 $&$  \rho_{13}$, $\lambda_7$, $s=12 $&$  \rho_{14}$, $\lambda_7$, $s=12 $\\
VAR D$\ell^1$ + P$\ell^1$             &$             \lambda_{27} $&$              \lambda_7 $&$              \lambda_6 $&$              \lambda_6 $\\
VAR D$\ell^1$ + P$\ell^2$             &$              \rho_8 $&$              \rho_6 $&$              \rho_6 $&$              \rho_6 $\\
VAR D$\ell^1$ + P TV                  &$              \mu_8 $&$              \mu_8 $&$              \mu_8 $&$              \mu_8 $\\
VAR D$\ell^2$ + P$\ell^1$             &$             \lambda_{40} $&$             \lambda_{40} $&$             \lambda_{40} $&$             \lambda_{12} $\\
VAR D$\ell^2$ + P$\ell^2$             &$             \rho_{10} $&$             \rho_{10} $&$             \rho_{12} $&$             \rho_{12} $\\
VAR D$\ell^2$ + P TV                  &$             \mu_{11} $&$             \mu_{15} $&$             \mu_{16} $&$             \mu_{15} $\\
 \hline
\end{tabular}
\caption{Optimal parameters creating the results for the shape phantom without whitening as displayed in Table~\ref{tab:non_whitened_shape_metrics}. $\rho_i=0.5^{i-1}$, $\lambda_i,\mu_i$ analogously and $\alpha_i=10^{-i}$.}
\label{tab:non_whitened_shape_parameters}
\end{table*}

\begin{table*}
\scriptsize
\centering
\rowcolors{1}{}{lightgray}
\begin{tabular}{ |l||c|c|c|c|  }
 \hline
 & $\tau=0$ & $\tau=1$ & $\tau=3$ & $\tau=5$\\
 \hline
 \rowcolor{white}
  & \multicolumn{4}{c|}{PSNRs} \\

DIP D$\ell^1$ + P-                    &$     \alpha_4$, $s=14000 $&$      \alpha_4$, $s=6000 $&$     \alpha_3$, $s=3500 $&$     \alpha_4$, $s=4000 $\\
KACZ D$\ell^2$ + P$\ell^2$            &$        \rho_{17}$, $s=16 $&$         \rho_{16}$, $s=5 $&$        \rho_{15}$, $s=6 $&$       \rho_{16}$, $s=29 $\\
KACZ D$\ell^2$ + P$\ell^1$            &$          \lambda_7$, $s=1 $&$          \lambda_7$, $s=1 $&$         \lambda_5$, $s=1 $&$         \lambda_4$, $s=1 $\\
KACZ+TSVD D$\ell^2$ + P$\ell^1$       &$  d=1024$, $\lambda_7$, $s=1 $&$  d=1024$, $\lambda_8$, $s=1 $&$  d=512$, $\lambda_7$, $s=1 $&$  d=512$, $\lambda_7$, $s=4 $\\
KACZ D$\ell^2$ + P$(\ell^1$+$\ell^2)$ &$    \rho_{19}$, $\lambda_8$, $s=2 $&$    \rho_{18}$, $\lambda_8$, $s=2 $&$   \rho_{16}$, $\lambda_8$, $s=2 $&$   \rho_{17}$, $\lambda_8$, $s=3 $\\
VAR D$\ell^1$ + P$\ell^1$             &$               \lambda_8 $&$               \lambda_7 $&$              \lambda_8 $&$              \lambda_8 $\\
VAR D$\ell^1$ + P$\ell^2$             &$               \rho_6 $&$               \rho_4 $&$              \rho_4 $&$              \rho_4 $\\
VAR D$\ell^1$ + P TV                  &$              \mu_{11} $&$              \mu_{10} $&$             \mu_{11} $&$             \mu_{10} $\\
VAR D$\ell^2$ + P$\ell^1$             &$              \lambda_{18} $&$              \lambda_{14} $&$             \lambda_{15} $&$             \lambda_{15} $\\
VAR D$\ell^2$ + P$\ell^2$             &$              \rho_{13} $&$              \rho_{12} $&$             \rho_{13} $&$             \rho_{13} $\\
VAR D$\ell^2$ + P TV                  &$              \mu_{15} $&$              \mu_{16} $&$             \mu_{16} $&$             \mu_{16} $\\

  & \multicolumn{4}{c|}{SSIMs} \\

DIP D$\ell^1$ + P-                    &$      \alpha_4$, $s=4500 $&$      \alpha_4$, $s=1600 $&$     \alpha_3$, $s=1800 $&$       \alpha_4$, $s=6000 $\\
KACZ D$\ell^2$ + P$\ell^2$            &$        \rho_{18}$, $s=12 $&$        \rho_{17}$, $s=14 $&$       \rho_{15}$, $s=14 $&$         \rho_{16}$, $s=17 $\\
KACZ D$\ell^2$ + P$\ell^1$            &$          \lambda_7$, $s=1 $&$          \lambda_7$, $s=1 $&$         \lambda_5$, $s=1 $&$           \lambda_4$, $s=1 $\\
KACZ+TSVD D$\ell^2$ + P$\ell^1$       &$  d=1024$, $\lambda_8$, $s=1 $&$  d=1024$, $\lambda_8$, $s=2 $&$  d=512$, $\lambda_8$, $s=2 $&$  d=256$, $\lambda_{10}$, $s=12 $\\
KACZ D$\ell^2$ + P$(\ell^1$+$\ell^2)$ &$    \rho_{20}$, $\lambda_8$, $s=4 $&$    \rho_{18}$, $\lambda_9$, $s=5 $&$   \rho_{17}$, $\lambda_9$, $s=6 $&$     \rho_{17}$, $\lambda_9$, $s=8 $\\
VAR D$\ell^1$ + P$\ell^1$             &$              \lambda_{17} $&$               \lambda_9 $&$              \lambda_9 $&$                \lambda_8 $\\
VAR D$\ell^1$ + P$\ell^2$             &$               \rho_6 $&$               \rho_5 $&$              \rho_4 $&$                \rho_4 $\\
VAR D$\ell^1$ + P TV                  &$              \mu_{12} $&$              \mu_{10} $&$              \mu_9 $&$               \mu_{10} $\\
VAR D$\ell^2$ + P$\ell^1$             &$              \lambda_{40} $&$              \lambda_{18} $&$             \lambda_{17} $&$               \lambda_{40} $\\
VAR D$\ell^2$ + P$\ell^2$             &$              \rho_{12} $&$              \rho_{12} $&$             \rho_{13} $&$               \rho_{14} $\\
VAR D$\ell^2$ + P TV                  &$              \mu_{14} $&$              \mu_{17} $&$             \mu_{17} $&$               \mu_{19} $\\
 \hline
\end{tabular}
\caption{Optimal parameters creating the results for the resolution phantom without whitening as displayed in Table~\ref{tab:non_whitened_resolution_metrics}. $\rho_i=0.5^{i-1}$, $\lambda_i,\mu_i$ analogously and $\alpha_i=10^{-i}$.}
\label{tab:non_whitened_resolution_parameters}
\end{table*}

\end{document}